\documentclass[10pt,final]{iopart} % Uncomment also \ioptwocol below
\usepackage{iopams}  
\usepackage{dcolumn,graphicx,color,booktabs,microtype,afterpage}
\usepackage{float}
\usepackage{amssymb}
\usepackage{nicefrac}
\usepackage{url}  % To allow also url-s with underscore
\usepackage[charter,greekuppercase=italicized]{mathdesign}
\usepackage{sidecap}
\usepackage[mathlines]{lineno}
\usepackage{cite}
\usepackage{color}
\usepackage[colorlinks,plainpages=false,linkcolor=blue,urlcolor=blue,citecolor=blue,pdfpagemode=UseNone,pdfstartview=FitBH]{hyperref}
\usepackage{gensymb}
\usepackage{microtype}
\graphicspath{{./}{figure/}}
\newcommand{\tcr}[1]{\textcolor{black}{#1}}

\usepackage[UKenglish,british]{babel}
\begin{document}
	
	\makeatletter\renewcommand{\ps@plain}{%
		\def\@evenhead{\hfill\itshape\rightmark}%
		\def\@oddhead{\itshape\leftmark\hfill}%
		\renewcommand{\@evenfoot}{\hfill\small{--~\thepage~--}\hfill}%
		\renewcommand{\@oddfoot}{\hfill\small{--~\thepage~--}\hfill}%
	}\makeatother\pagestyle{plain}

%\title
\topical[Experimental progress in Eu(Al,Ga)$_4$ topological antiferromagnets]{Experimental progress in Eu(Al,Ga)$_4$ topological antiferromagnets} 
\author{Tian\ Shang$^{1,*}$, Yang Xu$^{1}$, Shang\ Gao$^{2}$, Run\ Yang$^{3}$, Toni\ Shiroka$^{4,5}$, Ming~Shi$^{6}$}
\address{$^1$Key Laboratory of Polar Materials and Devices (MOE), School of Physics and Electronic Science, East China Normal University, Shanghai 200241, China}
\address{$^2$Department of Physics, University of Science and Technology of China, Hefei 230026, China }
\address{$^3$Key Laboratory of Quantum Materials and Devices of Ministry of Education, School of Physics, Southeast University, Nanjing 211189, China}
\address{$^4$Laboratory for Muon-Spin Spectroscopy, Paul Scherrer Institut, Villigen PSI, Switzerland} 
\address{$^5$Laboratorium f\"ur Festk\"orperphysik, ETH Z\"urich, CH-8093 Z\"urich, Switzerland}
\address{$^6$Center for Correlated Matter and School of Physics, Zhejiang University, Hangzhou 310058, China}

%\address{$^2$Laboratory for Multiscale Materials Experiments, Paul Scherrer Institut, CH-5232 Villigen PSI, Switzerland}
%\address{$^5$Laboratory for Muon-Spin Spectroscopy, Paul Scherrer Institut, CH-5232 Villigen PSI, Switzerland}
%\address{$^6$Laboratorium f\"ur Festk\"orperphysik, ETH Z\"urich, CH-8093 Zurich, Switzerland}

\eads{\mailto{tshang@phy.ecnu.edu.cn}} %\mailto{ming.shi@psi.ch}}	
%\mailto{yxu@phy.ecnu.edu.cn}, 
%\mailto {sgao@ustc.edu.cn}	

\date{\today}
	
\begin{abstract}
The non-trivial magnetic and electronic phases occurring in
topological magnets 
are often entangled, thus leading
to a variety of exotic physical properties. Recently, the BaAl$_4$-type
compounds have been extensively investigated to elucidate the topological features appearing in 
their real- and momentum spaces.
In particular, the topological Hall effect and the spin textures, typical of the
centrosymmetric Eu(Al,Ga)$_4$ family, have stimulated extensive experimental and theoretical research.
In this topical review, we discuss the latest findings regarding 
the Eu(Al,Ga)$_4$ topological antiferromagnets and related materials,  
arising from a vast array of experimental techniques.
We show that Eu(Al,Ga)$_4$ represents a suitable platform to explore the
interplay between lattice-, charge-, and spin degrees of freedom, and
associated emergent phenomena. Finally, we address some key questions open to future investigation.

\end{abstract}
\noindent{\it Keywords}: topological Hall effect, topological spin textures, magnetic skyrmions, Weyl semimetal, antiferromagnet, charge density wave, spin density wave\\

\submitto{\JPCM}

%\footnotetext{These authors contributed equally}
%\footnotetext{These authors contributed equally}
\maketitle
\ioptwocol
%

%\maketitle
	
\section{Introduction}
%
%%%%%%%%%%%%%%%%%%%%%%
In recent years, the topologically non-trivial magnetic- and
electronic structures have attracted an extraordinary attention~\cite{yan2012,lv2021,bernevig2022}. 
Introducing magnetic order into topological materials promotes exotic topological phases.
Unlike ferromagnets, the antiferromagnets offer far more
interesting cases of magnetic topological band structures.
These arise from the enormous number of possible spin configurations
and provide several interesting situations and, in principle, unforeseen
properties~\cite{Bonbien2022,Smejkal2022}.

Weyl semimetals represent one of the most interesting subclasses
of topological materials. They are characterized by linearly dispersed
electronic bands and often exhibit large Berry curvatures near the Fermi surface~\cite{yan2017,wang2017,armitage2018}. The Weyl-semimetal phase emerges only upon breaking the space inversion- or the time-reversal symmetry, the latter being realized either by applying an external magnetic field, or via intrinsic magnetic order or fluctuations~\cite{yan2017,wang2017,armitage2018,liu2019,Ma2019}.
In the magnetic topological materials, the non-trivial magnetic and electronic structures are strongly coupled, leading to a variety of exotic physical properties, such as the topological Hall effect (THE) or the anomalous Hall effect (AHE)~\cite{Xu2021,Suzuki2016,Chen2014,Nakatsuji2015,Sinova2015}, both %\tcr{of them} 
suitable for spintronic- or quantum devices~\cite{Nagaosa2010,Fert2017}. Due to their entanglement, the Weyl nodes and the associated Berry curvature can be effectively tuned by various external parameters, as e.g., % No successive "the"-s are need here. TS
the magnetic field~\cite{puphal2020,ueda2023}, chemical substitution~\cite{Fujishiro2019},
physical pressure~\cite{piva2023-1,sun2021-1}, or epitaxial strain~\cite{ohtsuki2019}. 

Topologically non-trivial spin textures are very promising for the 
next generation of logic and memory devices~\cite{Fert2017,Fert2013,Nagaosa2013}. 
Among the non-trivial spin textures, the magnetic skyrmions are the most renown ones~\cite{Muhlbauer2009,Yu2011,Yu2010,Seki2012,Kezsmarki2015,Tokunaga2015,Kurumaji2019,Hirschberger2019,Khanh2020,Li2019,Batista2016,Heinze2011,Ozawa2017,Ukleev2021}. Yet, other types of spin textures have also been discovered or proposed~\cite{Gobel2021}, as e.g., hedgehogs~\cite{Kanazawa2016,Fujishiro2019}, hopfions~\cite{Gobel2020}, merons~\cite{puphal2020}, and magnetic bubbles~\cite{Vistoli2019}.
Magnetic skyrmions are chiral spin structures with a whirling configuration,
typically in the nanoscale range.
As topologically protected phases, they cannot be
continuously deformed into other magnetic states. This robustness
against external perturbations makes them perfect for spintronic
devices~\cite{Fert2017,Fert2013,Nagaosa2013}. 
In general, all these
non-trivial spin textures can be stabilized
by the Dzyaloshinskii-Moriya interactions (DMIs), often observed
at the interfaces of thin films or in materials that lack an
inversion symmetry~\cite{Yu2011,Yu2010,Seki2012,Kezsmarki2015,Tokunaga2015,Singh2023}.
Conversely, magnetic materials with a centrosymmetric crystal structure
that still can host magnetic skyrmions are rare.
To date, only a few of such systems have been reported, including
Gd$_2$PdSi$_3$~\cite{Kurumaji2019}, Gd$_3$Ru$_4$Al$_{12}$~\cite{Hirschberger2019},
GdRu$_2$Si$_2$~\cite{Khanh2020}, Fe$_3$Sn$_2$~\cite{Li2019},
EuCd$_2$As$_2$~\cite{Xu2021,Wu2023}, Mn$_4$Ga$_2$Sn~\cite{Chakrabartty2022},
and, very recently, EuAl$_4$~\cite{Shang2021,takagi2022}. The last
compound, belonging to the BaAl$_4$ class, represents the topic of this review. 

In centrosymmetric systems, skyrmions can be stabilized, for instance,
by magnetic frustration (e.g., in Gd$_3$Ru$_{4}$Al$_{12}$, Gd$_2$PdSi$_3$,
and Fe$_3$Sn$_2$)~\cite{Hirschberger2019,Khanh2020,Li2019,Batista2016}, 
or by the competition between the magnetic interactions and magnetic
anisotropy (e.g., in GdRu$_2$Si$_2$)~\cite{Khanh2020}.
Despite adopting the same crystal structure as GdRu$_2$Si$_2$,
the magnetic anisotropy of EuAl$_4$ is only moderate. 
Consequently, in EuAl$_4$, skyrmions should be stabilized by
other mechanisms. For instance,
the magnetic dipolar interactions were
found to play a significant role in the formation of magnetic
skyrmions in Gd$_2$PdSi$_3$~\cite{Paddison2022}.
In addition, a four-spin interaction, mediated by itinerant electrons, 
has been proposed as an important ingredient for the formation of skyrmions
in centrosymmetric materials~\cite{Batista2016,Heinze2011,Ozawa2017}.  
Not long ago, the chiral magnet Co$_7$Zn$_7$Mn$_6$ was found to host a 
skyrmion phase far below the magnetic ordering temperature, where  
spin fluctuations are believed to be the key for stabilizing the
magnetic skyrmions~\cite{Ukleev2021}. EuAl$_4$ also exhibits robust spin
fluctuations against external magnetic field,  which analogously
might be crucial for the occurrence
of the topological Hall effect and of possible skyrmions
in it~\cite{Zhu2023}. 
Compared to noncentrosymmetric systems, in centrosymmetric materials
skyrmions exhibit the unique advantages of a tunable size 
and of spin helicity~\cite{Yu2014}. Nevertheless, the above
mechanisms cannot account for all the cases where skyrmions are observed
in centrosymmetric materials. Hence, their origin is not yet fully
understood and requires further investigation.  

In recent years, the Hall effect has become one of the
most used methods to investigate magnetic materials.
In general, the Hall effect involves the deflection of 
charge-carrier trajectories by the Lorentz force. In addition to an 
external magnetic field (the origin of the classical Hall effect), 
other sources can also act as an effective field, typically
associated with a non-zero Berry curvature. In magnetic materials, the
Hall resistivity $\rho_\mathrm{xy}$ can be expressed as  
$\rho_\mathrm{xy} = \rho_\mathrm{xy}^\mathrm{O} + \rho_\mathrm{xy}^\mathrm{A}$,
where $\rho_\mathrm{xy}^\mathrm{O}$ and $\rho_\mathrm{xy}^\mathrm{A}$
represent the ordinary- and the anomalous Hall resistivity, respectively.
Usually, the AHE is observed in magnetic materials with a finite
magnetization, such as ferro- or ferrimagnets. Here, it is
due to an intrinsic Karplus-Luttinger mechanism, or to extrinsic
mechanisms as the skew scattering and side jump~\cite{Nagaosa2010}.
In materials with Dirac- or Weyl points near the Fermi surface, a
finite $\rho_\mathrm{xy}^\mathrm{A}$ might arise from to\-po\-lo\-gi\-cal\-ly
non-trivial momentum-space features, a possibility attracting an
intense research interest~\cite{Xu2021,Nakatsuji2015,Suzuki2016,Onoda2004,Liang2015,Liang2017,Liang2018}. For instance, the noncollinear antiferromagnets Mn$_3$Sn, Mn$_3$Ge,
and Mn$_3$Ir~\cite{Nakatsuji2015,Ikhlas2017,Chen2014,Nayak2016}, 
the nonmagnetic- TaAs, TaP, and NbP~\cite{Caglieris2018,Watzman2018}
and the magnetic GdPtBi Weyl semimetals~\cite{Suzuki2016},
as well as the Cd$_3$As$_2$ and ZrTe$_5$ Dirac semimetals~\cite{Liang2015,Liang2017,Liang2018},
all exhibit $\rho_\mathrm{xy}^\mathrm{A}$ in a wide range of
temperatures and magnetic fields. Very recently, the AHE has been
theoretically proposed and experimentally observed in a new type of
magnetic materials, namely, the altermagnets~\cite{Smejkal2022b,Mazin2022,Gonzalez2023,Feng2022}.

The $\rho_\mathrm{xy}^\mathrm{A}$ term can be further split into two
sub-terms (owing to their different origins), %i.e.,
$\rho_\mathrm{xy} = \rho_\mathrm{xy}^\mathrm{O} + \rho_\mathrm{xy}^\mathrm{AHE} + \rho_\mathrm{xy}^\mathrm{THE}$. 
Here, $\rho_\mathrm{xy}^\mathrm{AHE}$ represents the \emph{conventional}
anomalous Hall term, mostly determined by the electrical resistivity
and magnetization. The last term represents the \emph{topology-} (THE)
induced contribution. The topological Hall effect has been frequently
used to search for and study the chiral spin textures in various materials. 
As opposed to the intrinsic AHE, which depends on the Berry curvature
in \emph{momentum} space, the THE is related to a Berry-phase accumulation in 
\emph{real} space due to chiral spin textures~\cite{Nagaosa2013}.
When passing through chiral spin textures, such as skyrmions, charge
carriers pick up an additional Berry phase and experience a local
emergent magnetic field $B_\mathrm{eff}$. The latter is proportional
to the scalar spin chirality $\chi_\mathrm{ijk} = \boldsymbol{S_\mathrm{i}} \cdot (\boldsymbol{S_\mathrm{j}} \times \boldsymbol{S_\mathrm{k}}$), with $\chi_\mathrm{ijk}$ the solid angle spanned by neighboring
spins $S_\mathrm{i}$, $S_\mathrm{j}$, and $S_\mathrm{k}$, which becomes
non-zero for non-coplanar spin textures. The transverse deflection of
charge carriers interacting with $B_\mathrm{eff}$ results in the topological
Hall resistivity $\rho_\mathrm{xy}^\mathrm{THE}$, whose amplitude is given
by $\rho_\mathrm{xy}^\mathrm{THE} = P \cdot R_0 \cdot (n_\mathrm{sk} \cdot \Phi_0)$~\cite{Neubauer2009,Raju2019}. Here, $P$ is the spin polarization of %the
charge carriers, $R_0$ is the Hall coefficient representing the effective charge density
contributing to the THE (usually taken as the classical Hall coefficient),
$B_\mathrm{eff} = n_\mathrm{sk} \cdot \Phi_0$ is the emergent field
associated with a given skyrmion density $n_\mathrm{sk}$, and $\Phi_0$ is the magnetic flux quantum.

Many other techniques have been successfully used to
visualize and study 
the chiral spin textures, e.g., neutron scattering~\cite{Tokunaga2015,Singh2023,takagi2022}, magnetic resonant elastic x-ray scattering~\cite{Kurumaji2019}, magnetic force microscopy~\cite{Raju2019}, magnetic exchange force microscopy~\cite{Grenz2017}, magnetic transmission x-ray microscopy~\cite{Woo2016,Soumyanarayanan2017}, Lorentz transmission electron microscopy~\cite{Yu2010,Chakrabartty2022}, spin-polarized low-energy electron microscopy~\cite{Chen2013}, and scanning tunneling microscropy~\cite{Romming2013}.
Even though such techniques can 
probe the topological spin textures, they cannot be used as information-reading
tools in real devices. By contrast, the plain detection of THE is simple and can be easily realized in real applications,
with big advantages compared to the large-scale- 
or microscopic techniques. Unfortunately, most
magnetic materials with chiral spin textures exhibit only
weak topological transport signals, i.e., $\rho_\mathrm{xy}^\mathrm{THE} \lesssim 1$\,{\textmu}{\textohm}cm.
Clearly, increasing the magnitude of $\rho_\mathrm{xy}^\mathrm{THE}$ 
remains the key to viable topological spintronic devices 
and the most important challenge so far. 

Exploring new magnetic skyrmion materials, especially those with a
centrosymmetric crystal structure, is 
a hot topic which drives 
the ongoing research on
topological spintronic devices. The discovery of a non-trivial band
topology and large magnetoresistance (MR) in the prototype compound
BaAl$_4$~\cite{wang2021}, as well as of THE in its magnetic
counterpart EuAl$_4$~\cite{Shang2021}, have stimulated intense interest
in this family of materials. The $A$Al$_4$ and $A$Ga$_4$ ($A$ = Ca, Sr, Ba, and Eu)
binary compounds exhibit a variety of physical properties, attributed to
the presence of structural-, charge-, and magnetic order phase
transitions. Similar to BaAl$_4$, also BaGa$_4$ exhibits metallic behavior
without undergoing any phase transitions, while SrAl$_4$ shows a
charge-density-wave (CDW) and a structural phase transition at $T_\mathrm{CDW} \sim 250$\,K
and $T_\mathrm{S} \sim 90$\,K, respectively~\cite{Nakamura2016}. Both
compounds are also expected to exhibit topological features, requiring
further investigations.
By replacing Ba (or Sr) with Eu, the $4f$ electrons
bring new intriguing aspects to the topology, as clearly illustrated by
recent work on EuAl$_4$ and EuGa$_4$~\cite{Shang2021,takagi2022,lei2023}.
EuAl$_4$ exhibits coexisting antiferromagnetic- (AFM) and CDW orders 
with onset temperatures $T_\mathrm{N} \sim 15.6$\,K and $T_\mathrm{CDW} \sim 140$\,K~\cite{Shang2021,Araki2013,Nakamura2014,Nakamura2015,Shimomura2019,Kobata2016} 
and undergoes a series of metamagnetic transitions in the AFM state~\cite{Shang2021,Nakamura2015}. 
Topological Hall resistivity was observed in the $\sim$1--2.5\,T
field range~\cite{Shang2021} and attributed to the formation of
magnetic skyrmions~\cite{takagi2022}. Hence, EuAl$_4$ represents a rare
case, where magnetic skyrmions not only arise in a centrosymmetric
structure, but also coexist with CDW order. EuGa$_4$ is also an
antiferromagnet (with $T_\mathrm{N} \sim 16.5$\,K), whose CDW order 
appears only under applied pressure (close to 0.75\,GPa), with $T_\mathrm{CDW}$ reaching $\sim 175$\,K 
at 2.3\,GPa~\cite{Nakamura2014,Nakamura2015,Nakamura2013}. It also exhibits
THE in the $\sim$5--7\,T field range in the AFM state, most likely also
related to topological spin textures (awaiting
experimental confirmation). Recently, EuGa$_4$ was confirmed
to be a magnetic Weyl nodal-ring semimetal, where the line nodes
form closed rings near the Fermi level~\cite{lei2023}. 

%
%==== figure =============================%	
\begin{figure*}[!tbh]
	%\begin{center}
	\includegraphics[width=0.80\textwidth]{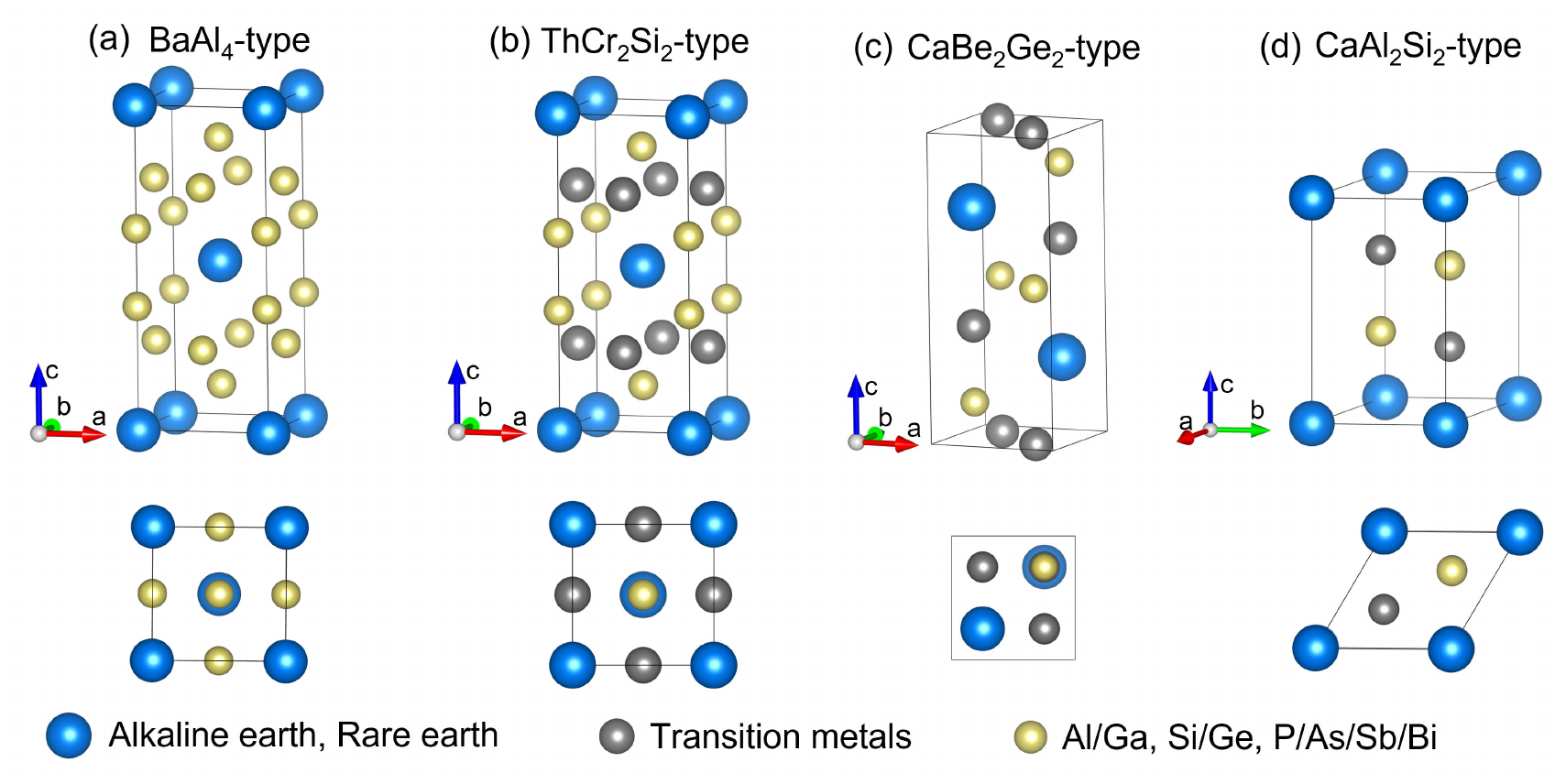}
	%\end{center}
	\centering
	\vspace{-2ex}%
	\caption{\label{fig:structure}Summary of crystal structures related to
	the alkaline-earth and rare-earth-based  $A$$T$$_2$$X$$_2$ materials:
	(a) BaAl$_4$-type, (b) ThCr$_2$Si$_2$-type, (c) CaBe$_2$Ge$_2$-type,
	and (d) CaAl$_2$Si$_2$-type structure. At the bottom of each panel,
	we depict the crystal structures viewed along the $c$-axis. The
	coordinates of atoms in each crystal structure are listed in
	table~\ref{tab:coordinates}.
	}
\end{figure*}
%=== end figure ==========================%
%

Eu(Al,Ga)$_4$ antiferromagnets represent one of the rare 
material classes to exhibit exotic physical properties originating
from topological aspects in both the real- and momentum space.
This topical review focuses mostly on recent experimental
investigations of the Eu(Al,Ga)$_4$ topological antiferromagnets and related materials. 
After this short introduction, in section~2, we briefly discuss
the crystal structures and CDW order (including its possible mechanisms)
in the $A$(Al,Ga)$_4$ family of materials. Section~3 describes the
magnetic and transport properties of Eu(Al,Ga)$_4$ single crystals and
the resulting magnetic phase diagrams.
The topological Hall effect is also discussed in this section. 
In section~4, we summarize the neutron and synchrotron resonant
x-ray scattering (RXS) results, aimed at determining the magnetic structures in the AFM state, 
in particular, of magnetic skyrmions in EuAl$_4$. In section~5,
we focus on the muon-spin rotation and relaxation ({\textmu}SR) study, 
in both the AFM and paramagnetic (PM) states of Eu(Al,Ga)$_4$ single
crystals, where strong and robust spin fluctuations have been found. 
In section~6, we review the current nuclear magnetic resonance (NMR)
studies, able to detect both the magnetic and charge order.  
In the following section~7, the magneto-optical spectroscopy studies on
Eu(Al,Ga)$_4$ single crystals are discussed, to show the effects
induced by CDW order. Before the final section, we discuss the electronic
band topology investigated by the angle-resolved photoemission spectroscopy
(ARPES) and quantum oscillations (QO). Finally, we conclude by outlining
some possible future directions in this interesting field of research.

\section{Crystal structure and charge-density-wave order\label{sec:structure}} 
\subsection{Crystal structure}

There are mainly four different crystal structures that are related to the compounds with $A$$T$$_2$$X$$_2$ stoichiometry. These structures exhibit pronounced chemical flexibility, the $A$ site can be occupied by alkali metals, alkaline earth and rare-earth metals, first-column transition metals, actinide elements (e.g., U, Th, Pu); the $T$ site can be occupied mostly by transition metals and $p$-block elements; and the $X$ site can be occupied by $p$-block elements and some transition metals.
In general, the $A$$T$$_2$$X$$_2$ compounds adopt mainly four different crystal structures (see figure~\ref{fig:structure}). The BaAl$_4$-, ThCr$_2$Si$_2$-, and CaBe$_2$Ge$_2$-type crystal structures are tetragonal, while the CaAl$_2$Si$_2$-type structure is trigonal. 
The BaAl$_4$- and ThCr$_2$Si$_2$-type structures share the
high-symmetry space group $I4/mmm$ (No.~139). When both Wyckoff sites 4$d$ (0, 0.5, 0.25) and 4$e$ (0, 0, $z$) are occupied by the same atoms (see details in table~\ref{tab:coordinates}), the compounds adopt a BaAl$_4$-type crystal structure [see figure~\ref{fig:structure}(a)], e.g., $A$Al$_4$ and $A$Ga$_4$ ($A$ = Ca, Sr, Ba, and Eu)~\cite{Zhang2013,Bruzzone1965,Bruzzone1975,Zogg1979,Mooij1985,Tobash2005,Zevalkink2017}. Note that, close to 443\,K, CaAl$_4$ was
found to undergo a structural phase transition, from tetragonal- to monoclinic (space group $C$2/m, No.~12)~\cite{Miller1993}. In the EuAl$_{4-x}$Ga$_x$ and EuAl$_{4-x}$Zn$_x$ series of compounds, the Zn or Ga atoms replace the Al atoms both at the 4$d$ and 4$e$ sites~\cite{Verbovytskyy2011,Stavinoha2018}. While in $A$Al$_{4-x}$Si$_x$ or $A$Al$_{4-x}$Ge$_x$ ($A$ = Sr, Ba, Eu) compounds, the Si or Ge atoms prefer to occupy only the 4$e$ site~\cite{Zhang2013,Tobash2006}. These chemical substitutions can effectively tune the magnetic interactions in EuAl$_4$, yielding rich magnetic properties. For instance, the magnetic transition temperature increases from 16\,K in EuAl$_4$ to 25\,K in EuAl$_{2.75}$Zn$_{1.25}$~\cite{Shang2021,Verbovytskyy2011}. The magnetic properties are
discussed in detail in the following section.

\begin{table}
	\caption{\label{tab:coordinates}Atomic coordinates and site occupancy
	factors (SOF) of the four
	crystal structures shown in figure~\ref{fig:structure}.
	We took EuAl$_4$, EuZn$_2$Si$_2$, EuZn$_2$Sn$_2$, and EuAl$_2$Si$_2$
	as typical examples of the BaAl$_4$-, ThCr$_2$Si$_2$-, CaBe$_2$Ge$_2$-,
	and CaAl$_2$Si$_2$-type structures, respectively. Data %were
	taken from Refs.~\cite{Zhang2013,Grytsiv2002,Dhar2009,Schobinger1989}.}
	\vspace{6pt}
	\centering
	\begin{indented}
		\item[]\begin{tabular}{l*{5}{c}}
			\br
			\multicolumn{6}{l}{Structure: BaAl$_4$-type \quad Space group: $I4/mmm$, No.~139} \\ 
			\mr
			Atom & Wyckoff & $x$ & $y$ & $z$ & SOF \\
			\mr
			Eu1           & 2$a$  & 0.00000  & 0.00000  & 0.00000 	& 1 \\ %\rule{0pt}{2.6ex} \\
			Al1           & 4$d$  & 0.00000  & 0.50000  & 0.25000 	& 1 \\
			Al2           & 4$e$  & 0.00000  & 0.00000  & 0.38507 	& 1 \\
			\br
		\end{tabular}
	\end{indented}

	\vspace{6pt}
	\centering
	\begin{indented}
		\item[]\begin{tabular}{lccccc}
	     	\br
			\multicolumn{6}{l}{Structure: ThCr$_2$Si$_2$-type \quad Space group: $I4/mmm$, No.~139} \\ 
			\mr
			Atom & Wyckoff & $x$ & $y$ & $z$ & SOF \\
			\mr
			Eu1           & 2$c$  & 0.00000  & 0.00000  & 0.00000 	& 1 \\ %\rule{0pt}{2.6ex} \\
			Zn1           & 4$d$  & 0.00000  & 0.50000  & 0.25000 	& 1 \\
			Si1           & 4$e$  & 0.00000  & 0.00000  & 0.38560 	& 1 \\
			\br
		\end{tabular}
	\end{indented}
	
	\vspace{6pt}
	\centering
	\begin{indented}
		\item[]\begin{tabular}{lccccc}
		    \br
			\multicolumn{6}{l}{Structure: CaBe$_2$Ge$_2$-type \;\; Space group: $P4/nmm$, No.~129} \\
			\mr
			Atom & Wyckoff & $x$ & $y$ & $z$ & SOF \\
			\mr
			Eu1           & 2$c$  & 0.25000  & 0.25000  & 0.75271	& 1 \\ %\rule{0pt}{2.6ex} \\
			Zn1           & 2$c$  & 0.25000  & 0.25000  & 0.36280	& 1 \\
			Zn2           & 2$a$  & 0.75000  & 0.25000  & 0.00000 	& 1 \\
			Sn1           & 2$c$  & 0.25000  & 0.25000  & 0.13526	& 1 \\
			Sn2           & 2$b$  & 0.75000  & 0.25000  & 0.50000 	& 1 \\
			\br
		\end{tabular}
	\end{indented}
	
	\vspace{6pt}
	\centering
	\begin{indented}
		\item[]\begin{tabular}{lccccc}
			\br
			\multicolumn{6}{l}{Structure: CaAl$_2$Si$_2$-type \quad Space group: $P\overline{3}m1$, No.~164} \\ 
			\mr
			Atom & Wyckoff & $x$ & $y$ & $z$ & SOF \\
			\mr
			Eu1           & 1$a$  & 0.00000  & 0.00000  & 0.00000 	& 1 \\ %\rule{0pt}{2.6ex} \\
			Al1           & 2$d$  & 0.33330  & 0.66670  & 0.63000 	& 1 \\
			Si1           & 2$d$  & 0.33330  & 0.66670  & 0.26800 	& 1 \\
			\br
		\end{tabular}
	\end{indented} 
\end{table}

As shown in figure~\ref{fig:structure}(b), the ThCr$_2$Si$_2$-type
structure represents a ternary variant of the BaAl$_4$-type structure,
obtained by replacing two crystallographically unique Al sites (4$d$ and 4$e$),
which have distinct coordination environments for different atoms,
e.g., Cr and Si (see details in table~\ref{tab:coordinates}).
The ThCr$_2$Si$_2$-type structure represents one of the most ubiquitous
structural arrangements for ternary combinations of elements. It can also be viewed as a layered structure, where the Th-layer and CrSi
blocks stack along the crystallographic $c$-axis. The ThCr$_2$Si$_2$-type
$A$$T_2$$X_2$ compounds exhibit a rich variety of electronic properties,
including topological magnetic order, heavy-fermion features,
quantum phase transition, unconventional and high-$T_c$ superconductivity (SC),
as discussed in many detailed reviews~\cite{Lai2022,Pfleiderer2009,Brando2016,Lohneysen2007,Stewart2011,Shatruk2019,Szytula1989}.
As shown in figure~\ref{fig:structure}(c), the CaBe$_2$Ge$_2$-type structure
($P$4/$nmm$, No.~129) is also an ordered ternary variant of the
BaAl$_4$ prototype. Compared with ThCr$_2$Si$_2$-type, CaBe$_2$Ge$_2$
has a much lower symmetry, as clearly shown in
the bottom panels,
where the crystal structures viewed along the $c$-axis are presented.
Although there are fewer CaBe$_2$Ge$_2$-type than ThCr$_2$Si$_2$-type compounds,
these, too, exhibit a variety of interesting physical properties.
The CaAl$_2$Si$_2$-type structure ($P\overline{3}m1$, No.~164) is plotted in
figure~\ref{fig:structure}(d), which can be viewed also as a layered structure.
Although not a variant of the prototype BaAl$_4$ structure, the
CaAl$_2$Si$_2$-type compounds are the ideal platform for investigating
novel phenomena. For instance, in EuCd$_2$As$_2$, strong magnetic
fluctuations have been found far above the magnetic ordering temperature,
which leads to a Weyl-semimetal state~\cite{Ma2019}.  
Unconventional contributions to the anomalous Hall and anomalous Nernst
effects were observed both above and below the magnetic transition
temperature of EuCd$_2$As$_2$ and EuZn$_2$As$_2$~\cite{Xu2021,Yi2023},
indicating the existence of significant Berry curvatures. A colossal
magnetoresistance (MR) due to the suppression of magnetic fluctuations has
been reported in EuCd$_2$P$_2$~\cite{Wang2021b}. It is worth mentioning
that the BaNiSn$_3$-type structure ($I4mm$, No.~107) is also an
ordered ternary variant of the BaAl$_4$ prototype structure, which
lacks an inversion center and has the lowest symmetry among these
crystal structures. 

\begin{figure}[!tb]
	%\begin{center}
	\includegraphics[width=0.75\linewidth]{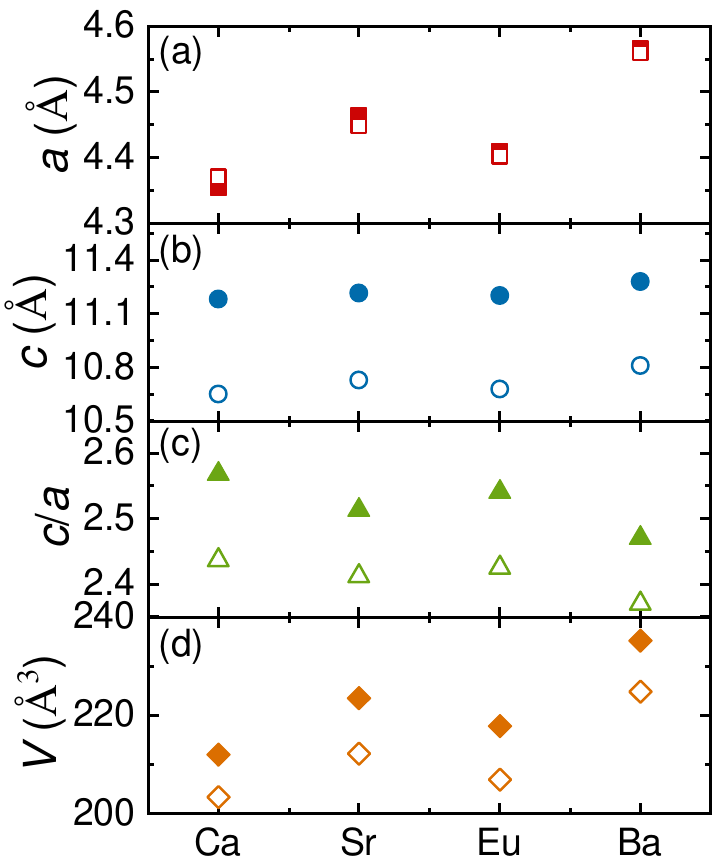}
	%\end{center}
	\centering
	\vspace{-1ex}%
	\caption{\label{fig:lattice}Lattice parameters of $A$Al$_4$ (solid symbols) and $A$Ga$_4$ (open symbols) compounds, where $A$ = Ca, Sr, Eu, and Ba. Data taken from Refs.~\cite{Zhang2013,Bruzzone1965,Bruzzone1975,Zogg1979,Mooij1985,Tobash2005,Zevalkink2017}. Note that, close to 443\,K, CaAl$_4$ undergoes a structural phase transition from tetragonal- to monoclinic (space group $C$2/$m$)~\cite{Miller1993}. Here the lattice parameters of tetragonal Ca(Al,Ga)$_4$ are presented.}
\end{figure}

It should be pointed out that, in this topical review, we focus
mostly on the $A$(Al,Ga)$_4$ ($A$ = Ca, Sr, Eu, Ba) compounds with BaAl$_4$-type structure.
Compounds with other ordered ternary variants of the BaAl$_4$ prototype
(i.e., the ThCir$_2$Si$_2$-, CaBe$_2$Ge$_2$-, and BaNiSn$_3$-type structures),
or with the CaAl$_2$Si$_2$-type structure will not be discussed further here. 
Figure~\ref{fig:lattice} summarizes the lattice parameters of the
$A$(Al,Ga)$_4$ compounds. The ionic radius increases from $\sim$1\,\AA{}
of Ca$^{2+}$ to $\sim$1.35\,\AA{} of the Ba$^{2+}$ ions,
resulting in an expansion of the unit-cell volume
[see figure~\ref{fig:lattice}(d)].
Thus, when moving from Ca(Al,Ga)$_4$ to Ba(Al,Ga)$_4$, the $a$- and $c$-axes
increase, while the $c$/$a$ ratio decreases slowly. Note also that
the $a$-axes are almost identical for $A$Al$_4$ and $A$Ga$_4$, while the
$c$-axes of $A$Al$_4$ are much larger than those of $A$Ga$_4$
[figure~\ref{fig:lattice}(a)-(b)]. The decrease of the $c$-axis
lattice constant in $A$Ga$_4$ is most likely due to the suppression of
the Al-Ga(Al)-Al bond angle on increasing the Ga-content~\cite{Stavinoha2018}.
As a consequence, the electronic properties of $A$(Al,Ga)$_4$ can be
tuned by the Al/Ga substitution.

%
%==== figure =============================%
%=== end figure =======================
%
%
%
\subsection{Charge-density-wave order}
\begin{figure}[!tb]
	%\begin{center}
	\includegraphics[width=0.85\linewidth]{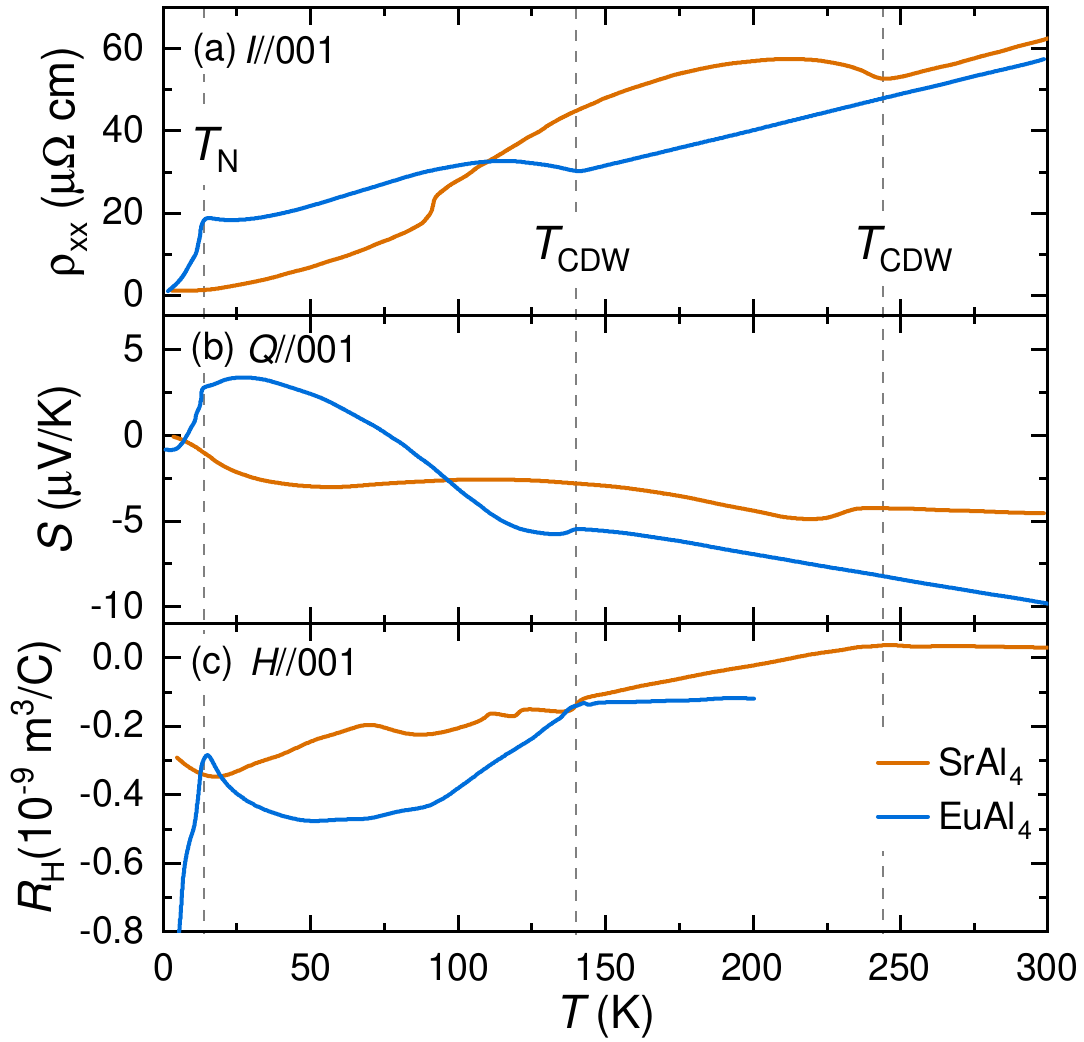}
	%\end{center}
	\centering
	\vspace{-2ex}%
	\caption{\label{fig:rho}Temperature dependence of the electrical resistivity $\rho_\mathrm{xx}(T)$ (a),  thermoelectric power $S(T)$ (b),  and Hall coefficient $R_\mathrm{H}(T)$ (c) for SrAl$_4$ and EuAl$_4$, respectively. The dashed lines mark the charge-density-wave- and antiferromagnetic transitions. Data were taken from Refs.~\cite{Araki2013,Nakamura2015,Nakamura2016}.}
\end{figure}

The SrAl$_4$ and EuAl$_4$ topological semimetals have been found to
exhibit a CDW transition in a wide temperature range.
Conversely, CaAl$_4$, BaAl$_4$, as well as $A$Ga$_4$ ($A$ = Ca, Sr, Eu, Ba), 
exhibit a typical metallic behavior, without undergoing any
CDW transition below room temperature~\cite{Nakamura2016}. The CDW order
in SrAl$_4$ and EuAl$_4$, with $T_\mathrm{CDW} \sim 243$ and
140\,K, respectively, is clearly reflected in their temperature-dependent
electrical resistivity $\rho_\mathrm{xx}(T)$ [see figure~\ref{fig:rho}(a)]. 
The distinct anomalies at $T_\mathrm{CDW}$ and  $T_\mathrm{N}$ are also
reflected in the temperature-dependent thermoelectric power $S(T)$ and
Hall coefficient $R_\mathrm{H}(T)$ [see figure~\ref{fig:rho}(b)-(c)].
The enhanced resistivity due to a partial gap opening at
the Fermi surface below $T_\mathrm{CDW}$ is a typical feature of the
CDW materials. The CDW order also leads to a decrease in the carrier
density that is proportional to 1/$R_\mathrm{H}$. In SrAl$_4$,
there is also another anomaly in the resistivity at $T_\mathrm{S} \sim 90$\,K,
showing a significant hysteresis [figure~\ref{fig:rho}(a)]~\cite{Nakamura2016}.
Such a first-order transition is most likely attributed to a change
of the crystal structure, from tetragonal ($I$4/$mmm$) to monoclinic ($C$2/$m$),
as observed also in CaAl$_4$ at $T_\mathrm{S} \sim 443$\,K~\cite{Miller1993}.
It could be interesting to search for
possible critical phenomena related to such a structural phase transition
in SrAl$_{4-x}$Ga$_x$ or Sr$_{1-x}$Ba$_x$Al$_4$. 
In fact, substituting Al with Si suppresses the structural
phase transition at $T_\mathrm{S}$, allowing superconductivity to emerge,
with a $T_c$ of 2.1 and 2.6\,K in SrAl$_{1.5}$Si$_{0.5}$ and SrAl$_2$Si$_2$,
respectively~\cite{Zevalkink2017}. Differently from SrAl$_4$,
EuAl$_4$ exhibits an additional antiferromagnetic order at $T_\mathrm{N} \sim 15.6$\,K,
and undergoes a series of metamagnetic transitions in the AFM state
in an applied magnetic field [figure~\ref{fig:rho}(a)]~\cite{Shang2021,Nakamura2015}.

%==== figure =============================%
%=== end figure =======================

%
%==== figure =============================%
\begin{figure}[!htp]
	%\begin{center}
	\includegraphics[width=0.9\linewidth]{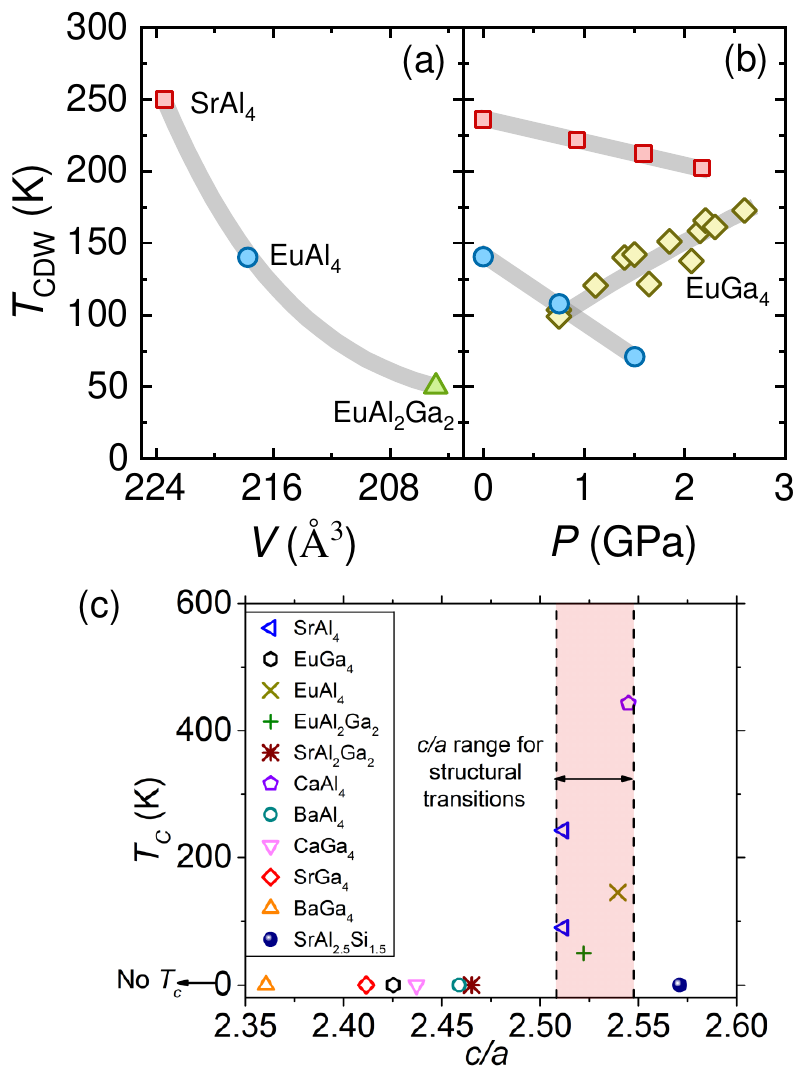}
	%\end{center}
	\centering
	\vspace{-2ex}%
	\caption{\label{fig:TCDW}(a) Summary of charge-density-wave ordering temperature $T_\mathrm{CDW}$ versus unit-cell volume for $A$(Al,Ga)$_4$. (b) The pressure-dependent $T_\mathrm{CDW}$ for SrAl$_4$, EuAl$_4$, and EuGa$_4$, respectively.
		Data were taken from Refs.~\cite{Nakamura2016,Nakamura2015,Stavinoha2018}. (c) $T_\mathrm{CDW}$ vs. $c$/$a$ ratio for $A$(Al,Ga)$_4$, which was reproduced from Ref.~\cite{Ramakrishnan2023}.}
\end{figure}
%=== end figure =======================
%
 
Figure~\ref{fig:TCDW} summarizes the CDW ordering temperature $T_\mathrm{CDW}$
versus the unit cell volume and external physical pressure. As the
unit cell volume decreases, the $T_\mathrm{CDW}$ decreases as well.
The substitution of Al with Ga further decreases the unit cell volume
[see details in figure~\ref{fig:lattice}(c)] and, in EuAl$_2$Ga$_2$,
$T_\mathrm{CDW}$ is suppressed down to 50\,K~\cite{Stavinoha2018}.
Upon further increasing the Ga-content, the CDW order disappears.
In figure~\ref{fig:TCDW}(a), BaAl$_4$ with $V$ = 235~\AA{} (not shown)
is located at the left of SrAl$_4$. Although no CDW transition
has been detected in BaAl$_4$ below room temperature~\cite{Nakamura2016},
according to the $T_\mathrm{CDW}(V)$ plot, if BaAl$_4$ has a CDW transition, its $T_\mathrm{CDW}$ might be
above 400\,K, a circumstance requiring further investigation. %The
$T_\mathrm{CDW}$ is also very sensitive to external physical pressure,
being linearly suppressed by it,
reaching 70\,K at $1.5$\,GPa in EuAl$_4$ and 200\,K at $2.2$\,GPa in SrAl$_4$, respectively~\cite{Nakamura2016}. 
In EuAl$_4$, the CDW transition becomes invisible when the external pressure
exceeds 2.2\,GPa~\cite{Nakamura2015}. Unlike the $A$Al$_4$ compounds, EuGa$_4$ shows an opposite
pressure-dependent $T_\mathrm{CDW}$. At ambient pressure, EuGa$_4$
undergoes only an AFM transition at $T_\mathrm{N} \sim 16$\,K~\cite{Nakamura2016,Nakamura2015,Stavinoha2018}. 
At $P =  0.75$\,GPa, however, also a CDW order appears at
$T_\mathrm{CDW} \sim 104$\,K [figure~\ref{fig:TCDW}(b)]~\cite{Nakamura2015}.
Upon a further increase of pressure, the CDW order moves to
higher temperatures, with $T_\mathrm{CDW}$ reaching $\sim$172\,K at $P$ = 2.6\,GPa. 
Interestingly, a CDW seems to occur only for $2.51 < c/a < 2.54$
[figure~\ref{fig:TCDW}(c)].  
It is not yet clear why EuGa$_4$ exhibits a different pressure-dependent
$T_\mathrm{CDW}$ compared with SrAl$_4$ and EuAl$_4$. Although the mechanism
of CDW in $A$(Al,Ga)$_4$ is not yet fully understood, 
a clear discrepancy in the Fermi surfaces of EuGa$_4$ and EuAl$_4$,
might account for their different pressure-dependent
$T_\mathrm{CDW}$s~\cite{lei2023,Nakamura2014,Kobata2016,Nakamura2013}.

%
%==== figure =============================%
\begin{figure}[!htp]
	%\begin{center}
	\includegraphics[width = 0.99\linewidth]{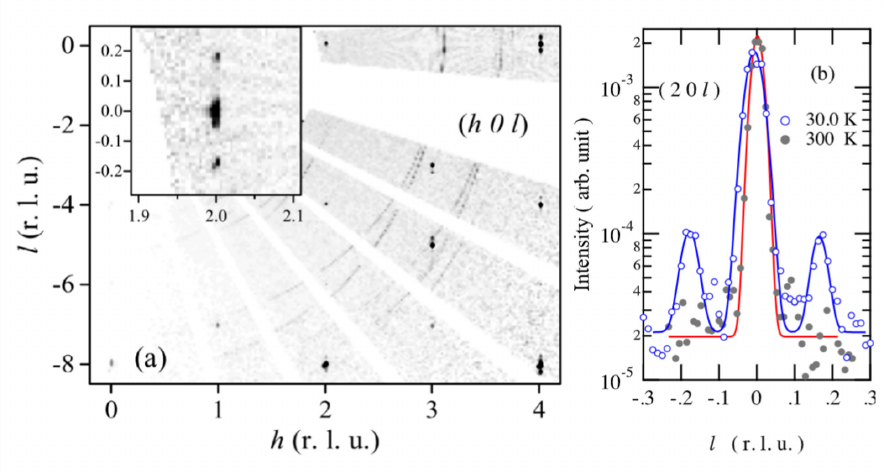}
	%\end{center}
	\centering
	\vspace{-2ex}%
	\caption{\label{fig:CDW}(a) Neutron scattering intensity map of the reciprocal ($h$,0,$l$) plane collected at 30\,K for EuAl$_4$. The inset shows an enlarged map around (2,0,0). (b) A cut along (0,0,$l$) direction across the nuclear peak at $\boldsymbol{Q}$ = (2,0,0) at 30\,K ($T$ < $T_\mathrm{CDW}$) and 300\,K ($T$ > $T_\mathrm{CDW}$). 
		Figures were reproduced from Ref.~\cite{Kaneko2021}.}
\end{figure}
%=== end figure ======================
%

Since the localized 4$f$ electrons of Eu atoms are well below the Fermi
level~\cite{Kobata2016}, the CDW order is mainly attributed to the orbitals
of Al atoms. This is in good agreement with the observed CDW transition
at $T_\mathrm{CDW} =  243$\,K of SrAl$_4$, which has no 4$f$
electrons~\cite{Nakamura2016}. The CDW state has been investigated by
Kaneko \emph{et al.}\ using single-crystal time-of-flight neutron
scattering~\cite{Kaneko2021}. Below $T_\mathrm{CDW}$, additional superlattice peaks appear close to the nuclear Bragg peaks [see figure~\ref{fig:CDW}(b)]. Such superlattice peaks
are then split along the (00$l$) direction [see enlarged plot in the
inset of figure~\ref{fig:CDW}(a)].
Figure~\ref{fig:CDW}(b) shows a cross-sectional profile along the (00$l$)
direction through the (2,0,0) Bragg peak, measured above and below
$T_\mathrm{CDW}$. According to the Gaussian fits, the satellites 
appear at $l \sim 0.19$. Thus, the CDW order in EuAl$_4$ 
is characterized by structural satellite peaks with an
incommensurate wave vector $\boldsymbol{q}_\mathrm{CDW} = (0,0,\delta)$,
with $\delta \sim 0.19$ at 30\,K. Note that the $\delta$ value
changes with temperature.

\begin{figure*}[!htp]
	%\begin{center}
	\includegraphics[width = 0.85\linewidth]{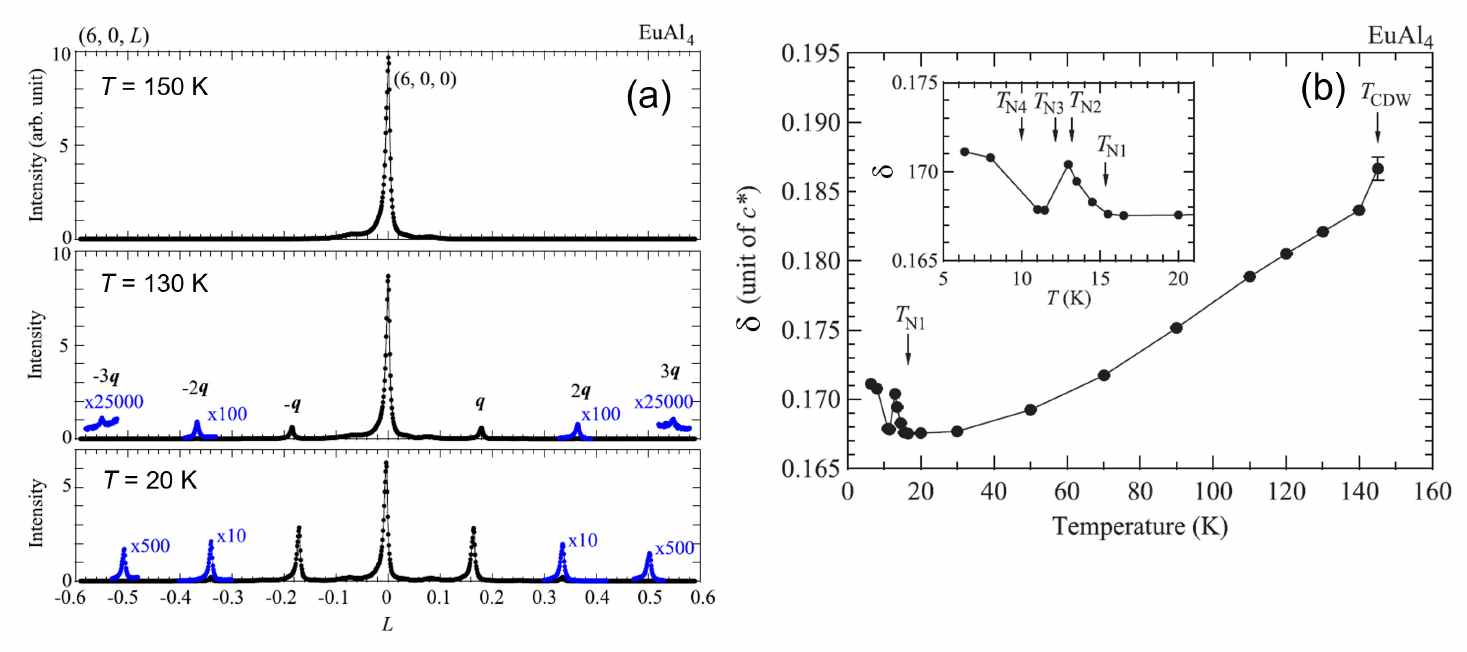}
	%\end{center}
	\centering
	\vspace{-2ex}%
	\caption{\label{fig:CDW_2} Synchrotron x-ray diffraction profiles along the $c^*$ direction passing through the (6,0,0) collected at 150 K ($T > T_\mathrm{CDW}$), 130\,K and 20\,K ($T < T_\mathrm{CDW}$). 
		The satellite peaks denoted by $\pm nq$ are characterized by the modulation wave vector $\boldsymbol{q}_\mathrm{CDW}$ = (0,0,$\delta$) with $\delta$ $\sim$ 0.18. (b) Temperature dependence of the $\delta$ value in $\boldsymbol{q}_\mathrm{CDW}$ = (0,0,$\delta$). 
		The inset shows an enlarged view at low temperatures. Figures were reproduced from Ref.~\cite{Shimomura2019}.}
\end{figure*}

By using synchrotron radiation, Shimomura and Ramakrishnan
\emph{et al.}\ further confirmed the CDW order in EuAl$_4$ via
single-crystal x-ray diffraction. Satellite peaks, characterized by
an incommensurate wave vector $\boldsymbol{q}_\mathrm{CDW}$ = (0,0,$\delta$) 
with $\delta \sim 0.18$, appear below 145\,K [see
figure~\ref{fig:CDW_2}(a)]~\cite{Shimomura2019,Ramakrishnan2022}, thus confirming the occurrence
of a CDW order. 
As the temperature decreases, the satellite peaks grow and shift slightly 
to lower $\delta$ values, while persisting even in the AFM state. 
Curiously, as the temperature drops from $T_\mathrm{CDW}$ to $T_\mathrm{N}$,
$\delta$ not only decreases monotonically, but it also exhibits an
inflection at $T_\mathrm{N}$ [figure~\ref{fig:CDW_2}(b)]. The latter 
implies a sizable coupling between the magnetic- and charge orders in EuAl$_4$. 

Structural refinements indicate that the symmetry of the incommensurately
modulated crystal structure is orthorhombic~\cite{Ramakrishnan2022},
while both the lattice and atomic coordinates of the basic structure
remain tetragonal in the CDW-ordered state. Very recently,
Korshunov \emph{et al.}\ found that both tetragonal and orthorhombic
modulations are compatible with the x-ray data~\cite{Korshunov2024}.
In the CDW-ordered state, the satellite peaks are absent along the (00$l$)
direction, while they show up either upon increasing the
in-plane component (i.e., $h$ or $k$) or on decreasing the out-of-plane
component $l$~\cite{Kaneko2021}. Such features indicate that structural
modulations arise mostly from the atomic displacements within the $ab$-plane. 
Moreover, the Al2-Al2 and Al1-Al2 distances are hardly modulated,
while the Al1-Al1 distance shows the largest modulation (see details
on the Al1 and Al2 atoms in table~\ref{tab:coordinates})~\cite{Ramakrishnan2022}.
The strong modulation of Al1-Al1 distance suggests that the CDW occurs
on the layers of Al1 atoms.

%
%==== figure =============================%
%=== end figure =======================
%

Now we discuss the possible mechanisms for the formation of CDW order in BaAl$_4$-type materials. 
The traditional view of CDW order comes from the Peierls’ instability
with a perfect Fermi surface nesting (FSN) in a 1D system to induce
a Kohn anomaly or soft phonon mode, which results in a structural
phase transition at low temperature~\cite{Gruner2018}.
The charge-density modulation is usually accompanied by a periodic
lattice distortion or a modulation of the atomic positions.
%Fermi surface nesting in the low-dimensional system can lead to the formation of CDW order, where  charge density modulation is accompanied by a periodic lattice distortion or atomic positions~\cite{Gruner2018}.
%the nesting vector of the periodic structure becomes the wave vector of the CDW and also the modulation of atomic positions. 
The modulation wave vector can be commensurate or incommensurate with respect to the crystal structure. For example, the modulation vector of EuAl$_4$ is $\boldsymbol{q}_\mathrm{CDW}$ = (0,0,$\delta$), with $\delta$ $\sim$ 0.18, thus,
its CDW order is incommensurate. Recently, the strong electron-phonon coupling has been proposed as
the mechanism at the origin of CDW in 2D and 3D materials~\cite{Johannes2008,Zhu2015,Zhu2017}. 
According to de Haas–van Alphen and ARPES experiments, as well as to
electronic band-structure calculations, a substantial nesting
of the Fermi surface is not evident in EuAl$_4$~\cite{Nakamura2014,Nakamura2015,Kobata2016}. 
%and thus, the mechanism of  
%CDW is very unlikely the FSN.  
Nakamura and Kobata \emph{et al.}\ claimed that  
the Dirac-like dispersion of the valence and conduction
bands form Fermi surface shells resulting in an imperfect FSN along the
$\Gamma$-$Z$ direction in the $A$Al$_4$ ($A$ = Sr, Eu, Ba) compounds,
but not in $A$Ga$_4$~\cite{Nakamura2014,Nakamura2015,Kobata2016}.
In the Al compounds (here, in SrAl$_4$ and EuAl$_4$), a large electron-phonon coupling 
at small $q$-vectors has been found in the transverse acoustic
mode along the $\Gamma$-$Z$ direction, whose softening is thought to
provide the driving force for the CDW. 
However, as this softening is absent in BaAl$_4$, the origin of CDW
order in BaAl$_4$-type materials is most likely the strong
electron-phonon coupling interaction. This might explain the
absence of a CDW order in the other closely related isostructural
and isovalent compounds. Note that, since in EuAl$_4$ the CDW order
is coupled to the AFM order (see details below), such coupling
adds further complexity to the origins of the spin- and
charge order.  

%
%==== figure =============================%
\begin{figure*}[!htp]
	%\begin{center}
	\includegraphics[width = 0.9\linewidth]{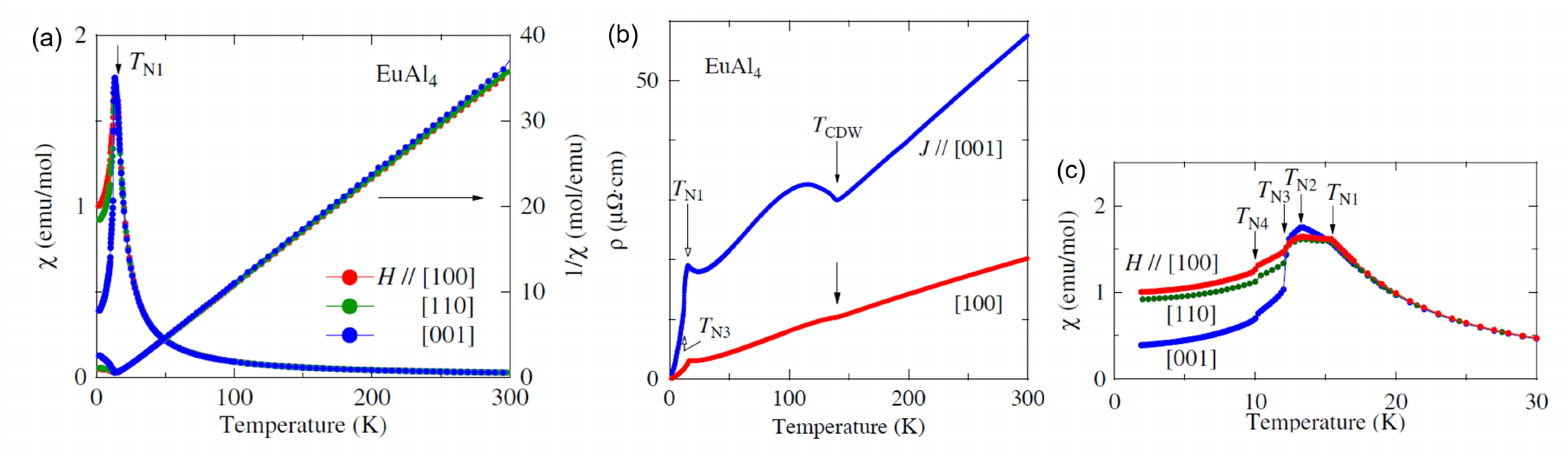}
	%\end{center}
	\centering
	\vspace{-2ex}%
	\caption{\label{fig:EuAl4_chi_rho}\tcr{Temperature dependence of the
	magnetic susceptibility $\chi(T)$ (a) and electrical resistivity
	$\rho_\mathrm{xx}(T)$ (b) of an EuAl$_4$ single crystal. (c) Enlarged
	plot of $\chi(T)$ at low temperature, with the arrows indicating 
	the various AFM transitions of the Eu$^{2+}$ ions.
	Arrows in (b) indicate the first AFM transition at
	$T_\mathrm{N1} \sim 15.4$\,K and the CDW transition
	at $T_\mathrm{CDW}$ $\sim$ 140\,K.
	The magnetic susceptibility was collected in a magnetic field
	$\mu_0H$ = 1\,T, applied both parallel ($H \parallel$ [001])
	%$\chi_\mathrm{c}$)
	and perpendicular ($H \parallel$ [100] or $H \parallel$ [110])
	%($\chi_\mathrm{ab}$ or $\chi_\mathrm{a}$) 
	to the $c$-axis. 
	Electrical resistivity was measured in a zero-field condition,
	with the electric current applied both parallel ($J \parallel$ [001])
	%($\rho_\mathrm{xx}^\mathrm{c}$)
	and perpendicular ($J \parallel$ [100])
	%($\rho_\mathrm{xx}^\mathrm{ab}$) 
	to the $c$-axis.
	%The $\chi_\mathrm{c}$ and $\rho_\mathrm{xx}^\mathrm{c}$ data
	%were taken from Ref.~\cite{Shang2021}, while the $\chi_\mathrm{ab}$
	%and $\rho_\mathrm{xx}^\mathrm{ab}$ data are original. 
	Figures reproduced from Ref.~\cite{Nakamura2015}.}}
\end{figure*}
%=== end figure =======================
%
%
\section{Magnetic and transport properties\label{sec:trsnsport}}

\subsection{Magnetic phase diagrams of EuAl$_4$ and EuGa$_4$}

%
%==== figure =============================%
\begin{figure}[!htp]
	%\begin{center}
	\includegraphics[width = 0.85\linewidth]{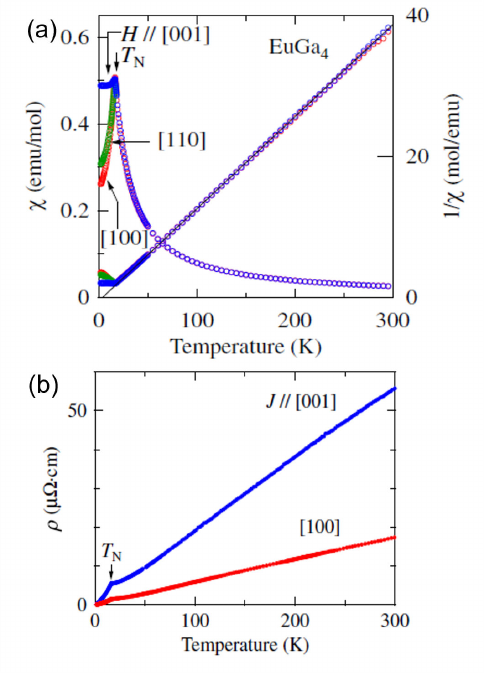}
	%\end{center}
	\centering
	\vspace{-2ex}%
	\caption{\label{fig:EuGa4_chi_rho}\tcr{Temperature dependence of the
	magnetic susceptibility $\chi(T)$ (a) and electrical resistivity
	$\rho_\mathrm{xx}(T)$ (b) of an EuGa$_4$ single crystal. 
	%The inset in (a) shows the enlarged plot at low $T$.
	The magnetic susceptibility was collected by applying the magnetic field
	%in a magnetic field
	%$\mu_0H$ = 0.1\,T, applied both 
	both parallel ($H \parallel$ [001])
	%$\chi_\mathrm{c}$)
	and perpendicular ($H \parallel$ [100] or $H \parallel$ [110])
	%($\chi_\mathrm{ab}$ or $\chi_\mathrm{a}$) 
	to the $c$-axis. 
	%    parallel ($\chi_\mathrm{c}$)
	%and perpendicular ($\chi_\mathrm{ab}$) to the $c$-axis.
	The electrical
	resistivity was measured in zero-field condition with the electric
	current applied both parallel ($J \parallel$ [001]) 
	%($\rho_\mathrm{xx}^\mathrm{c}$) 
	and
	perpendicular ($J \parallel$ [100])
	%($\rho_\mathrm{xx}^\mathrm{ab}$) 
	to the  $c$-axis.
	The arrows indicate the AFM transition at $T_\mathrm{N}$ $\sim$ 16.5\,K.
	%The $\chi_\mathrm{c}$ and $\rho_\mathrm{xx}^\mathrm{c}$ data
%	were taken from Ref.~\cite{Zhang2022}, while the $\chi_\mathrm{ab}$
	%and $\rho_\mathrm{xx}^\mathrm{ab}$ data are original. 
	Figures reproduced from Ref.~\cite{Nakamura2013}.}}
\end{figure}
%=== end figure =======================
%
In the Eu(Al,Ga)$_4$ family of compounds, the spectroscopic and neutron/resonant x-ray studies designated to find topological features in both the momentum- and real space were motivated by the observation of a topological Hall effect~\cite{Shang2021,Zhang2022,Moya2023}. 
The transport studies, in turn, were inspired by basic magnetic property
characterizations which found possible magnetic ground states, suitable for hosting
these topological features. Echoing this logic, in this section, we
first discuss the magnetic phase diagrams of the Eu(Al,Ga)$_4$ family,
followed by discussions of the transport properties and their
implications on the topological aspects. Note that, regarding the
magnetic properties, only basic characterizations will be included here
for the purpose of establishing the magnetic phase diagram.
A detailed description of the microscopic spin textures is
discussed in a later section.  

%
%==== figure =============================%
\begin{figure}[!tp]
	%\begin{center}
	\includegraphics[width = 1.0\linewidth]{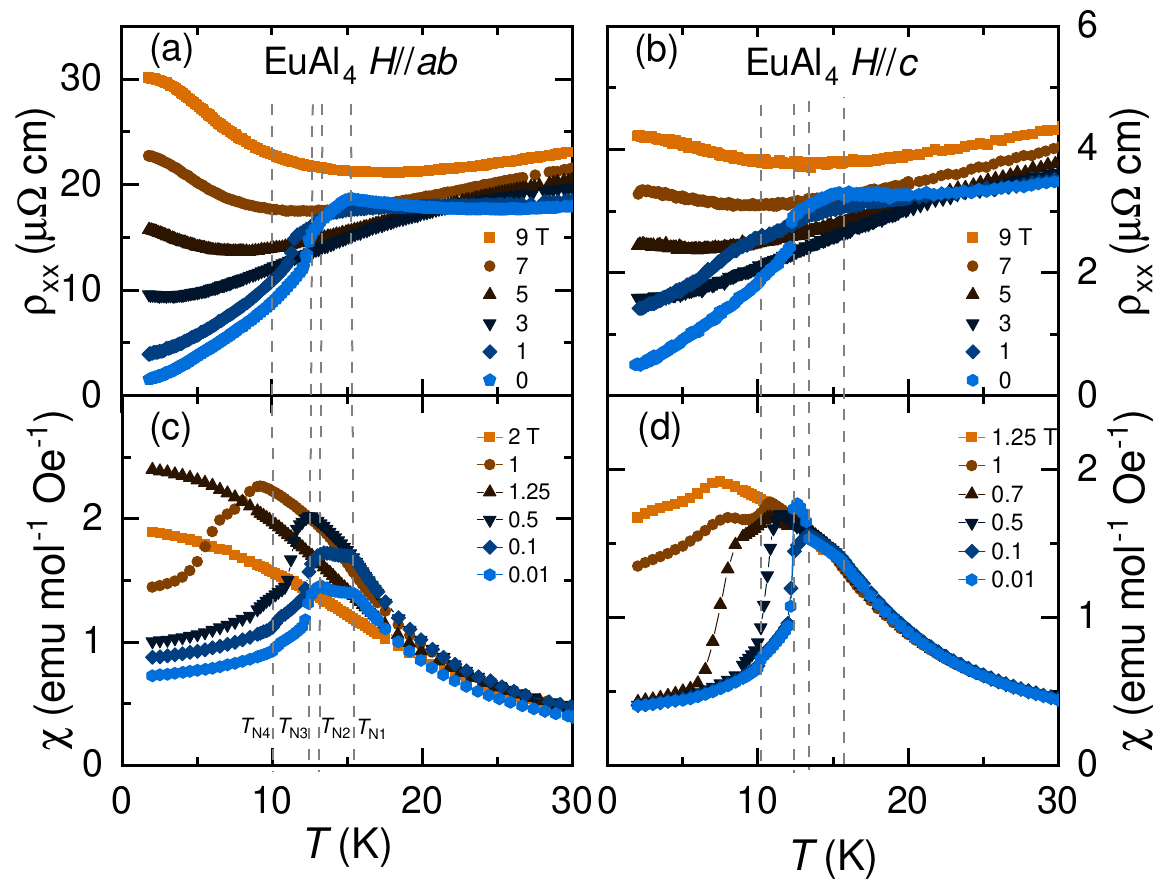}
	%\end{center}
	\centering
	\vspace{-4ex}%
	\caption{\label{fig:EuAl4_rho_H}Temperature dependence of the electrical resistivity $\rho_\mathrm{xx}(T)$ collected under various magnetic fields applied within the $ab$-plane (a) and along the $c$-axis (b) for EuAl$_4$. 
		Temperature dependence of the magnetic susceptibility $\chi(T)$ collected under various magnetic fields applied within the $ab$-plane (c) and along the $c$-axis (d) for EuAl$_4$.
		%The inset in (a) shows the enlarged plot at low-$T$ range. The magnetic susceptibility was measured by applying a magnetic field $\mu_0H$ = 0.1\,T, both parallel ($\chi_\mathrm{c}$) and perpendicular ($\chi_\mathrm{ab}$) to the $c$-axis. 
		%The electrical resistivity was measured in zero field with the current applied both parallel ($\rho_\mathrm{xx}^\mathrm{c}$) and perpendicular ($\rho_\mathrm{xx}^\mathrm{ab}$) to the  $c$-axis.
		The dashed lines mark the AFM transitions for $\chi(T)$ collected in a field of 0.01\,T. Data in panels (b) and (d) were taken from Ref.~\cite{Shang2021}, while data in panels (a) and (c) are original.}
\end{figure}
%=== end figure =======================
%

\begin{figure}[!tp]
	%\begin{center}
	\includegraphics[width = 1.03\linewidth]{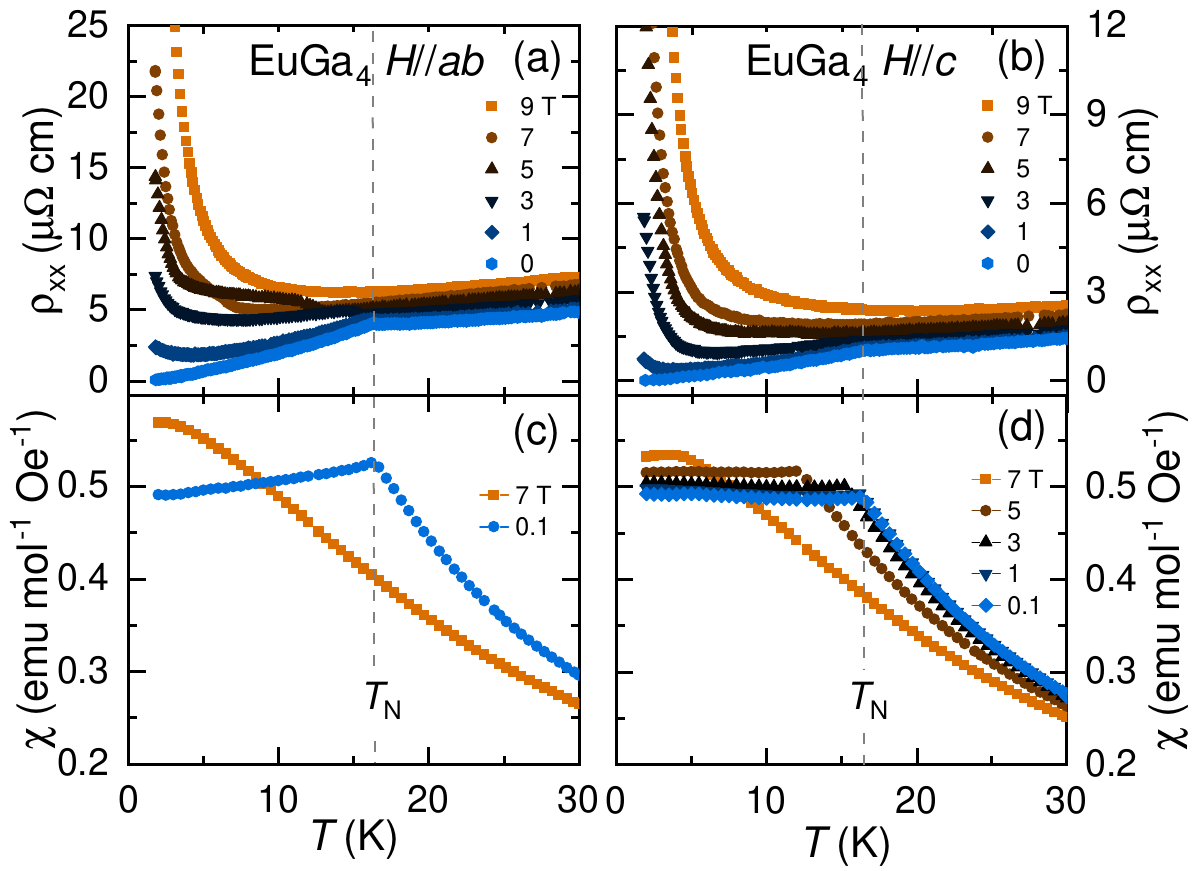}
	%\end{center}
	\centering
	\vspace{-3.5ex}%
	\caption{\label{fig:EuGa4_rho_H}Temperature dependence of the electrical resistivity $\rho_\mathrm{xx}(T)$ collected under various magnetic fields applied within the $ab$-plane (a) and along the $c$-axis (b) for EuGa$_4$. 
		Temperature dependence of the magnetic susceptibility $\chi(T)$ collected under various magnetic fields applied within the $ab$-plane (c) and along the $c$-axis (d) for EuGa$_4$.
		%The inset in (a) shows the enlarged plot at low-$T$ range. The magnetic susceptibility was measured by applying a magnetic field $\mu_0H$ = 0.1\,T, both parallel ($\chi_\mathrm{c}$) and perpendicular ($\chi_\mathrm{ab}$) to the $c$-axis. 
		%The electrical resistivity was measured in zero field with the current applied both parallel ($\rho_\mathrm{xx}^\mathrm{c}$) and perpendicular ($\rho_\mathrm{xx}^\mathrm{ab}$) to the  $c$-axis.
		The dashed lines mark the AFM transitions for $\chi(T)$ collected in a field of 0.1\,T. Data in panels (b) and (d) were taken from Ref.~\cite{Zhang2022}, while data in panels (a) and (c) are original.}
\end{figure}

\tcr{The temperature-dependent magnetic susceptibility $\chi(T)$ and the longitudinal electrical
resistivity $\rho_\mathrm{xx}(T)$ up to room temperature, 
%collected by
%varying the temperature under various magnetic fields applied
%along the $c$-axis and within the $ab$-plane, 
are summarized in figure~\ref{fig:EuAl4_chi_rho} and figure~\ref{fig:EuGa4_chi_rho} for
EuAl$_4$ and EuGa$_4$, respectively~\cite{Nakamura2015,Nakamura2013}. While the low-$T$ $\chi(T)$ and $\rho_\mathrm{xx}(T)$ collected under various magnetic fields applied along the $c$-axis and within the $ab$-plane 
are summarized in figure~\ref{fig:EuAl4_rho_H} and figure~\ref{fig:EuGa4_rho_H} for EuAl$_4$ and EuGa$_4$, respectively~\cite{Shang2021,Zhang2022}.} In EuAl$_4$, four successive antiferromagnetic transitions can be
identified  at $T_\mathrm{N1}$ $\sim$ 15.4\,K, $T_\mathrm{N2}$ $\sim$ 13.2\,K,
$T_\mathrm{N3}$ $\sim$ 12.2\,K, and $T_\mathrm{N4}$ $\sim$ 10\,K in
the $\chi(T)$ [see figure~\ref{fig:EuAl4_chi_rho}(c)]. These
transition temperatures are consistent between the
various studies~\cite{Shang2021,Nakamura2014,Nakamura2015}, and are
clearly reflected also in the thermal expansion coefficient and
the heat-capacity data~\cite{Meier2022}, implying a robust
magnetic order against crystal defects in EuAl$_4$. By contrast, EuGa$_4$
exhibits only one AFM transition at $T_\mathrm{N}$ $\sim$ 16.5\,K [see figure~\ref{fig:EuGa4_chi_rho}]. These
transitions are also evident in the $\rho_\mathrm{xx}(T)$ curves and
show a similar field dependence for both $H \parallel ab$ and
$H \parallel c$, i.e., the AFM transitions shift toward lower
temperature as the magnetic field increases (figure~\ref{fig:EuAl4_rho_H}
and figure~\ref{fig:EuGa4_rho_H}). Though the $\chi(T)$ shows a
negligible anisotropy in the paramagnetic (PM) state, this is
significant in $\rho_\mathrm{xx}(T)$ [see figures~\ref{fig:EuAl4_chi_rho}(b)
and \ref{fig:EuGa4_chi_rho}(b)]. In both compounds, the
negligible bifurcation of the zero-field-cooling- and field-cooling
magnetic susceptibilities confirms the AFM nature of the magnetic
transitions~\cite{Shang2021,Nakamura2014,Nakamura2015,Zhang2022,Moya2023}.

\begin{figure}[!tp]
	%\begin{center}
	\includegraphics[width = 1.0\linewidth]{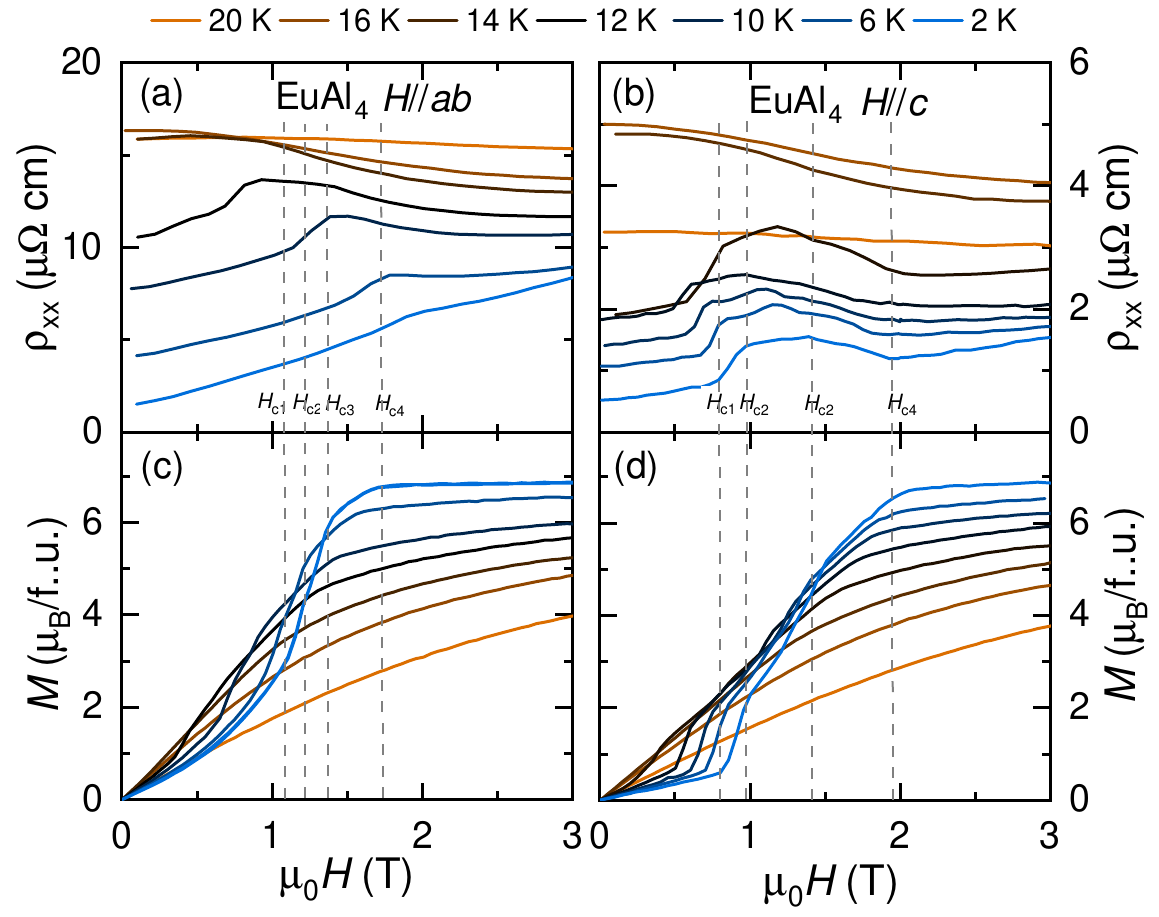}
	%\end{center}
	\centering
	\vspace{-3ex}%
	\caption{\label{fig:EuAl4_Hscan}Field-dependent electrical resistivity
	$\rho_\mathrm{xx}(H)$ collected at various temperatures, with the
	magnetic field applied within the $ab$-plane (a) and along the
	$c$-axis (b), for EuAl$_4$. 
	Field-dependent magnetization $M(H)$ collected at various temperatures,
	with the magnetic fields applied within the $ab$-plane (c) and along
	the $c$-axis (d), for EuAl$_4$.
	Dashed lines mark the saturation field ($H_\mathrm{c4}$) and the
	three critical fields ($H_\mathrm{c1}$, $H_\mathrm{c1}$, and $H_\mathrm{c3}$),
	where EuAl$_4$ undergoes metamagnetic transitions (magnetization
	data at 2\,K). Data in panels (b) and (d) were taken from
	Ref.~\cite{Shang2021}, while data in panels (a) and (c) are original.}
\end{figure}

\begin{figure}[!tp]
	%\begin{center}
	\includegraphics[width = 1.0\linewidth]{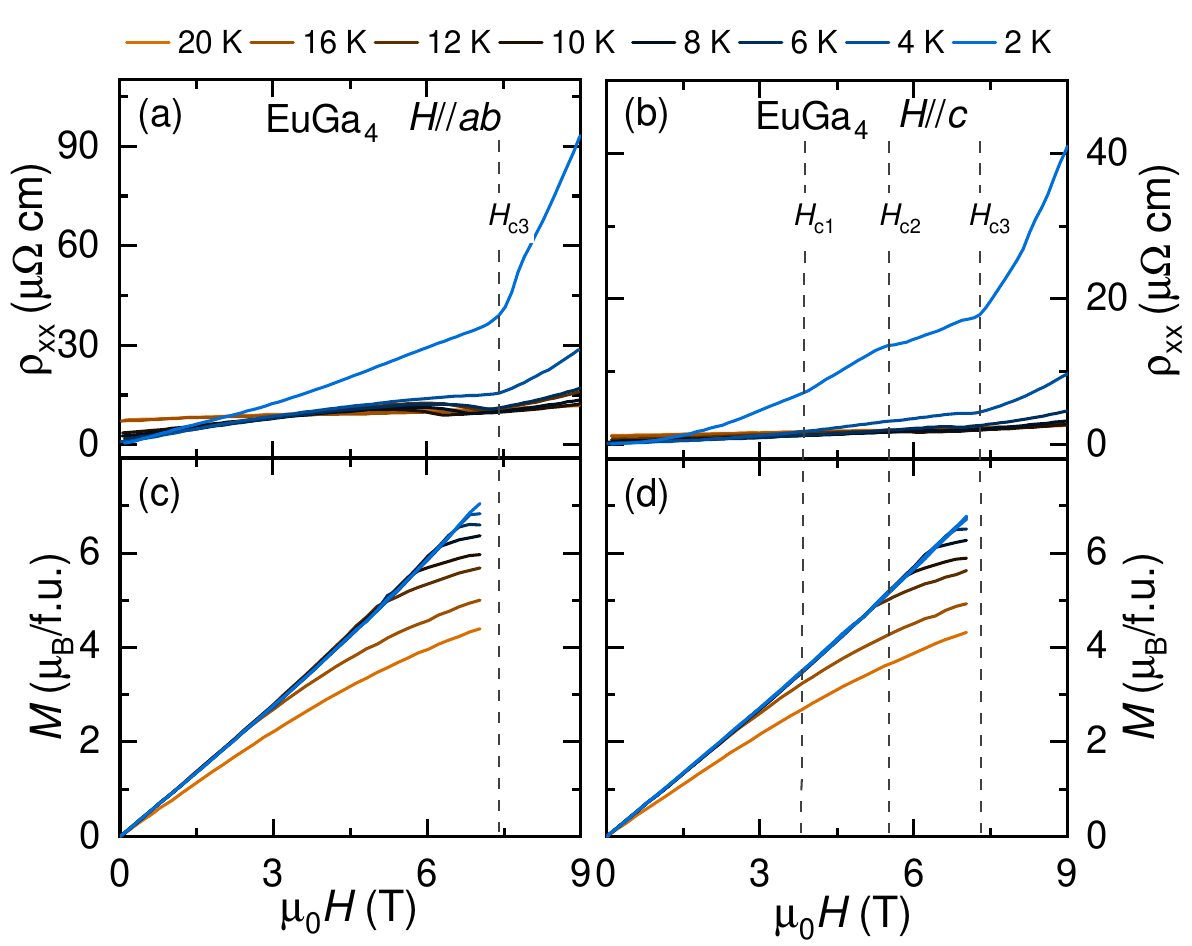}
	%\end{center}
	\centering
	\vspace{-3.5ex}%
	\caption{\label{fig:EuGa4_Hscan} Field-dependent electrical resistivity
	$\rho_\mathrm{xx}(H)$ collected at various temperatures, with the
	magnetic fields applied within the $ab$-plane (a) and along the
	$c$-axis (b), for EuGa$_4$.
	Field-dependent magnetization $M(H)$ collected at various temperatures,
	with the magnetic fields applied within the $ab$-plane (c) and along
	the $c$-axis (d), for EuGa$_4$.
	Dashed lines mark the transitions at $H_\mathrm{c1}$, $H_\mathrm{c2}$,
	and $H_\mathrm{c3}$, for resistivity data collected at 2\,K. For
	$H \parallel ab$, only $H_\mathrm{c3}$ can be identified. Data in
	panels (b) and (d) were taken from Ref.~\cite{Zhang2022}, while data in panels (a) and (c) are original.}
\end{figure}

\begin{figure}[!tp]
	%\begin{center}
	\includegraphics[width=0.9\linewidth]{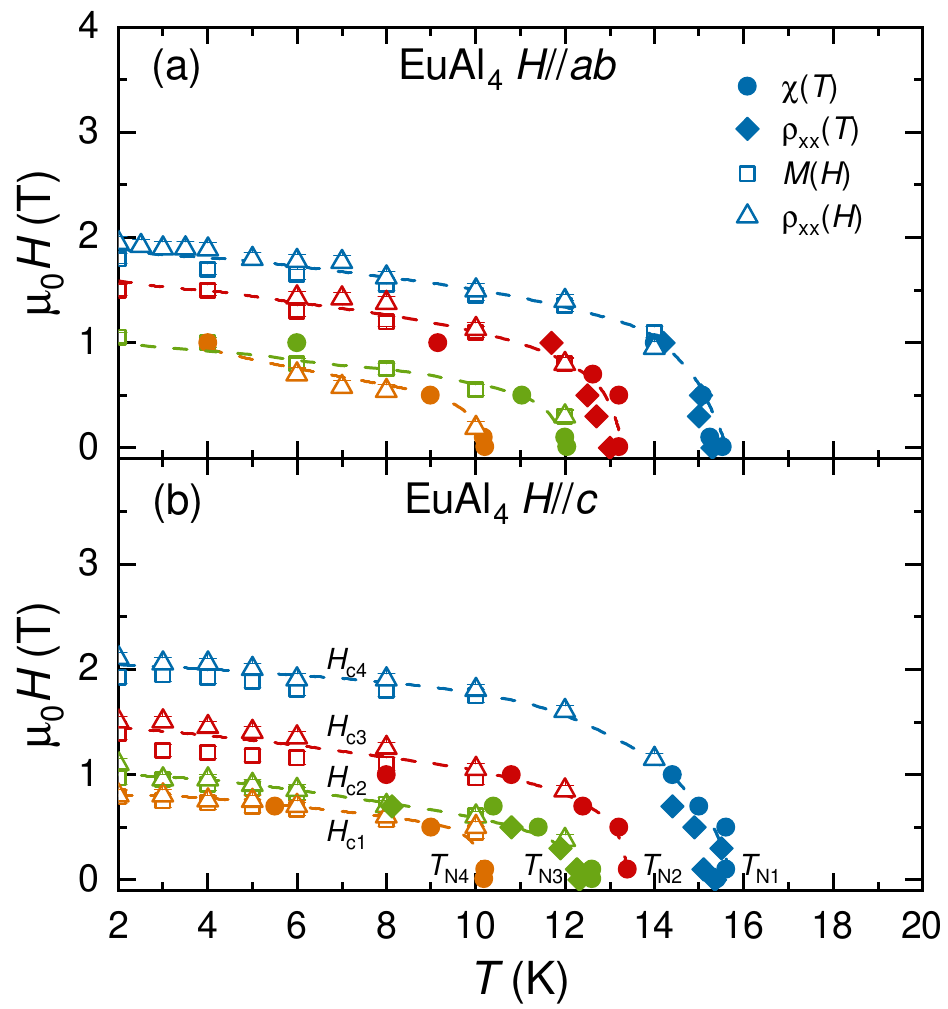}
	%\end{center}
	\centering
	\vspace{-2ex}%
	\caption{\label{fig:EuAl4_diagram}Magnetic phase diagram of an EuAl$_4$
		single crystal, with the field applied in the $ab$-plane (a)
		and along the $c$-axis (b). The critical temperatures ($T_\mathrm{N1}$
		to $T_\mathrm{N4}$) are determined from $\chi(T,H)$ (circles) and $\rho_\mathrm{xx}(T,H)$ (diamonds), while the critical
		fields ($H_\mathrm{c1}$ to $H_\mathrm{c4}$) are determined from $M(H,T)$ (squares) and $\rho_\mathrm{xx}(H,T)$ (triangles). 
		The dashed lines are guides to the eyes. Data in panel (a) are original, while data in panel (b) were taken from Ref.~\cite{Shang2021}.}
\end{figure}

Field-dependent electrical resistivity $\rho_\mathrm{xx}(H)$ and
magnetization $M(H)$ collected at various temperatures, covering both
the AFM- and PM states, are summarized in figure~\ref{fig:EuAl4_Hscan}
and figure~\ref{fig:EuGa4_Hscan} for EuAl$_4$ and EuGa$_4$, respectively~\cite{Shang2021,Zhang2022}. 
At $T = 2$\,K, EuAl$_4$ undergoes three metamagnetic transitions at
$\mu_0H_\mathrm{c1}$ $\sim$ 0.8\,T, $\mu_0H_\mathrm{c2}$ $\sim$ 1.1\,T,
and $\mu_0H_\mathrm{c3}$ $\sim$ 1.5\,T, before saturating at
$\mu_0H_\mathrm{c4}$ $\sim$ 2.1\,T (values for $H \parallel c$).
These metamagnetic transitions occur at slightly different critical
fields for $H \parallel ab$ (figure~\ref{fig:EuAl4_diagram}). The
$H_\mathrm{c1}$ shows up only at $T >$ 4\,K for $H \parallel ab$.
EuGa$_4$ exhibits a significantly different field response
compared with EuAl$_4$. For $H \parallel ab$, only a saturation field
can be identified, with an onset slightly above $\sim$ 7.2\,T
(at 1.3\,K), which is suppressed to $\sim$ 5.3\,T at 12\,K~\cite{Nakamura2013}.
While for $H \parallel c$, three transitions at $\mu_0 H_\mathrm{c1}$ $\sim$ 3.8\,T,
$\mu_0 H_\mathrm{c2}$ $\sim$ 5.6\,T, and $\mu_0 H_\mathrm{c3}$ $\sim$ 7.1\,T
can be tracked in $\rho_\mathrm{xx}(H)$ at $T$ = 2\,K, with $H_\mathrm{c3}$
close to the saturation field in $M(H)$. Zhang \emph{et al.}\ proposed
that the transitions at $H_\mathrm{c1}$ and $H_\mathrm{c2}$, missed in
previous reports, might correspond to metamagnetic transitions,
similar to those in EuAl$_4$~\cite{Zhang2022}. Different from EuAl$_4$,
these metamagnetic transitions do not show up in the magnetization
data (figure~\ref{fig:EuGa4_Hscan}), a fact requiring further
investigation. In both compounds, the saturation magnetization
($M_\mathrm{s}$ $\sim$ 6.8\,$\mu_\mathrm{B}$ for EuAl$_4$ and
$M_\mathrm{s}$ $\sim$ 6.9\,$\mu_\mathrm{B}$ for EuGa$_4$) is consistent
with 7.0\,$\mu_\mathrm{B}$, the expected value for Eu$^{2+}$ ions with
$J = 7/2$. Another prominent feature in $\rho_\mathrm{xx}(H)$ of EuGa$_4$
is the large, non-saturating magnetoresistance, which is
discussed further on in this section.

%at $T$ = 2\,K, EuGa$_4$ exhibits two metamagnetic transitions at $\mu_0H_\mathrm{c1}$ $\sim$ 3.8\,T and $\mu_0H_\mathrm{c2}$ $\sim$ 5.6\,T, before saturating at $\mu_0H_\mathrm{c4}$ $\sim$ 7.1\,T.

%For EuGa$_4$, only a saturation behavior can be identified, which onsets slightly above $\sim$ 7.2\,T at 1.3\,K~\cite{Nakamura2013} and is suppressed to $\sim$ 5.3\,T at 12\,K. 
%negligible anisotropy for in-plane and out-of-plane fields. For both compounds, there is negligible bifurcation between the zero-field-cooling- and field-cooling magnetic susceptibilities, confirming the AFM nature of the magnetic transitions. A Curie-Weiss ft to the inverse susceptibility yields an effective magnetic moment $\mu_\mathrm{eff}$ $\sim$ 7.77\,$\mu_\mathrm{B}$ and a paramagnetic Curie temperature $\theta_\mathrm{p}$ $\sim$ 14.5\,K for EuAl$_4$, and $\mu_\mathrm{eff}$ $\sim$ 7.90\,$\mu_\mathrm{B}$ and $\theta_\mathrm{p}$ $\sim$ 2\,K for EuGa$_4$, respectively. The yielded effective moment is close to the free ion value of Eu$^{2+}$ (7.94\,$\mu_\mathrm{B}$).

%
%==== figure =============================%
%=== end figure =======================
%
%

To summarize briefly, three observations can be made from the comparison between the magnetic
properties of EuAl$_4$ and EuGa$_4$ single crystals. (1) The $M(H)$ of
EuAl$_4$ shows three metamagnetic transitions, with a clear hysteresis
in the AFM state~\cite{Shang2021}, while the magnetization of EuGa$_4$
shows only a smooth saturation. Such difference might hint at possible
different origins of the transport anomalies, as will be discussed later
in this section. (2) For both compounds, $M(H)$ shows a sublinear behavior
in the PM state, suggesting the presence of distinct magnetic fluctuations
near the AFM order, as will be discussed in the {\textmu}SR section.
(3) For both compounds, $\chi(T)$ and $M(H)$ measured with $H \parallel ab$
and $H \parallel c$ show only moderate differences, pointing to a weak
magnetic anisotropy. The latter has important implications
regarding the mechanism of the formation of topological spin textures,
as will be discussed later in this section.

%7For EuAl$_4$, the metamagnetic transitions evident in $M(H)$ can also be tracked in $\rho_\mathrm{xx}(H)$. 
%For EuGa$_4$, when the field is applied along the $c$-axis, in contrast to the smooth saturation behavior of $M(H)$, three transitions at $\mu_0 H_\mathrm{c1}$ $\sim$ 3.8\,T, $\mu_0 H_\mathrm{c2}$ $\sim$ 5.6\,T, and $\mu_0 H_\mathrm{c3}$ $\sim$ 7.1\,T can be tracked in $\rho_\mathrm{xx}(H,\mathrm{2\,K})$, with $H_\mathrm{c3}$ close to the saturation field in $M(H)$. Zhang \textit {et~al.} proposed that the transitions at $H_\mathrm{c1}$ and $H_\mathrm{c2}$, missed in previous reports, might correspond to metamagnetic transitions, as in the case of EuAl$_4$~\cite{Zhang2022}. Another prominent feature in $\rho_\mathrm{xx}(H)$ is the large, nonsaturating magnetoresistance (MR), which will be discussed in detail later in this section.

%==== figure =============================%
%=== end figure =======================
%

\begin{figure}[!tp]
	%\begin{center}
	\includegraphics[width=0.9\linewidth]{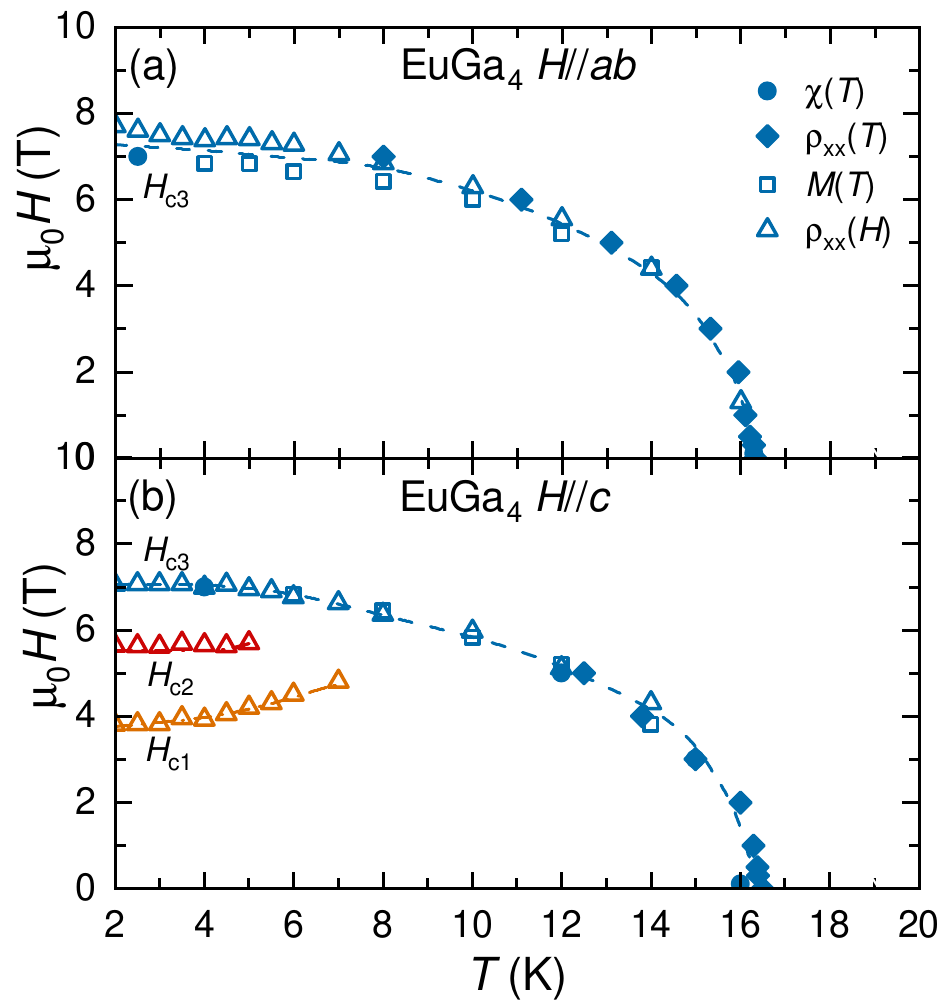}
	%\end{center}
	\centering
	\vspace{-2ex}%
	\caption{\label{fig:EuGa4_diagram}Magnetic phase diagram of an EuGa$_4$ single crystal,
		with the field applied in the $ab$-plane (a) and along the $c$-axis (b). The critical temperatures $T_\mathrm{N}$ 
		are determined from $\chi(T,H)$ (circles) and $\rho_\mathrm{xx}(T,H)$ (diamonds), while the critical
		fields ($H_\mathrm{c1}$ to $H_\mathrm{c3}$) are determined from $M(H,T)$ (squares) and $\rho_\mathrm{xx}(H,T)$ (triangles). 
		The dashed lines are guides to the eyes. Data in panel (a) are original, while data in panel (b) were taken from Ref.~\cite{Zhang2022}.}
\end{figure}

The magnetic phase diagrams of EuAl$_4$ and EuGa$_4$, as established
from the aforementioned datasets for
$H \parallel ab$ and $H \parallel c$, are summarized in figure~\ref{fig:EuAl4_diagram}
and figure~\ref{fig:EuGa4_diagram}, respectively.
For EuAl$_4$, the magnetic phase diagrams in the AFM state are
almost identical for both field orientations and type of measurements.
Conversely, for EuGa$_4$, evidence about
possible metamagnetic transitions is only visible
in the electrical-resistivity data with an out-of-plane field. For an
in-plane field, the $T_\mathrm{N}$ of EuGa$_4$ is smoothly suppressed to
lower temperature, without undergoing any metamagnetic transitions.
We also note that, in EuAl$_4$, additional magnetic phases
have been identified at $T < 6$\,K and 1.3\,T $< H <$ 1.8\,T through
heat-capacity- and dilatometry measurements in a magnetic field applied
along the $c$-axis~\cite{Meier2022}. 

%
%==== figure =============================%
%=== end figure =======================
%	

\subsection{Large magnetoresistance in EuAl$_4$ and EuGa$_4$}
\begin{figure}[t]
	%\begin{center}
	\includegraphics[width = 0.93\linewidth]{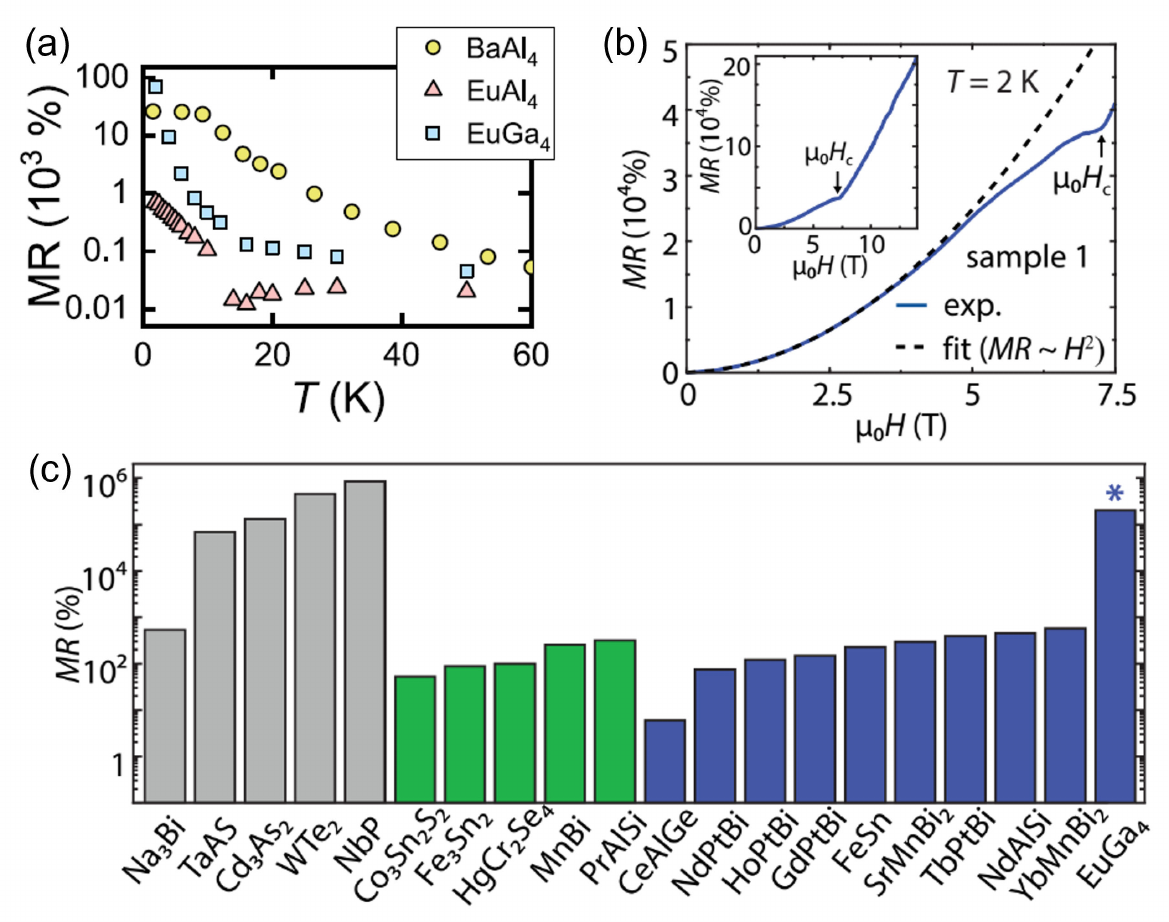}
	%\end{center}
	\centering
	\vspace{-2ex}%
	\caption{\label{fig:MR}Giant MR in BaAl$_4$-type materials. (a) Temperature dependence of MR of EuAl$_4$, EuGa$_4$, and BaAl$_4$ single crystals
	in a field of 9\,T. (b) MR of EuGa$_4$ measured at 2\,K. The arrow indicates the saturation field; the dashed line is a fit to $H^2$. Inset: MR data up to 14\,T. (c) Comparison of the MR for various non-magnetic (gray)
	and magnetic (FM in green, AFM in blue) semimetals. Panels (a) were reproduced from Ref.~\cite{Zhang2022}, while panels (b) and (c) were reproduced from Ref.~\cite{lei2023}.}
\end{figure}

As already mentioned above,
the MR reaches $\sim$800\% and  $\sim 7$ $\times$ 10$^4$\% at $T=2$\,K
in a field of 9\,T in EuAl$_4$ and EuGa$_4$, respectively [figure~\ref{fig:EuAl4_Hscan},
\ref{fig:EuGa4_Hscan}, and \ref{fig:MR}(a)].
Such values are comparable to the ``extremely large MR'' of the nonmagnetic BaAl$_4$~\cite{wang2021}, the latter representing a record among the BaAl$_4$-type materials. 
Recently, Lei \emph {et al.}\ extended the measurement temperature- and magnetic-field
range of EuGa$_4$ and found that its MR 
reaches $\sim 5$ $\times$ $10^5$\% at $\mu_0H$ = 40\,T,
with no signs 
of saturation~\cite{lei2023}. While comparable to the MR of
nonmagnetic semimetals, the MR of EuGa$_4$ is the largest among
the topological magnetic semimetals [figure~\ref{fig:MR}(c)].
It was also shown that, in the AFM state, the MR is
well described by a quadratic field dependence (i.e., $H^2$)
for $H \le$ 3.5\,T, while it levels off as the field tends towards its saturation value, a scenario typical of
uncompensated semimetals [figure~\ref{fig:MR}(b)]. Upon further increasing the magnetic field (up to the spin-polarized state),
the MR shows an abrupt upturn. Then, it keeps increasing without saturating.

%
%==== figure =============================%
%=== end figure =======================
%
Let us now discuss the mechanisms underlying the large MR, 
especially that of EuGa$_4$. The large MR of the nonmagnetic BaAl$_4$ and its comparable value
in EuAl$_4$ and EuGa$_4$, exclude the magnetic Eu$^{2+}$ ions from the
causes of MR.  
This is also corroborated by the fact that the MR of neither
EuAl$_4$ nor EuGa$_4$ exhibit clear anomalies at the AFM transition temperature [figure~\ref{fig:MR}(a)]. Field-induced spin fluctuations can also be excluded since, even in the spin-polarized state, where fluctuations are usually suppressed, the MR of both EuAl$_4$ and EuGa$_4$ continues to
increase~\cite{Shang2021,lei2023}. Further, since the Eu$^{2+}$ ions form a perfect square lattice (figure~\ref{fig:structure}), magnetic frustrations should be absent and, hence, play no significant role. In nonmagnetic semimetals, two scenarios, one including the charge compensation and another the guiding-center motion of charge carriers, 
are often proposed to explain the observed large MR~\cite{Leahy2018}.
The latter mechanism is usually relevant in topological semimetals with a linear dispersion, e.g., a Dirac spectrum. Charge compensation is not achieved in BaAl$_4$~\cite{wang2021}, but de Haas–van Alphen studies and electronic band-structure calculations suggest its potential realization in the PM state of EuGa$_4$~\cite{lei2023,Nakamura2014}. In the AFM state of EuGa$_4$, the very large MR hinders the accurate determination of the carrier type and density from the Hall resistivity. Considering
the comparable MR of nonmagnetic BaAl$_4$ (with a Dirac dispersion),
the MR of EuGa$_4$ can most likely be attributed to its non-trivial topological electronic bands. 
Taking into account the above factors, together with the anomalous transverse transport (to be discussed in the next subsection), Zhang \emph{et al.}\ proposed that the interesting features in the MR of EuGa$_4$ (i.e., a large nonsaturating value and anomalies at 
metamagnetic transitions) may result from a combination of charge compensation, non-trivial band topology, and topological spin textures~\cite{Zhang2022}. Recently, Lei \emph{et al.}\ argued that the charge compensation scenario is inconsistent with the non-saturated MR at a magnetic field up to 40\,T, and that the extremely large MR of EuGa$_4$ is a consequence of the Weyl nodal rings close to the Fermi level, protected by a mirror symmetry in the spin-polarized state~\cite{lei2023}. 
The symmetry analysis shows that the AFM state of EuGa$_4$, with or without moment canting, does not provide the required protection for the existence of Weyl nodal-ring states, and the recovery of mirror symmetry $m_\mathrm{z}$ in the spin-polarized state is necessary. This means that a field-induced topological phase transition, accompanying the AFM state to the spin-polarized state, must take place. Such transition explains also the upturn in MR upon entering the spin-polarized state.

%
%==== figure =============================%
%=== end figure =======================
%

Finally, we discuss the MR of EuGa$_4$ within a theoretical framework that attempts
to address the large non-saturating MR observed in many nonmagnetic semimetals~\cite{Leahy2018}. In this framework, the monotonic increase of electrical resistivity at large
fields and low temperatures, the small Hall angle (i.e., $\rho_\mathrm{xy}$ $\ll$ $\rho_\mathrm{xx}$), as well as the superlinear MR, all seem to suggest
%
%
%==== figure =============================%
that EuGa$_4$ is closer to the ``trivial'' semimetals, where the large MR stems merely from charge compensation instead of linear dispersion, inconsistent with a non-trivial band topology. However, the field-dependent Hall resistivity $\rho_\mathrm{xy}(H)$ of EuGa$_4$ shows complex features at low temperatures (see next subsection), in contradiction with the behavior expected from ``trivial'' semimetals. Clearly, although appealing, the above theoretical framework cannot account
for all the subtleties of the MR of EuGa$_4$.

\subsection{Topological Hall effect in EuAl$_4$ and EuGa$_4$}

\begin{figure}[!tp]
	%\begin{center}
	\includegraphics[width = 1.0\linewidth]{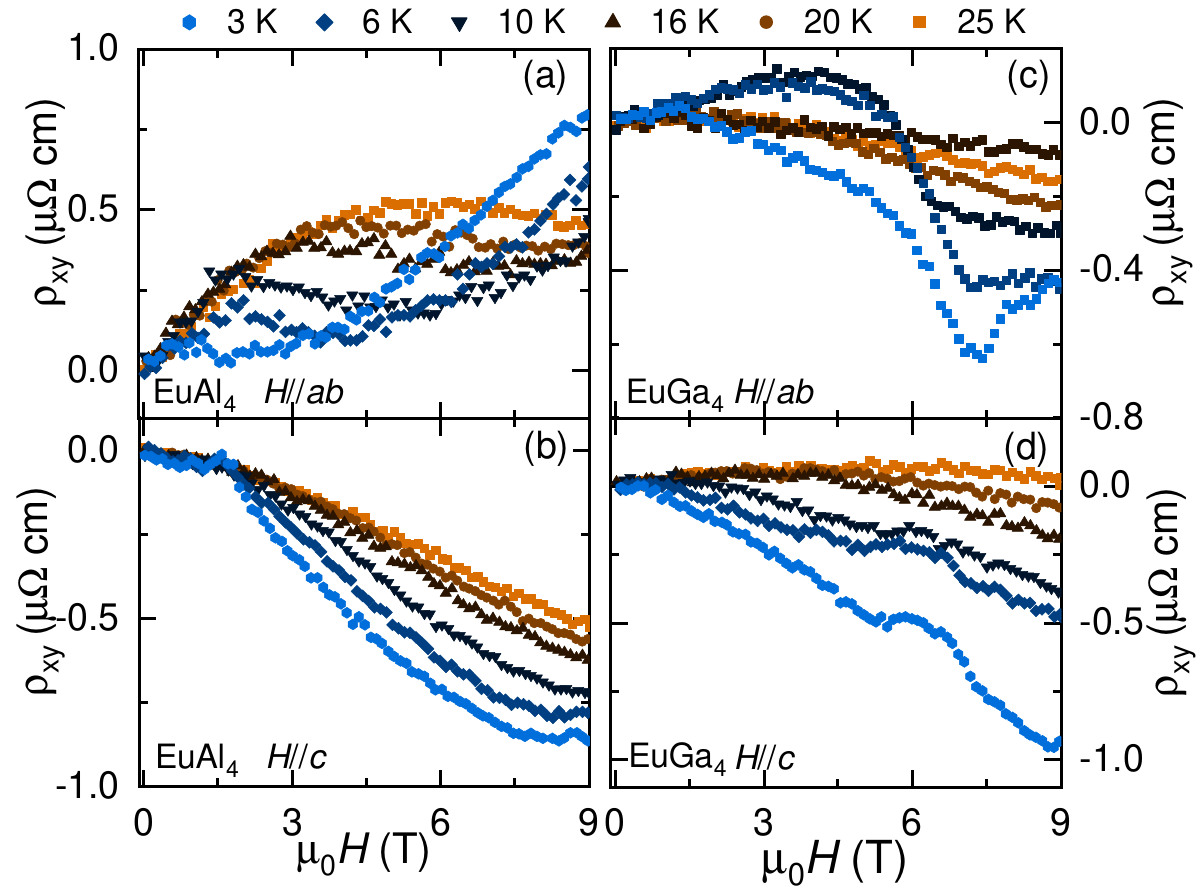}
	%\end{center}
	\centering
	\vspace{-2ex}%
	\caption{\label{fig:EuA4_Hall}Field dependence of the Hall resistivity $\rho_\mathrm{xy}(H,T)$ of EuAl$_4$ collected at various temperatures with the applied magnetic fields in the $ab$-plane (a) and along the $c$-axis (b).
	The analogous results for EuGa$_4$ are shown in panels (c) and (d). Data in panels (a) and (c) are original, while data in panels (b) and (d) were taken from Refs.~\cite{Shang2021,Zhang2022}.}
\end{figure}

We now turn to the transverse transport properties, embodied in the Hall resistivity $\rho_\mathrm{xy}$. Recently, Shang \emph{et al.}\ and Zhang \emph{et al.}\ investigated in detail the Hall resistivity of
EuAl$_4$ and EuGa$_4$~\cite{Shang2021,Zhang2022}. Their field-dependent Hall-resistivity data $\rho_\mathrm{xy}(H)$,
collected at various temperatures with $H \parallel ab$ and $H \parallel c$,
are summarized in figure~\ref{fig:EuA4_Hall}. In both cases, in the AFM state, $\rho_\mathrm{xy}(H)$ exhibits
a hump-like anomaly for $H \parallel c$ [panels~(b) and (d)].
Such a hump is reminiscent of the topological Hall resistivity $\rho_\mathrm{xy}^\mathrm{THE}$ arising from topological spin textures, e.g., magnetic skyrmions~\cite{Fujishiro2019,Kurumaji2019,Kanazawa2016,Gobel2020,Vistoli2019,Neubauer2009,Gayles2015,Kanazawa2011,Franz2014,Schulz2012,Qin2019,Matsuno2016}.
As mentioned in the introduction, the measured $\rho_\mathrm{xy}(H)$
can be decomposed into three components, 
$\rho_\mathrm{xy}^\mathrm{O}$, $\rho_\mathrm{xy}^\mathrm{AHE}$, and $\rho_\mathrm{xy}^\mathrm{THE}$. Here, $\rho_\mathrm{xy}^\mathrm{O}$ is
the \emph{ordinary} Hall effect, while $\rho_\mathrm{xy}^\mathrm{AHE}$ represents the \emph{conventional} anomalous Hall term, mostly determined by the electrical resistivity and magnetization. Finally, $\rho_\mathrm{xy}^\mathrm{THE}$ represents the \emph{unconventional} anomalous Hall term, also known as the topological Hall term, closely related to the Berry
curvature in the real- and momentum space.
Real-space Berry curvature is often carried by topological spin textures,
with the most notable example being skyrmions in noncentrosymmetric MnSi and related compounds~\cite{Neubauer2009,Gayles2015,Kanazawa2011,Franz2014}. The momentum-space Berry curvature, instead, underlies the famous quantum anomalous Hall effect and is nowadays often invoked to explain the anomalous Hall responses of noncollinear antiferromagnets and magnetic Weyl semimetals~\cite{Xu2021,Suzuki2016,Nakatsuji2015,Ikhlas2017,Liu2018,Manna2018}.

The topological Hall resistivity is distinct from the ordinary and anomalous Hall resistivity, the latter two being proportional to the applied magnetic field $H$ and magnetization $M(H)$, respectively~\cite{Nagaosa2010}. Experimentally, $\rho_\mathrm{xy}^\mathrm{THE}(H)$ can be identified as the residual between the measured Hall resistivity $\rho_\mathrm{xy}(H)$ and a fit to the ordinary and anomalous Hall resistivities, namely $\rho_\mathrm{xy}^\mathrm{fit}(H)$ = $R_0$$H$ + $R_\mathrm{S}$$M(H)$, where $R_0$ and $R_\mathrm{S}$ are the ordinary and anomalous Hall coefficients. The anomalous Hall resistivity $\rho_\mathrm{xy}^\mathrm{AHE}$ can be described as a combined linear- and quadratic dependence on the longitudinal electrical resistivity [i.e., $a \rho_\mathrm{xx}(H) + b \rho_\mathrm{xx}^2(H)$], corresponding to the skew scattering and side-jump mechanisms, respectively~\cite{Nagaosa2010}. In most metallic compounds, $\rho_\mathrm{xx}(H)$ is relatively small. Consequently, modeling their $\rho_\mathrm{xx}(H)$ with $R_\mathrm{S} = a\rho_\mathrm{xx}(H) + b\rho_\mathrm{xx}^2(H)$, or simply with $R_\mathrm{S}$ = constant has a negligible influence on the estimated $\rho_\mathrm{xy}^\mathrm{THE}$.

%
%==== figure =============================%
%=== end figure ======================
%

\begin{figure*}[!tp]
	%\begin{center}
	\includegraphics[width = 0.95\linewidth]{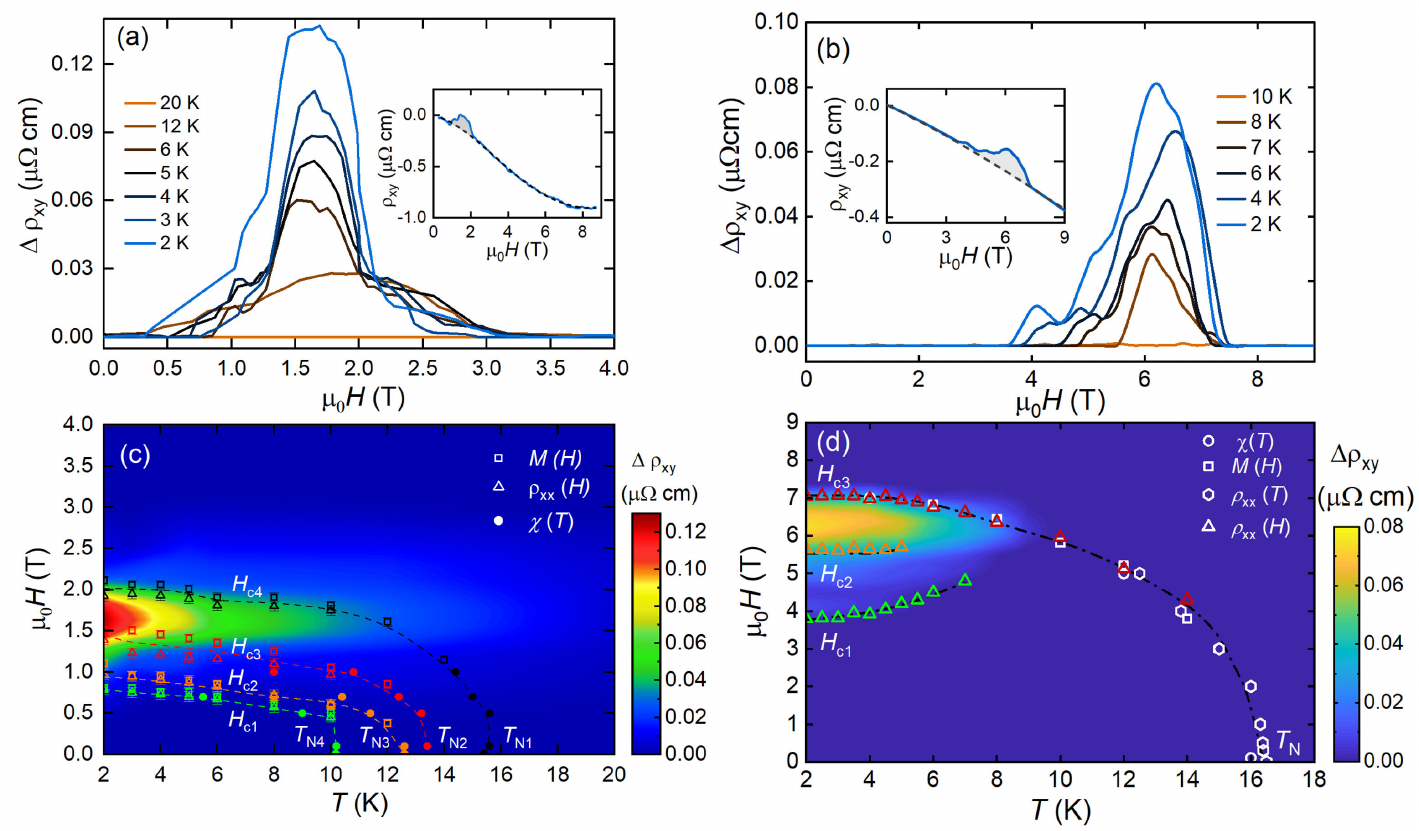}
	%\end{center}
	\centering
	\vspace{-2ex}%
	\caption{\label{fig:EuA4_Hall_diagram}Field dependence of the extracted Hall resistivity $\Delta\rho_\mathrm{xy}(H)$ of EuAl$_4$ (a) and EuGa$_4$ (b) at various temperatures. Insets show the $\rho_\mathrm{xy}(H)$ data at 2\,K of EuAl$_4$ and EuGa$_4$, with the dashed lines being polynomial fits.
	Magnetic phase diagram of EuAl$_4$ (c) and EuGa$_4$ (d), with the field applied
	along the $c$-axis.	The background color in panels (c) and (d) represents the magnitude of $\Delta\rho_\mathrm{xy}(H)$ at various temperatures. 
	Figures were reproduced from Refs.~\cite{Shang2021,Zhang2022}.}
\end{figure*}

In Eu(Al,Ga)$_4$ single crystals, owing to their mul\-ti\-band nature
and giant MR, the decomposition procedure mentioned above is not straightforward. Their multiband nature is evident 
in the nonlinear behavior
of $\rho_\mathrm{xy}(H)$ at large fields (figure~\ref{fig:EuA4_Hall}), 
similar to that of nonmagnetic BaAl$_4$~\cite{wang2021}. Since the multiband effect is more evident for $H \parallel ab$, the field-dependent $\rho_\mathrm{xy}(H)$ shows complex features [figure~\ref{fig:EuA4_Hall}(a) and (c)], both preventing the estimation
of the topological term. In what follows, we focus mostly
on the case $H \parallel c$. 
To circumvent the multiband effect, Shang \emph{et al.}\ and
Zhang \emph{et al.}\ subtracted a polynomial background from the measured
$\rho_\mathrm{xy}(H)$, to obtain a hump-like anomaly $\Delta\rho_\mathrm{xy}$ [see insets in figure~\ref{fig:EuA4_Hall_diagram}(a) and (b)]~\cite{Shang2021}. In EuAl$_4$, the obtained $\Delta\rho_\mathrm{xy}$ is most prominent at temperatures below $T_\mathrm{N3}$ and in the field range between $H_\mathrm{c3}$ and $H_\mathrm{c4}$ [see figure~\ref{fig:EuA4_Hall_diagram}(a) and (c)]. 
While, in EuGa$_4$, $\Delta\rho_\mathrm{xy}$ is clearly observed
below 10\,K between $H_\mathrm{c2}$ and $H_\mathrm{c3}$ [see figure~\ref{fig:EuA4_Hall_diagram}(b) and (d)]. The convenience of such
decomposition procedure, however, comes at a price: the complexity is
transferred to the interpretation of $\Delta\rho_\mathrm{xy}$, as the
relation between $\Delta\rho_\mathrm{xy}$ and $\rho_\mathrm{xy}^\mathrm{THE}$
is complex.
$\Delta\rho_\mathrm{xy}$ may indeed contain the
$\rho_\mathrm{xy}^\mathrm{THE}$ contribution, but it may also contain that from $\rho_\mathrm{xy}^\mathrm{AHE}$, or even $\rho_\mathrm{xy}^\mathrm{O}$.
By using the different functional forms that $\rho_\mathrm{xy}^\mathrm{AHE}$
may take --- which depend on the specific mechanism underlying the conventional anomalous Hall effect, i.e., intrinsic, side-jump, or skew scattering --- Shang \emph{et al.}\ showed that the extracted $\Delta\rho_\mathrm{xy}$ may correspond to $\rho_\mathrm{xy}^\mathrm{THE}$, or to the lower/upper limit of it~\cite{Shang2021}. Considering the region where a finite $\Delta\rho_\mathrm{xy}$ shows up in the phase diagram [figure~\ref{fig:EuA4_Hall_diagram}(c) and (d)], a topological Hall effect arising from the topological spin textures
was inferred, although a trivial origin cannot be fully excluded~\cite{Shang2021,Zhang2022}.

%
%==== figure =============================%
%=== end figure ======================
%	

%==== figure =============================%
\begin{figure*}[!htp]
	%\begin{center}
	\includegraphics[width = 0.7\linewidth]{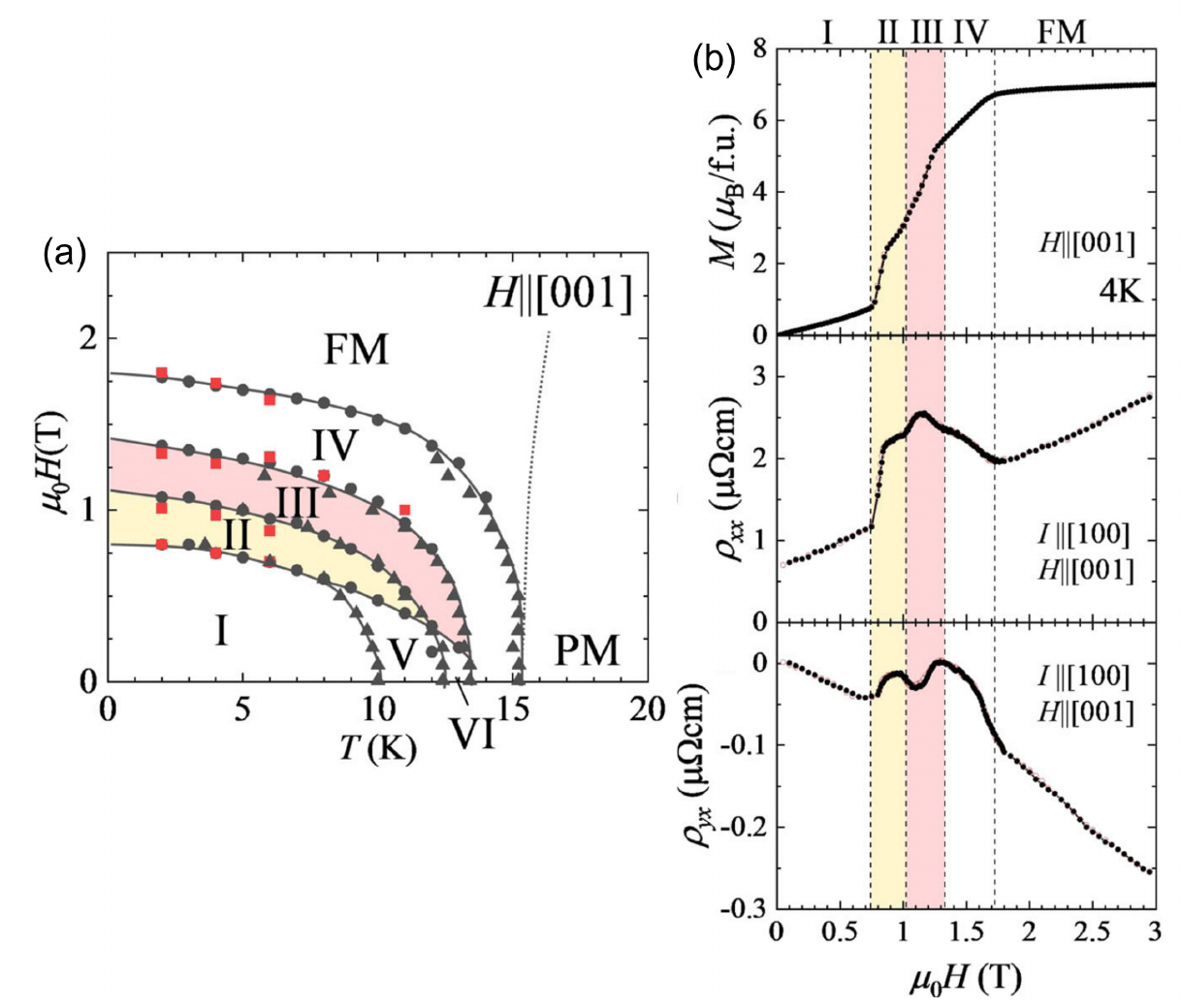}
	%\end{center}
	\centering
	\vspace{-2ex}%
	\caption{\label{fig:Skyrmion_diagram}Magnetic phase diagram of EuAl$_4$ for $H \parallel c$. Regions II and III correspond to a  rhombic and a square skyrmion lattice, respectively. (b) Magnetic-field dependence of the magnetization, longitudinal resistivity, and Hall resistivity collected at 4\,K. Figures were reproduced from Ref.~\cite{takagi2022}.}
\end{figure*}
%=== end figure ======================

%
%
%==== figure =============================%
\begin{figure*}[!tp]
	%\begin{center}
	\includegraphics[width = 1.0\linewidth]{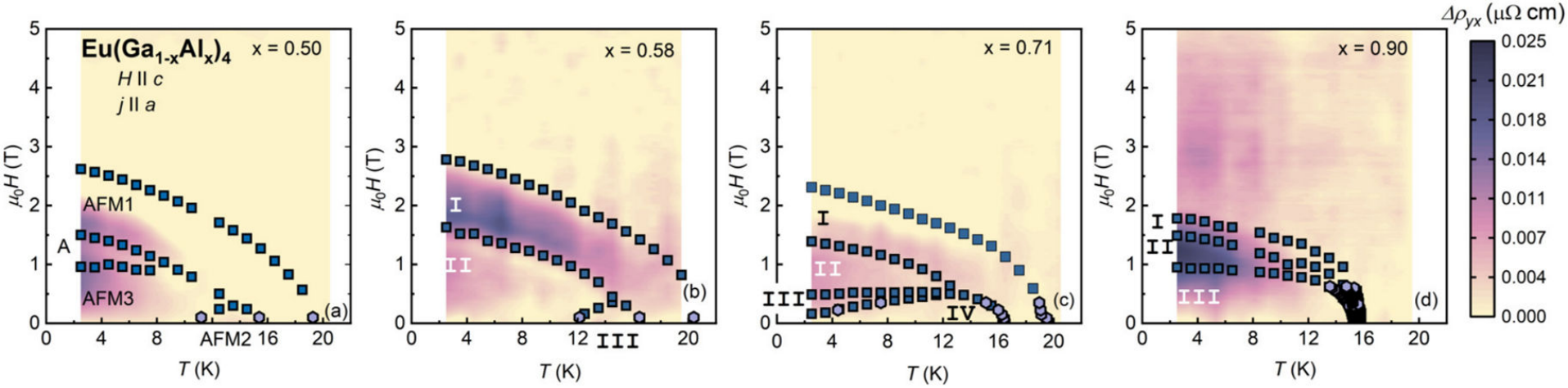}
	%\end{center}
	\centering
	\vspace{-5ex}%
	\caption{\label{fig:AlGa_Hall_diagram} 
	Magnetic phase diagrams of Eu(Ga$_{1-x}$Al$_x$)$_4$ single crystals, with the field applied along the $c$-axis.	The background color represents the magnitude of $\Delta\rho_\mathrm{xy}(H)$ at various temperatures. 
	Figures were reproduced from Ref.~\cite{Moya2023}.}
\end{figure*}
%=== end figure =======================%
%	
%

Since the centrosymmetric EuAl$_4$ does not host conventional Dzyaloshinskii-Moriya interactions, for long thought to be the key
to the formation of topological spin textures like skyrmions~\cite{Yu2011,Yu2010,Seki2012,Kezsmarki2015,Tokunaga2015,Singh2023},
the appearance of a topological Hall effect in EuAl$_4$ immediately
raised the interest of the scientific community.
The recent discovery of magnetic skyrmions in the Gd-based intermetallics with a centrosymmetric crystal structure, including Gd$_2$PdSi$_3$, Gd$_3$Ru$_4$Al$_{12}$, and GdRu$_2$Si$_2$, stimulated the search for new mechanisms
of skyrmion formation beyond the conventional ones~\cite{Kurumaji2019,Hirschberger2019,Khanh2020}. 
In centrosymmetric magnetic materials, skyrmions can be stabilized by magnetic frustration, or by the competition between magnetic interactions and anisotropies~\cite{Batista2016}. As discussed above, the magnetic anisotropy of EuAl$_4$ is moderate. 
Shang \emph{et al.}\ thus speculated that, if skyrmions indeed exist in EuAl$_4$, they might be stabilized by the latter mechanism, similar to GdRu$_2$Si$_2$ having the same crystal structure~\cite{Khanh2020,Shang2021}. Indeed, by using small-angle neutron and resonant elastic x-ray scattering, Takagi \emph{et al.}\ identified square- and rhombic lattices of magnetic skyrmions in EuAl$_4$~\cite{takagi2022} (see details in the next section).
As shown in figure~\ref{fig:Skyrmion_diagram}, a square lattice of
skyrmions exists between the phase boundaries that would correspond to the $H_\mathrm{c2}$ and $H_\mathrm{c3}$ defined in the magnetic phase diagram, while a rhombic lattice of skyrmions is found between $H_\mathrm{c1}$ and $H_\mathrm{c2}$ (figure~\ref{fig:EuA4_Hall_diagram}). These distinctive topological magnetic phases are accompanied by a multistep reorientation of the magnetic modulation vector. The skyrmions in EuAl$_4$ are argued to be stabilized by the interplay between the Ruderman-Kittel-Kasuya-Yosida (RKKY) interaction and the frustrated
itinerant interactions~\cite{takagi2022,Hayami2022}.

By comparing the magnetic phase diagrams in figures~\ref{fig:EuA4_Hall_diagram} and \ref{fig:Skyrmion_diagram}, slight
differences can be identified. Skyrmion lattices are evident
in the field region between $H_\mathrm{c1}$ and $H_\mathrm{c3}$, while $\Delta\rho_\mathrm{xy}$ is most prominent between $H_\mathrm{c3}$ and $H_\mathrm{c4}$,
where the spin state is a double-$Q$ vortex lattice described by the
superposition of two sinusoidally-modulated in-plane spin components
(see details in the next section)~\cite{takagi2022}. 
Such discrepancy cannot be attributed to the sample dependence since,
in the work of Takagi \emph{et al.}, $\rho_\mathrm{xy}(H)$ exhibit a hump-like
anomaly in a field range comparable to that in the work of Shang \emph{et al.}\ [figure~\ref{fig:EuA4_Hall_diagram}(a) and figure~\ref{fig:Skyrmion_diagram}(b)]~\cite{Shang2021,takagi2022}.
However, this discrepancy is plausible
considering the following arguments. First, as discussed above, $\Delta\rho_\mathrm{xy}$ may not be identical to the topological Hall term $\rho_\mathrm{xy}^\mathrm{THE}$, but rather an approximate measure of it. Second, similar to BaAl$_4$, momentum-space Berry curvature sources may also play a role in EuAl$_4$, acting in conjunction with topological spin textures in the AFM state. Third,
it has been proposed that, in addition to the different Hall-resistivity terms mentioned above, there is a so-called chiral Hall effect
related to the noncollinear magnetism, which exists in one-dimensional magnetic textures, such as domain walls and spin spirals~\cite{Lux2020}. This effect may provide a sizable contribution to the total $\rho_\mathrm{xy}$ as well, bringing in additional complexities.

\subsection{Magnetotransport properties of substituted Eu(Al,Ga)$_4$}

We discuss now briefly the magnetic and transport pro\-per\-ties of 
substituted Eu(Al,Ga)$_4$ single crystals. Moya \emph{et al.}\ investigated
an Eu(Ga$_{1-x}$Al$_x$)$_4$ series of crystals with $x$ = 0.15, 0.24, 0.31, 0.39, 0.50, 0.58, 0.71, and 0.90~\cite{Moya2023,Moya2022}. They found evidence of a topological Hall effect for $x \geq 0.5$, indicative of
the emergence of non-coplanar spin textures (figure~\ref{fig:AlGa_Hall_diagram}). At intermediate compositions ($0.24 < x < 0.39$), although the magnetization and longitudinal-resistivity data reveal multiple
magnetic orders, the Hall resistivity varies smoothly with field. 
Hence, the topological term is weak or negligible.
On the Ga-rich side, e.g, $x = 0.15$, only one magnetic phase
transition could be detected. In the spin-polarized state, a large
anomalous Hall effect was found for all compositions and
attributed to the appearance of Weyl nodes generated by the field-induced spin splitting of a Dirac point. Figure~\ref{fig:AlGa_Hall_diagram} presents some representative magnetic phase diagrams for Eu(Ga$_{1-x}$Al$_x$)$_4$ single crystals. 
Consistent with the previous discussion, these results clearly
reveal the coexistence of real- and momentum-space topology in a
single series of materials, in this case, in Eu(Ga$_{1-x}$Al$_x$)$_4$.

%==== figure =============================%
\begin{figure*}[!htp]
	%\begin{center}
	\includegraphics[width=0.9\linewidth]{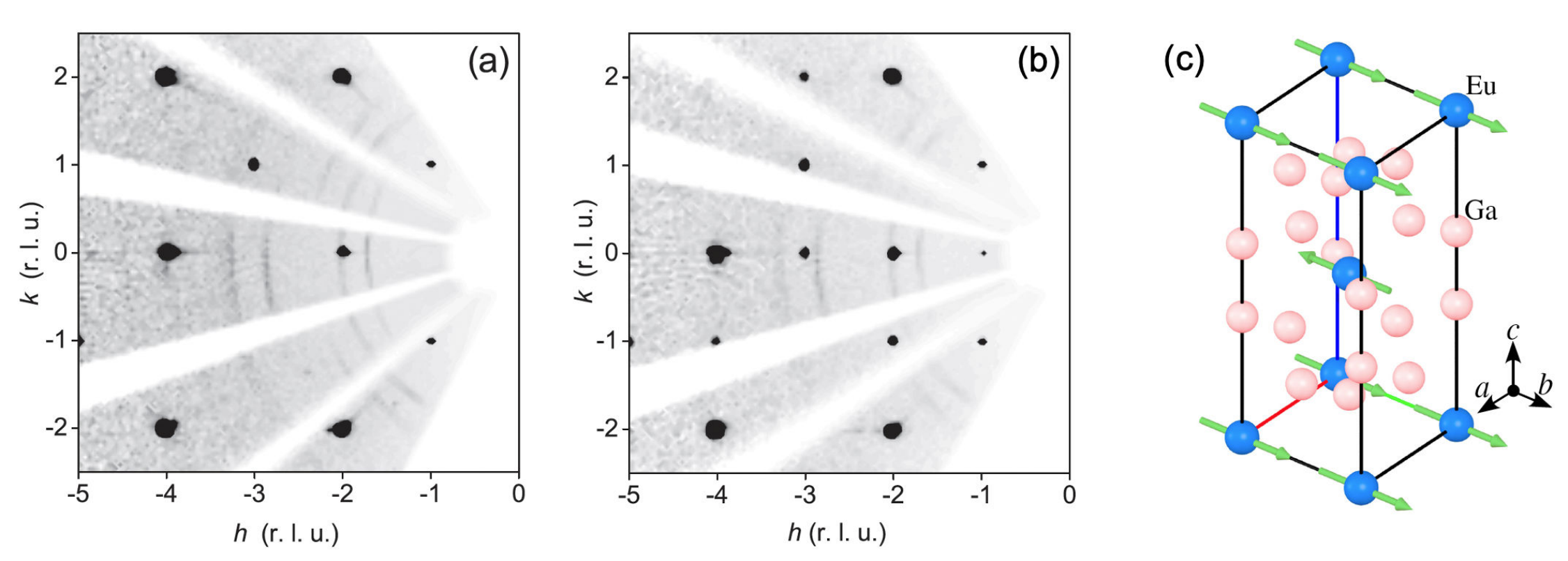}
	%\end{center}
	\centering
	\vspace{-2ex}%
	\caption{\label{fig:n_EuGa4} Neutron diffraction intensities of EuGa$_4$ in the $(h~k~0)$ reciprocal lattice plane collected at 30\,K (a) and 4.5\,K (b). (c) Magnetic structure of EuGa$_4$ as determined by single-crystal neutron diffraction measurements. Figures were reproduced from Ref.~\cite{Kawasaki2016}.}
\end{figure*}
%=== end figure ==========================

Wang \emph{et~al.}\ studied the effects of Si and Cu substitution on the magnetism of EuAl$_4$~\cite{Wang2023b}. They found that Cu or Si doping enhances the ferromagnetic interaction and that, upon Cu substitution,
the direction of easy magnetization changes from the $ab$-plane to the
$c$-axis. Multiple metamagnetic transitions were found in EuAl$_{3.35}$Si$_{0.65}$ for $H \parallel c$, while EuAl$_{3.82}$Cu$_{0.18}$ shows ferromagnetic features for both  $H \parallel c$ and $H \parallel ab$. It is interesting to check if the Si or Cu substitution could tune the topological features in real- and momentum space.

\section{Neutron and synchrotron resonant x-ray diffraction\label{sec:neutron}}

Considering the large neutron-absorption cross section of Eu nuclei,
inelastic neutron scattering experiments are challenging. Nevertheless, elastic neutron scattering experiments have been successfully performed on Eu(Al,Ga)$_4$ single crystals to solve their spin textures~\cite{takagi2022,Kaneko2021,Kawasaki2016}. To mitigate the neutron absorption, a relatively high incident neutron energy was employed for the measurements, and the samples were cut or polished into a slab shape with a typical thickness of $\sim 0.2$--0.6\,mm.

Since EuGa$_4$ exhibits only one AFM transition, we first discuss the magnetic structure of EuGa$_4$. Consistent with the magnetic- and transport properties of EuGa$_4$ single crystal (see section~\ref{sec:trsnsport}), the zero-field magnetic structure is rather simple. Comparing the neutron diffraction intensities collected above and below $T_\mathrm{N}$ [figure~\ref{fig:n_EuGa4}(a)-(b)], magnetic Bragg peaks indexed by a propagation vector $\boldsymbol{q} = (0,0,0)$ were identified in the AFM phase~\cite{Kawasaki2016}. After the absorption correction, neutron diffractions could be described by an A-type AFM magnetic structure, with the moments aligned ferromagnetically in the $ab$-plane, but stacked antiferromagnetically along the $c$-axis [figure~\ref{fig:n_EuGa4}(c)]. The temperature evolution of the ordered magnetic moment follows a scaled Brillouin function with $S = 7/2$, consistent with the zero orbital angular momentum of the Eu$^{2+}$ ions.

Different from EuGa$_4$, richer phase transitions have been observed in EuAl$_4$.  The CDW order below $T \sim 140$\,K has been confirmed also by neutron diffraction. Figure~\ref{fig:CDW} presents the neutron diffraction intensity map for EuAl$_4$ in the $(h,0,l)$ reciprocal lattice plane measured at 30\,K. Satellite peaks indexed by $\boldsymbol{q}_\mathrm{CDW} = (0,0,0.19)$ were observed below $T_\mathrm{CDW}$, indicating the development of a CDW superstucture along the $c$-axis. The details regarding the CDW order can be found in section~\ref{sec:structure}; in this section we focus only on the magnetic order.

\begin{figure}[!tp]
	%\begin{center}
	\includegraphics[width=1.0\linewidth]{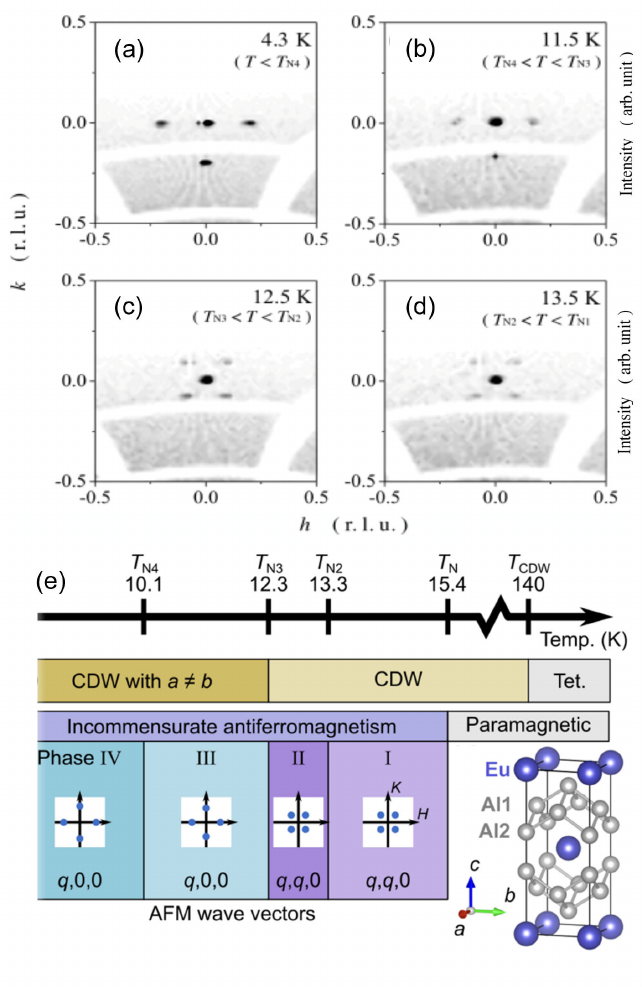}
	%\end{center}
	\centering
	\vspace{-4ex}%
	\caption{\label{fig:n_old}Neutron diffraction intensity maps of EuAl$_4$ in the $(h~k~4)$ reciprocal lattice plane measured at (a) 4.3\,K, (b) 11.5\,K, (c) 12.5\,K, and (d) 13.5\,K. (e) Summary of the low-temperature phases of an EuAl$_4$ single crystal. Figures reproduced from Refs.~\cite{Kaneko2021,Meier2022}.}
\end{figure}

The successive magnetic transitions occurring at low temperature
were also observed by Kaneko \emph{et al.}\ through neutron diffraction measurements~\cite{Kaneko2021}. Figure~\ref{fig:n_old} summarizes the temperature evolution of the diffraction patterns in the $(h,k,4)$ reciprocal lattice plane. At $T$ = 4.3\,K ($T < T_\mathrm{N4}$, phase IV) and 11.5\,K~($T_\mathrm{N4} < T < T_\mathrm{N3}$, phase III), magnetic Bragg peaks appear at $\boldsymbol{q_1}$ = ($\delta_1$,0,0), where $\delta_1$ varies from 0.194(5) to 0.17(1)
with increasing temperature. Upon heating the sample to  $T$ = 12.5\,K ($T_\mathrm{N3} < T < T_\mathrm{N2}$, phase II) and 13.5\,K ($T_\mathrm{N2} < T < T_\mathrm{N1}$, phase I), magnetic Bragg peaks appear at $\boldsymbol{q_2}$ = ($\delta_2$,$\delta_2$,0), where $\delta_2$ = 0.085(4) and  0.086(4) for 12.5 and 13.5\,K, respectively. No visible change in $\delta_2$ was found at $T_\mathrm{N2}$
beyond experimental accuracy. By contrast, a clear peak shift was observed between 4.3 and 11.5\,K through $T_\mathrm{N4}$. Since no higher-order harmonics were observed, it was proposed that a transition from an am\-pli\-tude-mo\-du\-la\-ted-ty\-pe to an equal-moment type structure may explain the transitions at $T_\mathrm{N2}$ and $T_\mathrm{N4}$. \tcr{The magnetic transition temperatures  $T_\mathrm{N1}$ to $T_\mathrm{N4}$ are summarized in figure~\ref{fig:n_old}(e).}
Note that, a quantitative analysis of neutron diffraction data is rather difficult owing to the strong wavelength-dependent neutron absorption. Further neutron diffraction experiments on Eu(Al,Ga)$_4$ single crystals with isotope substitution will be necessary to reveal their magnetic structures.

%==== figure =============================%
%=== end figure ==========================

\begin{figure}[!htp]
	%\begin{center}
	\vspace{-1ex}%
	\includegraphics[width=1.05\linewidth]{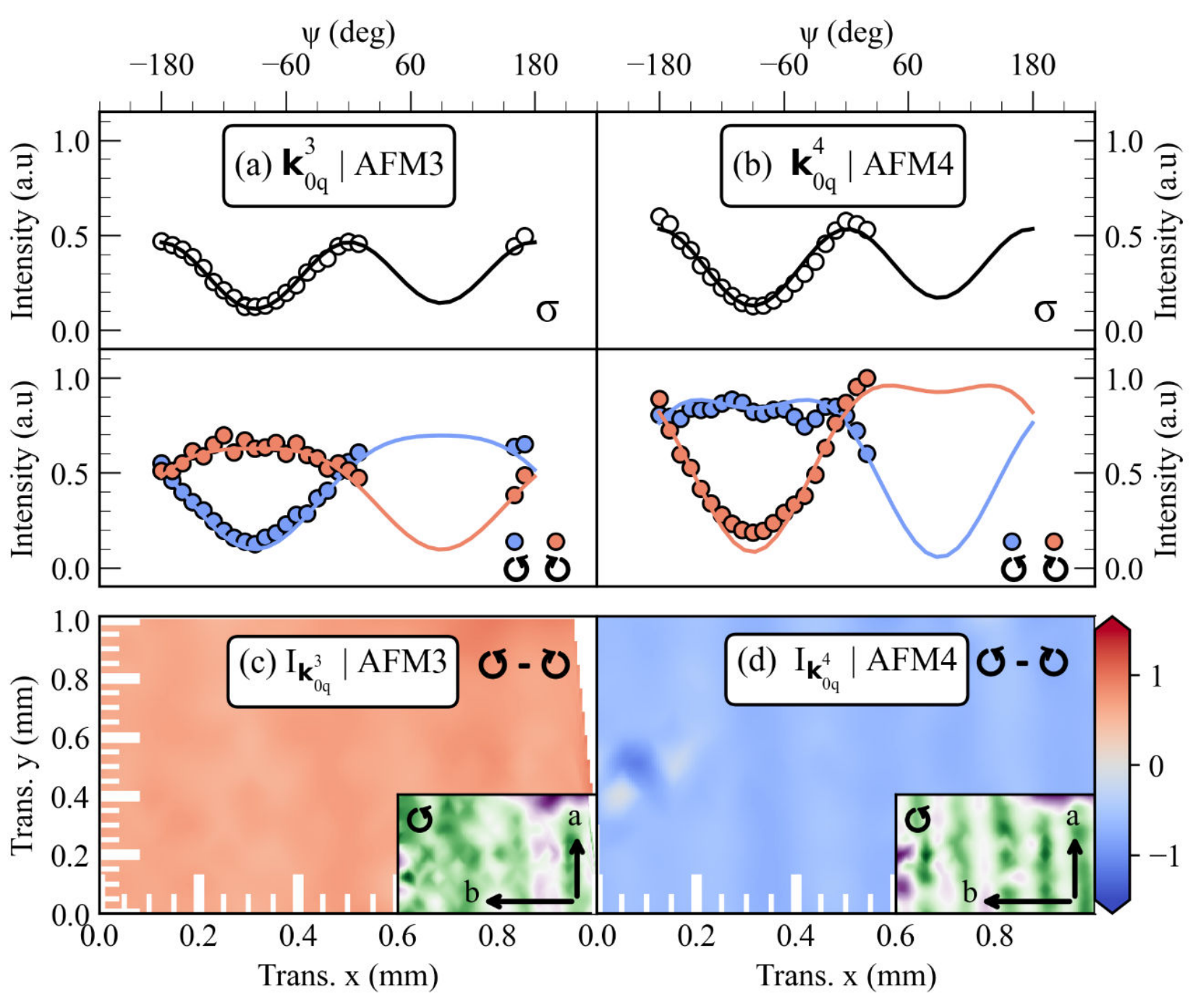}
	%\end{center}
	\centering
	\vspace{-3ex}%
	\caption{\label{fig:EuAL_REXS}Scattered intensity from the $k_\mathrm{0q}$ satellite of the (0,0,8) re\-flec\-tion collected as a function of azimuth in the phase-III at 10.6\,K (a) and in phase-IV at 7\,K (b). Here, $k_\mathrm{0q}^3$and $k_\mathrm{0q}^4$ are defined similarly to $q_1$ = ($\delta_1$,0,0) in figure~\ref{fig:n_old}. Data were collected using the incident light with 
	linear horizontal ($\sigma$), circular left (CL, $\circlearrowleft$), circular right (CR, $\circlearrowright$) polarization, respectively. Maps showing the variation in the difference in intensity of the $k_\mathrm{0q}$ satellite collected with incident CL and CR light at 10.6\,K in phase-III (c)
	and at 7\,K in phase-IV (d). Insets show only the maps of the CL light. Figures were reproduced from Ref.~\cite{Vibhakar2024}.}
\end{figure}

Recent resonant elastic x-ray scattering (REXS) measurements in EuAl$_4$ found similar results to the single-crystal neutron diffraction studies, but Vibhakar \emph{et al.}\ found also additional magnetic propagation vectors with $q$ = (0.175,0,0) and (0,0.178,0) below $T_\mathrm{N2}$ [i.e., phase II in figure~\ref{fig:n_old}(e)]~\cite{Vibhakar2024}. According to the new REXS data, the magnetic structures of all four different AFM phases 
of EuAl$_4$ are single-$k$ rather than multi-$k$ structures. In phase I ($T_\mathrm{N2} < T < T_\mathrm{N1}$), the Eu moments form a spin-density-wave (SDW) order, where the moments are oriented in the $ab$-plane and perpendicular to the $q$ vector (figure~\ref{fig:n_old}). In phase II ($T_\mathrm{N3} < T < T_\mathrm{N2}$), another SDW order develops, with the moments aligned along the $c$-axis, which coexists with the first SDW order. Upon further decreasing the temperature, in phase III ($T_\mathrm{N4} < T < T_\mathrm{N3}$), a magnetic helix structure with a single chirality is stabilized. Finally, in phase IV
($T < T_\mathrm{N4}$), the chirality of the magnetic helix structure is reversed, but the crystal remains a single chiral domain. As shown in figure~\ref{fig:EuAL_REXS}, when cooling the EuAl$_4$ crystal from phase III to phase IV,
%
%
%==== figure =============================%
%=== end figure ==========================
%
%
the sign of difference in %
intensity reverses, demonstrating the opposite magnetic chirality of
these two AFM phases.  
Further group-theory analysis suggests that, for $T < T_\mathrm{N3}$,
the symmetry is reduced to a polar monoclinic one, which is necessary
to explain the observed asymmetry in the chiral states of the magnetic
helix and the spin chiral reversal~\cite{Vibhakar2024}.

\begin{figure*}[!t]
	%\begin{center}
	\includegraphics[width=0.95\linewidth]{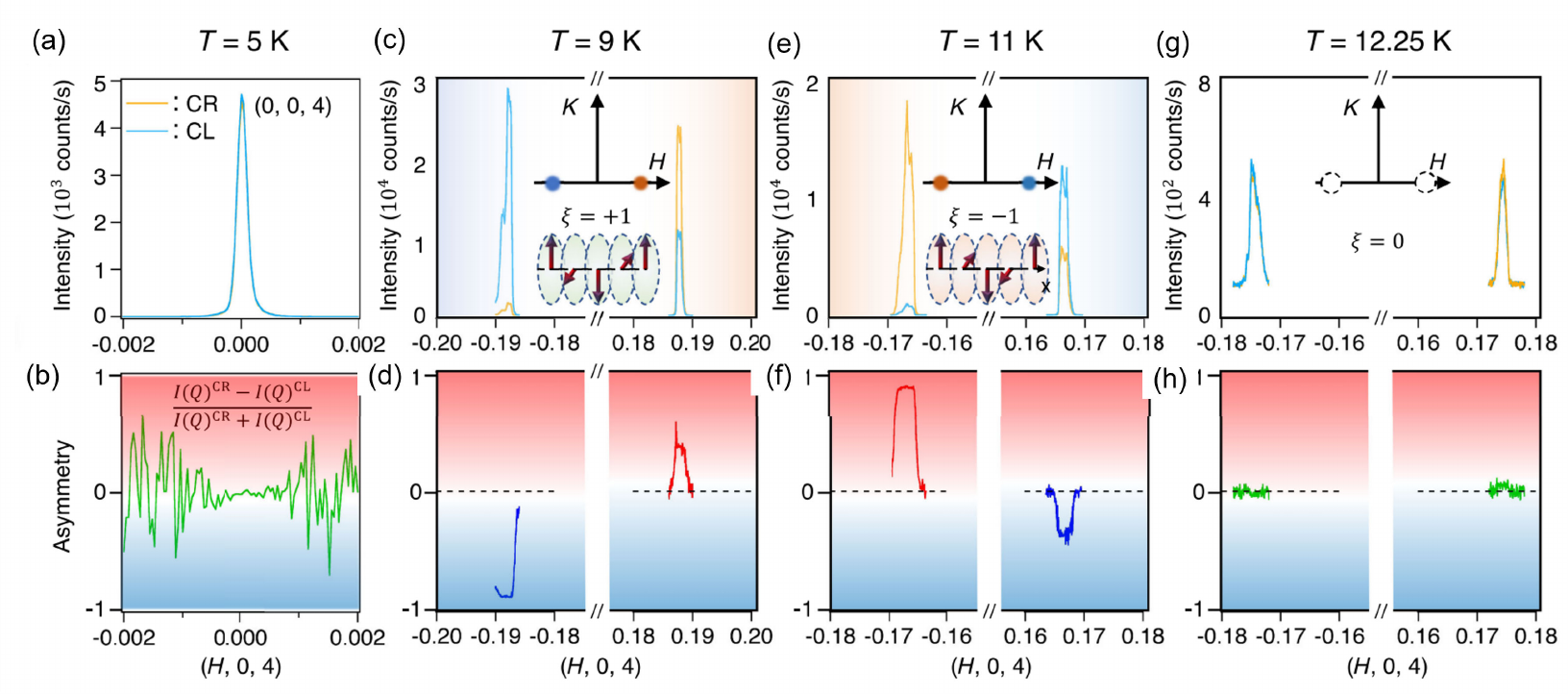}
	%\end{center}
	\centering
	\vspace{-4ex}%
	\caption{\label{fig:EuAl_CD}Circular dichroism of the structural
	and magnetic Bragg peaks. Yellow and cyan curves in (a), (c), (e), and (g) represent CR and CL incident photon polarizations, respectively.
	Red, green, and blue curves in (b), (d), (f), and (h) represent positive, zero, and negative asymmetry of the CD $F(Q)$. The definition of $F(Q)$ is shown in panel (b), where $I(Q)^\mathrm{CR}$ and $I(Q)^\mathrm{CL}$ represent the x-ray intensities obtained under a CR and CL incident photon energy, respectively. Figures were reproduced from Ref.~\cite{Miao2024}.}
\end{figure*}

The chiral reversal between phase III and IV was further confirmed by
recent circular-dichroism (CD) mea\-sure\-ments of the Bragg peaks via resonant magnetic x-ray scattering (XRMS). CD XRMS is a direct experimental probe of chiral electronic orders. Miao \emph{et al.}\ performed CD XRMS
on an EuAl$_4$ single crystal in its different magnetic phases~\cite{Miao2024}. At $T$ = 5\,K, the $I(Q)^\mathrm{CR}$ and $I(Q)^\mathrm{CL}$ are almost identical, resulting in a zero asymmetry of CD [$F(Q) \sim 0$] [figure~\ref{fig:EuAl_CD}(a)-(b)]. Upon increasing the temperature to 9\,K (phase IV), a giant CD at both $\boldsymbol{q}_\mathrm{SDW}$ and $-\boldsymbol{q}_\mathrm{SDW}$ emerges, implying that the helical SDW below $T_\mathrm{N4}$ is chiral, with a spin chirality $\chi$ =  1 [figure~\ref{fig:EuAl_CD}(c)-(d)]. In phase III ($T$ = 11\,K), the CD changes sign, which indicates a spontaneous chirality flipping from $\chi$ = 1 to $\chi$ = --1 [figure~\ref{fig:EuAl_CD}(e)-(f)]. The chiral density is back to nearly zero upon warming up the sample to phase II. These results are highly consistent with the REXS data shown in figure~\ref{fig:EuAL_REXS}.

%==== figure =============================%
%=== end figure ==========================

%==== figure =============================%
\begin{figure*}[!t]
	%\begin{center}
	\includegraphics[width=0.85\linewidth]{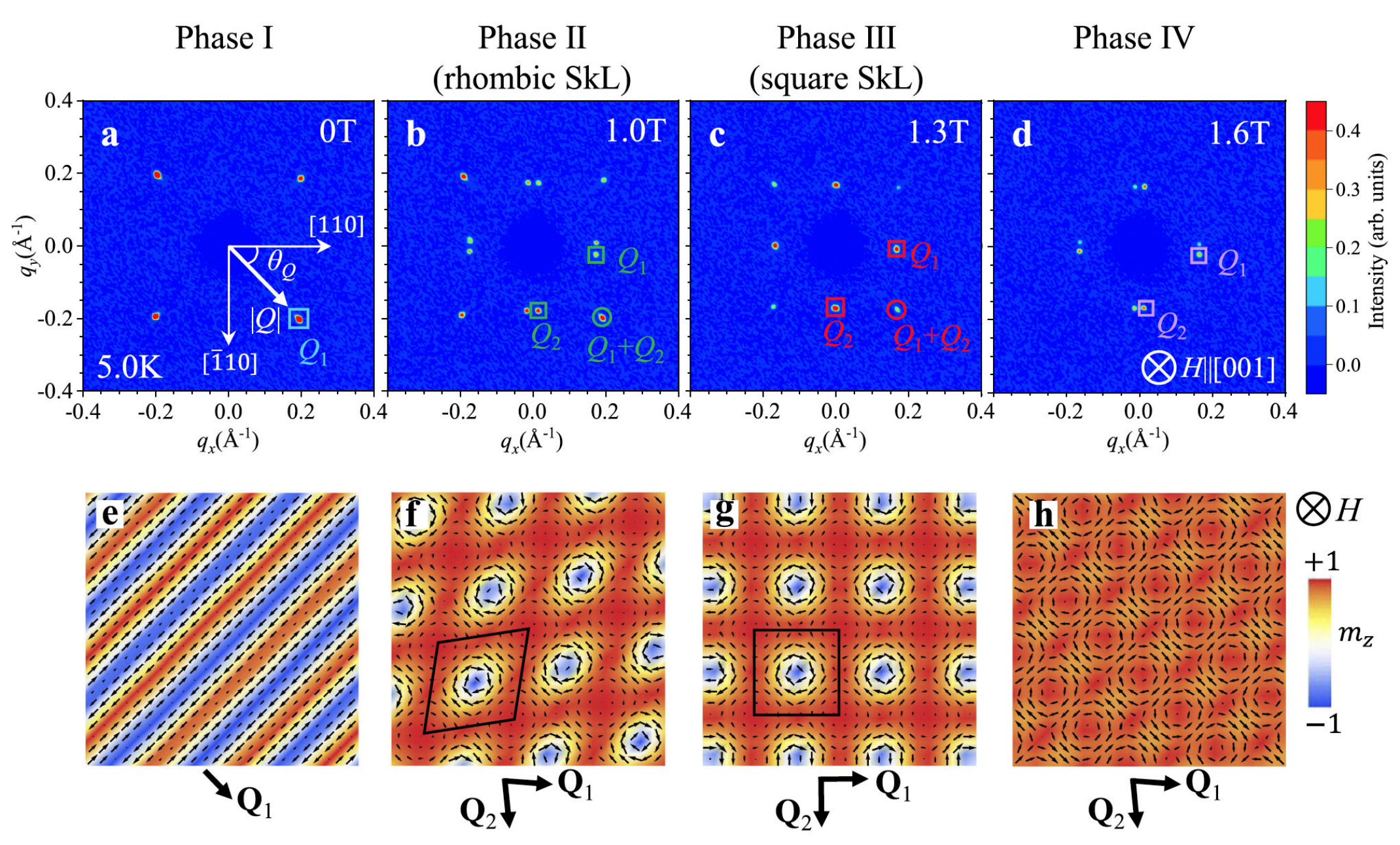}
	%\end{center}
	\centering
	\vspace{-2ex}%
	\caption{\label{fig:n_sans}Magnetic-field dependence of SANS patterns for EuAl$_4$. (a)-(d) Typical patterns collected at $T$ = 5\,K under various magnetic fields 
	for $H \parallel c$. (e)-(h) Schematics of the screw, rhombic skyrmion lattice, square skyrmion lattice, and vortex-lattice spin textures, respectively. Each phase is characterized by distinctive orientations of the fundamental magnetic-modulation vectors $Q_1$ and $Q_2$. Background color represents the out-of-plane component of the local magnetic moment $m_z$. Here, phases I-IV are defined differently from those in Fig.~\ref{fig:Skyrmion_diagram}. Figures were reproduced from Ref.~\cite{takagi2022}. }
\end{figure*}
%=== end figure ========================== 

%==== figure =============================%
\begin{figure}[!htp]
	%\begin{center}
	\includegraphics[width=1.0\linewidth]{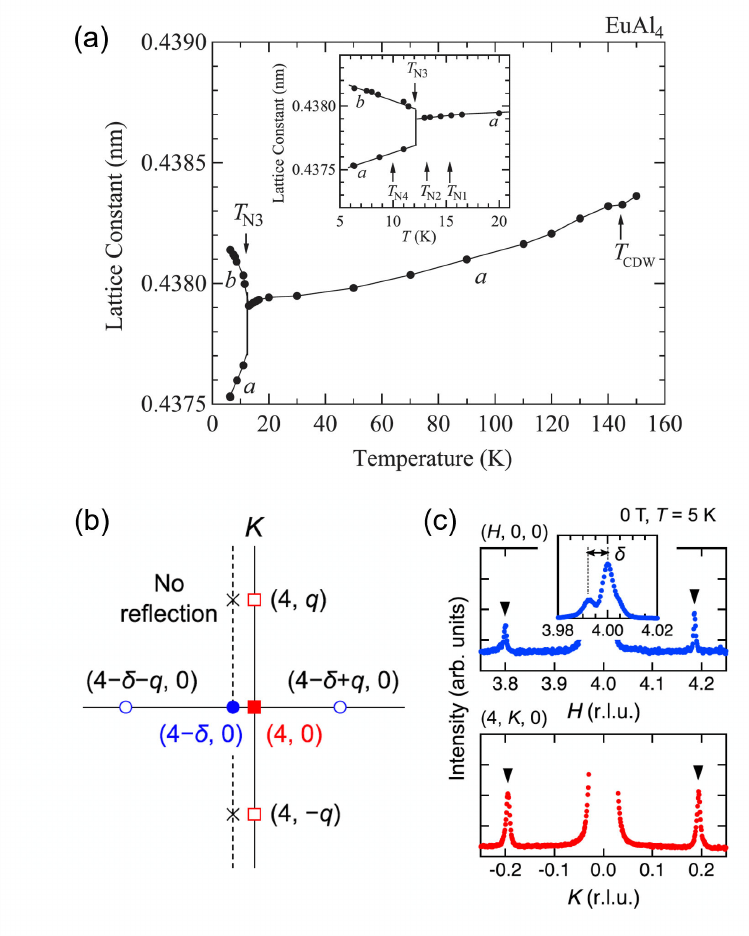}
	%\end{center}
	\centering
	\vspace{-5ex}%
	\caption{\label{fig:rxd}(a) Lattice constants vs.\ temperature
		de\-ter\-mined via syn\-chro\-tron x-ray diffraction measurements. (b)  Schematic plots of the observed fundamental- and magnetic Bragg reflections on the scattering plane at $T$ = 5\,K in zero field. (c) Intensity profiles of the $(h,0,0)$ and $(4,k,0)$ scans at $T$ = 5\,K in zero field. The inset enlarges the splitting of the fundamental (400) reflection. Figures were reproduced from Refs.~\cite{Shimomura2019,Gen2023}.}
\end{figure}
%=== end figure ==========================

When applying the magnetic field along the $c$-axis, EuAl$_4$ undergoes a series of
metamagnetic transitions, and a topological Hall resistivity appears
between such transitions (figure~\ref{fig:EuA4_Hall_diagram}), indicating the presence of possible topological spin textures in EuAl$_4$. To clarify this issue, Takagi \emph{et al.}\ performed small-angle neutron scattering (SANS) measurements
in the different magnetic phases of an EuAl$_4$ single crystal (figure~\ref{fig:n_sans})~\cite{takagi2022}. The SANS patterns were collected at $T = 5$\,K, in an
applied magnetic field of 0, 1.0, 1.3, and 1.6\,T, corresponding to
the phases I to IV in figure~\ref{fig:Skyrmion_diagram}. Note that,
the numbering of the various phases in Refs.~\cite{Kaneko2021,Meier2022,Vibhakar2024} is reversed with respect to Ref.~\cite{takagi2022}. In the absence of a magnetic field, the fundamental modulation vector is $\boldsymbol{Q_1}$ = (0.19,0,0), and the polarization analysis confirms the helical spin texture [figure~\ref{fig:n_sans}(e)], consistent with REXS results~\cite{Vibhakar2024}. In a 1.0\,T field, the fundamental modulation vector changes to (0.073,0.097,0), leading to pairs of magnetic Bragg peaks due to the domain effect. Surprisingly, strong $Q_1$+$Q_2$ reflections have been observed along the (100) direction, which was also confirmed through REXS~\cite{takagi2022}. The existence of  $Q_1$+$Q_2$ reflections reveals the double-$Q$ character of the spin textures, in contrast to the single-$Q$ structure in the absence of a magnetic field. 
A polarization analysis of the SANS data confirms that both $Q_1$ and $Q_2$ modulations are of helical character, as in the zero-field case. Since the modulation vectors $Q_1$ and $Q_2$ are not perpendicular to each other, the spin textures  can be represented as a rhombic skyrmion lattice [figure~\ref{fig:n_sans}(f)]. By further increasing the magnetic field to 1.6\,T, the spin textures remain double-$Q$, leading to a square lattice of magnetic skyrmions in phase III [figure~\ref{fig:n_sans}(g)] and vortices in phase IV [figure~\ref{fig:n_sans}(h)], respectively. Consequently, the
presence of magnetic skyrmions leads to a THE in EuAl$_4$.

%==== figure =============================%
\begin{figure}[!htp]
	%\begin{center}
	\includegraphics[width=1.0\linewidth]{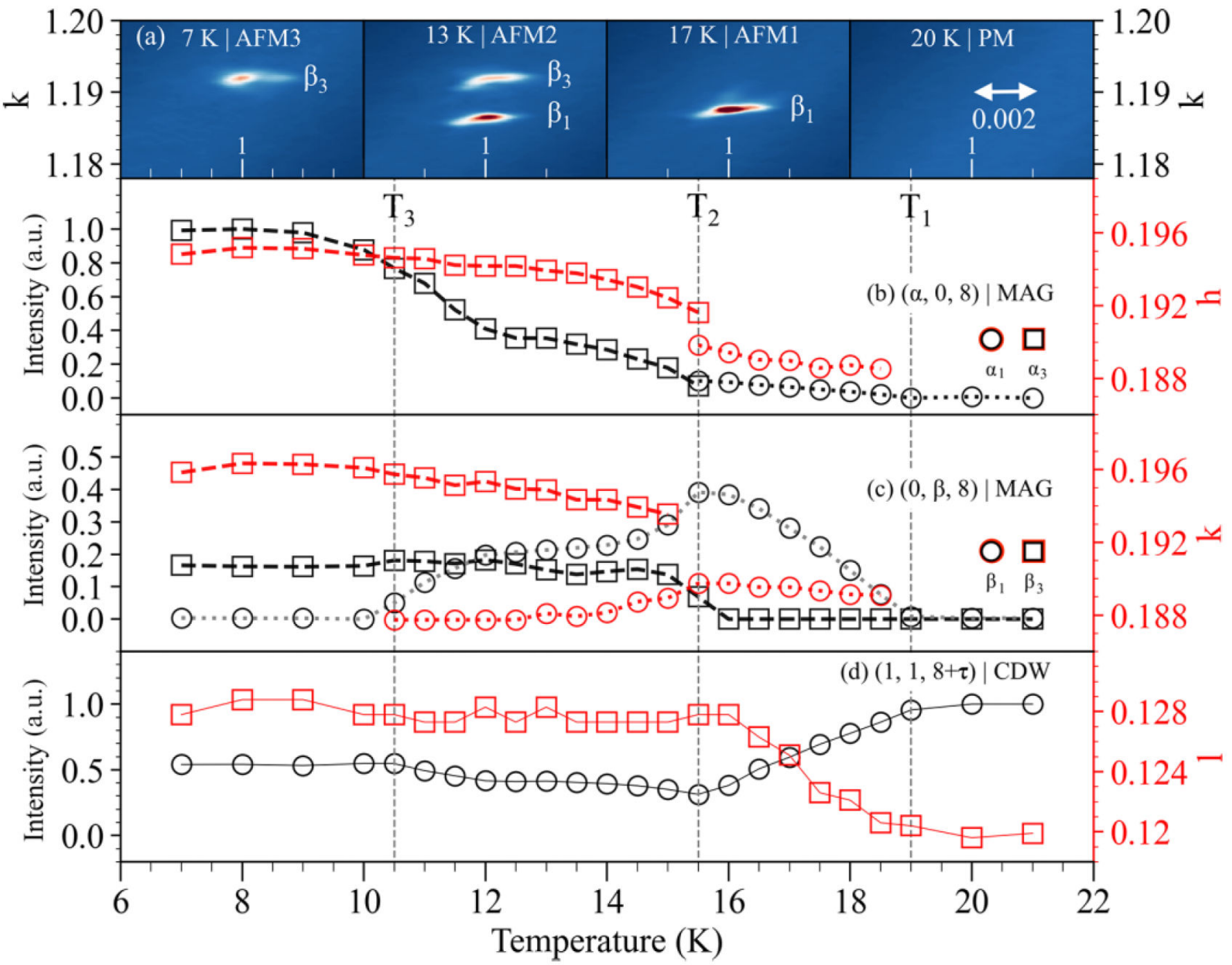}
	%\end{center}
	\centering
	\vspace{-3ex}%
	\caption{\label{fig:rxd_2}(a) Reciprocal space maps centered about
		the (1, 1+$\beta$, 8) reflection in the $hk$ plane collected in the
		PM, AFM1, AFM2, and AFM3 phases. Temperature dependence of the normalized integrated intensity and the magnitude of the propagation vector of (b) ($\alpha$, 0, 8), (c) (0, $\beta$, 8), and (d) (1, 1, 8+$\tau$) reflections.
		Figure was reproduced from Ref.~\cite{Vibhakar2023}.}
\end{figure}
%=== end figure ==========================

Although a complete understanding of the rich electronic phases of EuAl$_4$ is not yet available,
recent x-ray diffraction experiments suggest a sizable spin-lattice coupling, which plays an important role in determining the various spin textures in EuAl$_4$~\cite{Shimomura2019,Gen2023}. Shimomura \emph{et al.}\ performed non-resonant synchrotron x-ray diffraction measurements in a wide temperature range in EuAl$_4$. Except for the anomaly at $T_\mathrm{CDW}$, EuAl$_4$ exhibits a tetragonal-orthorhombic structural transition at $T_\mathrm{N3}$, where the unit cell is compressed and elongated along the $a$- and $b$-axis, respectively [see figure~\ref{fig:rxd}(a)]. However, no structural anomaly could be detected at the other magnetic transitions (i.e., at $T_\mathrm{N1}$, $T_\mathrm{N2}$, and $T_\mathrm{N4}$). The magnetic structure of EuAl$_4$ changes significantly
at $T_\mathrm{N3}$, from a SDW to a helix, while the change is moderate at $T_\mathrm{N2}$ and $T_\mathrm{N4}$. 
Although the mechanism underlying the lattice distortion at $T_\mathrm{N2}$ is not yet fully understood, it has a huge impact on the spin textures. Gen \emph{et al.}\ performed RXS experiments on EuAl$_4$ at $T$ = 5\,K [figure~\ref{fig:rxd}(b)-(c)]~\cite{Gen2023}. In the absence of a magnetic field, the magnetic modulation vectors are parallel to the elongated $b$-axis. The rhombic- and square skyrmion phases were found to possess a $\sim 0.1$\% and $\sim0.03$\% orthorhombic 
distortion in the $ab$ plane, respectively, implying a
sizable coupling between the spins and lattice in EuAl$_4$.

Vibhakar \emph{et al.}\ performed high-resolution resonant x-ray and neutron scattering studies on a EuAl$_2$Ga$_2$ single crystal~\cite{Vibhakar2023}. Similar to the EuAl$_4$ case, also in EuAl$_2$Ga$_2$, the CDW order ($T_\mathrm{CDW} = 50$\,K) and the AFM orders (at $T_\mathrm{N1} = 19.5$\,K, $T_\mathrm{N2} = 15$\,K, $T_\mathrm{N3} = 11$\,K) coexist~\cite{Moya2022}. The magnetic orders are denoted as AFM1 ($T_\mathrm{N1} < T < T_\mathrm{N2}$), AFM2 ($T_\mathrm{N3} < T < T_\mathrm{N2}$), and AFM3 ($T < T_\mathrm{N3}$), respectively. The magnetic propagation vectors of EuAl$_2$Ga$_2$ share some common features with EuAl$_4$. For instance, the intensity of CDW order is significantly reduced below $T_\mathrm{N1}$ and it becomes almost temperature-independent below $T_\mathrm{N2}$ (figure~\ref{fig:rxd_2}). Clearly, the formation of magnetic order suppresses the CDW order, demonstrating that the two electronic orders are in competition~\cite{Vibhakar2023}. Similar results have been found also in pure EuAl$_4$~\cite{Miao2024}, where the CDW couples strongly with the SDW and displays a rare commensurate-to-incommensurate transition at the chirality flipping temperature. Note that, the CDW order is formed by the \emph{itin\-er\-ant} electrons of Al atoms, while the \emph{localized} 4$f$ electrons of Eu atoms are responsible for the magnetic orders in Eu(Al,Ga)$_4$.
If the onset of the magnetic or\-der po\-la\-ri\-zes the itinerant electronic density, it could destabilize the CDW order. Another mechanism for the coupling between the SDW and CDW orders involves lattice strain. The magnetic order renormalizes the CDW order through the magnetoelastic coupling. This is clearly illustrated in figure~\ref{fig:rxd}(a), where the structural distortion appears at $T_\mathrm{N3}$ in EuAl$_4$. 
These observations suggest a strong coupling between the localized 4$f$ electrons and itinerant electrons in Eu(Al,Ga)$_4$, and a theoretical modeling for the spin correlations in Eu(Al,Ga)$_4$ should consider the perturbations caused by the lattice distortion and/or charge order, which might also play a significant role in the formation of the skyrmion lattice.

\section{Muon-spin rotation and relaxation\label{sec:muon}}
During the past decades, muon-spin rotation, relaxation, and resonance
(known as {\textmu}SR), has become one of the most powerful methods to
study the quantum materials at a microscopic level.
Due to the large muon magnetic moment ({\textmu}$_{\mu} = 8.89$\,{\textmu}$_\mathrm{N}$) and almost 100\% spin-polarized muon beams, muons can sense
extremely small internal fields ($\sim 10^{-2}$\,mT), and thus, probe local magnetic fields of either nuclear or electronic nature. 
In addition, since the muon is an elementary spin-1/2 particle, it acts as a purely magnetic probe, i.e., free of quadrupole interactions. All these features make {\textmu}SR an ideal technique for investigating the intrinsic magnetic properties of materials in the absence of external fields.
Depending on the reciprocal orientation of the external magnetic field $B$ with respect to the initial muon-spin direction $S$, in a {\textmu}SR experiment, two different configurations are possible: i) In transverse field
(TF) {\textmu}SR the externally applied field  $B$ is perpendicular to $S$ and the muon spin precesses around  $B$; ii) In a longitudinal field (LF) configuration the applied field is parallel to $S$, generally implying only a relaxing {\textmu}SR signal. In the absence of external magnetic fields, both configurations can be used to conduct zero-field (ZF)-{\textmu}SR measurements. In a weak transverse-field (wTF)-{\textmu}SR, where the applied
magnetic field is relatively weak (i.e., a few millitesla), one can
determine the temperature evolution of the magnetic volume fraction, as well as the magnetic ordering temperature. As for the ZF- and LF-{\textmu}SR, they can be used to study the temperature evolution of the magnetically ordered phase and the dynamics of spin fluctuations. 
Here, we outline briefly only the basics of the {\textmu}SR technique and direct the reader to other references for more detailed information~\cite{Lee1999,Yaouanc2011,Blundell2021,Amato2024,Amato1997,Blundell1999,Nuccio2014,Hillier2022}.

%==== figure =============================%
\begin{figure}[!tp]
	%\begin{center}
	\includegraphics[width=1.0\linewidth]{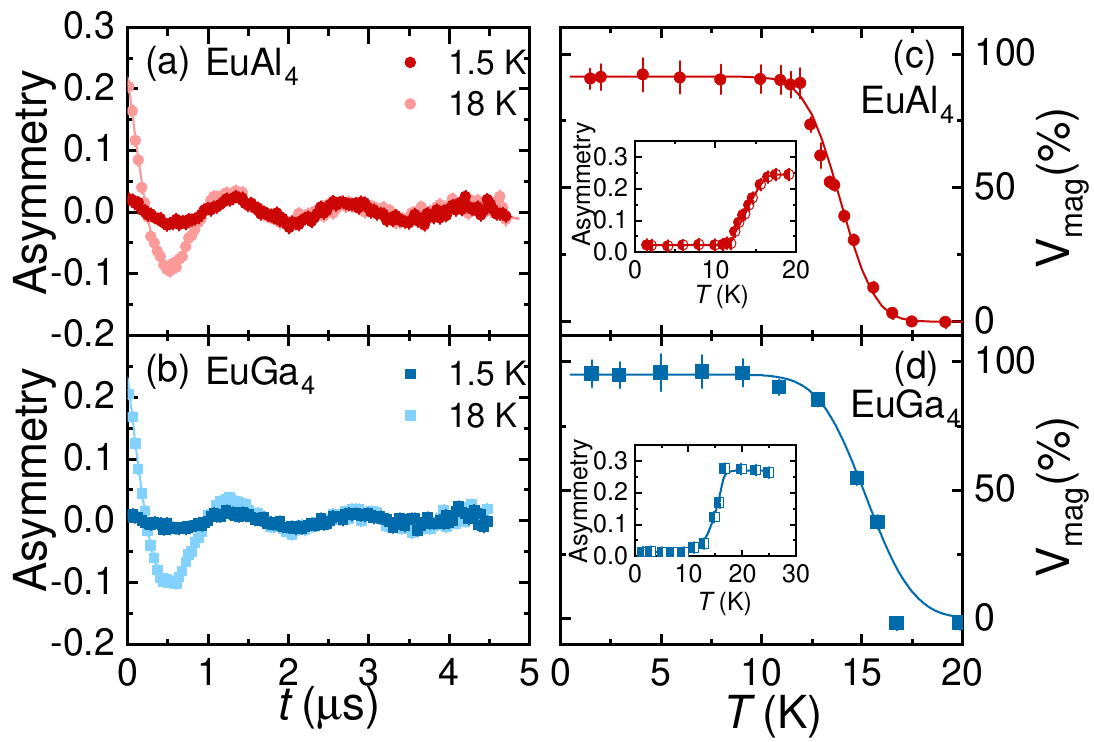}
	%\end{center}
	\centering
	\vspace{-4ex}%
	\caption{\label{fig:wTF-muSR}wTF-{\textmu}SR spectra of (a) EuAl$_4$ and
		(b) EuGa$_4$ single crystals, collected in the AFM (1.5\,K) and PM
		(18\,K) states in a weak transverse field of 5\,mT. The solid lines
		represent fits to 	$A_\mathrm{wTF}(t) = A_\mathrm{NM} \cos(\gamma_{\mu} B_\mathrm{int} t + \phi) \cdot e^{-\lambda t}$ (see details in Ref.~\cite{Zhu2023}). 	
		Temperature-dependent magnetic volume fraction for (c) EuAl$_4$ and (d) EuGa$_4$. The insets show the wTF-{\textmu}SR asymmetry $A_\mathrm{NM}$ vs temperature. 
		Data were taken from Ref.~\cite{Zhu2023}.}
\end{figure}
%=== end figure ==========================

Zhu \textit {et~al.} performed systematic {\textmu}SR measurements on
both EuAl$_4$ and EuGa$_4$ single crystals. Figure~\ref{fig:wTF-muSR}
sum\-ma\-rizes the wTF-{\textmu}SR results. 
In the PM state (i.e., at 18\,K), a weak transverse
field of 5\,mT leads to oscillations [figure~\ref{fig:wTF-muSR}(a)-(b)]. In the long-range ordered AFM state (i.e., at 1.5 K), the applied field is much smaller than the internal field.
As a consequence, upon entering the AFM state, muon spins precess with
frequencies that reflect the internal fields at the muon-stopping sites
rather than the weak applied field. Normally, the magnetic order leads
to a very fast muon-spin depolarization in the first tenths of {\textmu}s.
In the PM state, instead, all the implanted muons precess at the same frequency $\gamma_{\mu}$$B_\mathrm{int}$, 
determined by the external magnetic field. As the temperature approaches $T_\mathrm{N}$,
only the muons implanted in the remaining PM or non-magnetic (NM) phase precess at the
frequency $\gamma_{\mu}$$B_\mathrm{int}$, here reflected in a reduced
oscillation amplitude.  
The PM (or NM) sample fraction is determined by the oscillation
amplitude. In both EuAl$_4$ and EuGa$_4$,
$A_\mathrm{NM}$ starts to
decrease near the onset of AFM order  [see insets in
figure~\ref{fig:wTF-muSR}(c)-(d)]. The determined magnetic volume fraction $V_\mathrm{mag}$, 91\% for EuAl$_4$ and 95\% for EuGa$_4$, indicate a fully magnetically ordered state at low temperature. 
The transition temperatures $T_\mathrm{N}$ can also be tracked
from $V_\mathrm{mag}(T)$. Their onsets at $\sim 16.5$\,K and
$\sim 16.7$\,K for EuAl$_4$ and EuGa$_4$ [figure~\ref{fig:wTF-muSR}(c)-(d)], are in very good agreement with the magnetic susceptibility and electrical resistivity data (see details in section~\ref{sec:trsnsport}).

%==== figure =============================%
\begin{figure*}[!htp]
	%\begin{center}
	\includegraphics[width=1.0\linewidth]{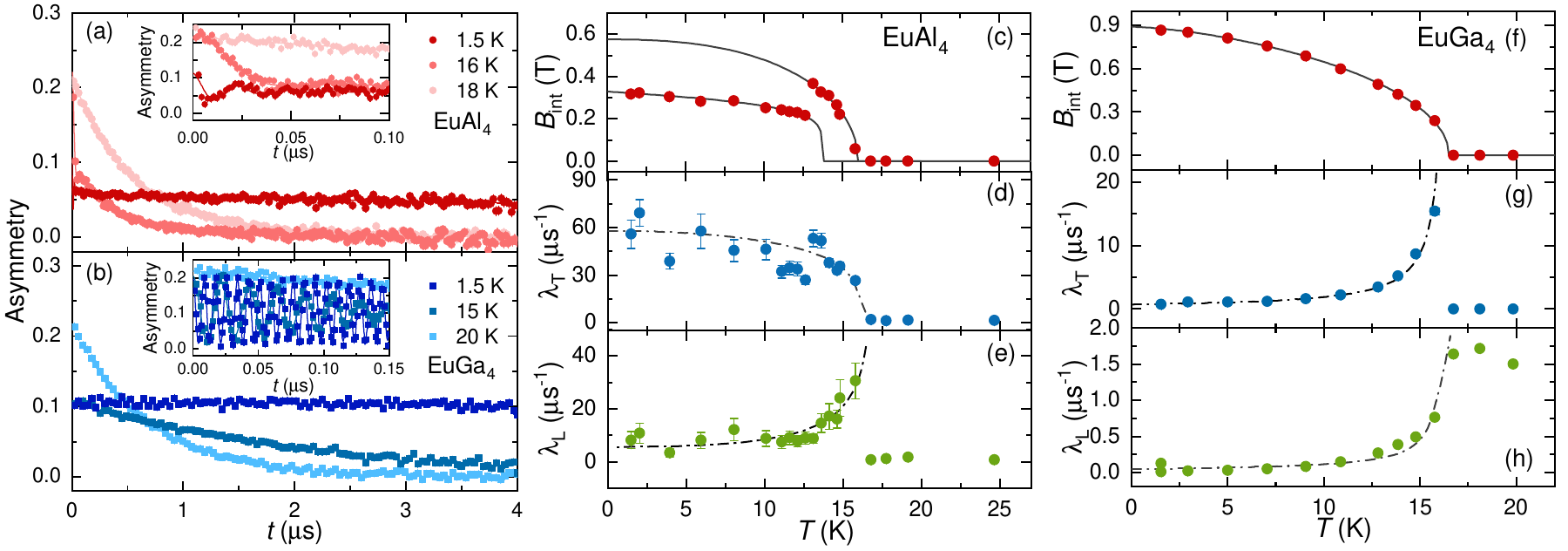}
	%\end{center}
	\centering
	\vspace{-3ex}%
	\caption{\label{fig:ZF-muSR} Representative ZF-{\textmu}SR spectra 
		collected in a transverse muon-spin configuration at temperatures covering both 
		the PM and AFM states of EuAl$_4$ (a) and EuGa$_4$ (b), respectively.	Insets highlight the short-time spectra, illustrating the 
		coherent oscillations caused by the long-range AFM order. 
		Temperature dependence of the internal field $B_\mathrm{int}(T)$ (c), transverse muon-spin relaxation rate (known also as damping rate) $\lambda_\mathrm{T}(T)$ (d), and longitudinal muon-spin relaxation rate  $\lambda_\mathrm{L}(T)$ (e) for EuAl$_4$, 
		as derived from ZF-{\textmu}SR analysis. The analogous results for EuGa$_4$ are shown in panels (f)-(h). The solid lines through the data represent fits to the theoretical models reported in Ref.~\cite{Zhu2023}, while the dash-dotted lines are guides to the eyes.
		Data were taken from Ref.~\cite{Zhu2023}.}
\end{figure*}
%=== end figure ==========================

As for the ZF-{\textmu}SR spectra, both crystals show typical
features of magnetically ordered materials (figure~\ref{fig:ZF-muSR})~\cite{Amato1997,Blundell1999,Hillier2022,Nuccio2014}. 
In the PM state, the {\textmu}SR spectra still exhibit a relatively fast
muon-spin depolarization ($\sim 2$\,{\textmu}s$^{-1}$), 
implying the existence of strong spin fluctuations. In EuAl$_4$
and EuGa$_4$, the nuclear moments lead only to a small muon-spin
relaxation rate, with typical values of less than
0.1\,{\textmu}s$^{-1}$~\cite{Fujita2020}. 
The {\textmu}SR spectra in the AFM state ($T \le T_\mathrm{N}$) are characterized by highly damped oscillations [see insets in figure~\ref{fig:ZF-muSR}(a)-(b)], superimposed on a slowly decaying relaxation, observable only at long times. 
There are two nonequivalent muon-stopping sites in EuAl$_4$, while a single muon-stopping site is sufficient to describe the ZF-{\textmu}SR spectra in
EuGa$_4$. In EuAl$_4$, muons stopping at the second site do not undergo
any precession, but show only a slow longitudinal relaxation, with a
similar temperature dependence to $\lambda_\mathrm{L}(T)$
[figure~\ref{fig:ZF-muSR}(e)].

EuAl$_4$ and EuGa$_4$ show rather different temperature-dependent $B_\mathrm{int}(T)$ [figure~\ref{fig:ZF-muSR}(c) and (f)]. Since $B_\mathrm{int}$ is directly proportional to  the magnetic moment, the evolution of $B_\mathrm{int}$ reflects that of the magnetic structure.
According to neutron scattering studies (see details in section~\ref{sec:neutron}), in EuAl$_4$, the magnetic $q$-vector changes from $\boldsymbol{q}_1$ = (0.085, 0.085, 0) at $T_\mathrm{N}$ = 13.5\,K to $\boldsymbol{q}_2$ = (0.170, 0, 0) at 11.5\,K and slightly to $\boldsymbol{q}_3$ = (0.194, 0, 0) at 4.3\,K~\cite{Kaneko2021}. Therefore, the drop of $B_\mathrm{int}$ at $\sim$13\,K is attributed to
a change in magnetic structure from $q_1$ to $q_2$. 
The modification of magnetic structure from $q_2$ to $q_3$ with the magnetic moments pointing at the same direction is too tiny to have a measurable effect on $B_\mathrm{int}$. Unlike EuAl$_4$, the AFM structure of EuGa$_4$ is rather simple [with a single magnetic vector, $\boldsymbol{q} = (0, 0, 0)$, down to 2\,K]~\cite{Kawasaki2016}, leading to a monotonic decrease in $B_\mathrm{int}(T)$ as the temperature increases. The significantly different internal field values are most likely attributed to the different muon-stopping sites or to different magnetic structures in EuAl$_4$ and EuGa$_4$. Further theoretical calculations, including 
the determination of the muon-stopping sites and of hyperfine fields, might be helpful to better appreciate the differences between
EuAl$_4$ and EuGa$_4$.

%==== figure =============================%
\begin{figure}[!htp]
	%\begin{center}
	\includegraphics[width=1.0\linewidth]{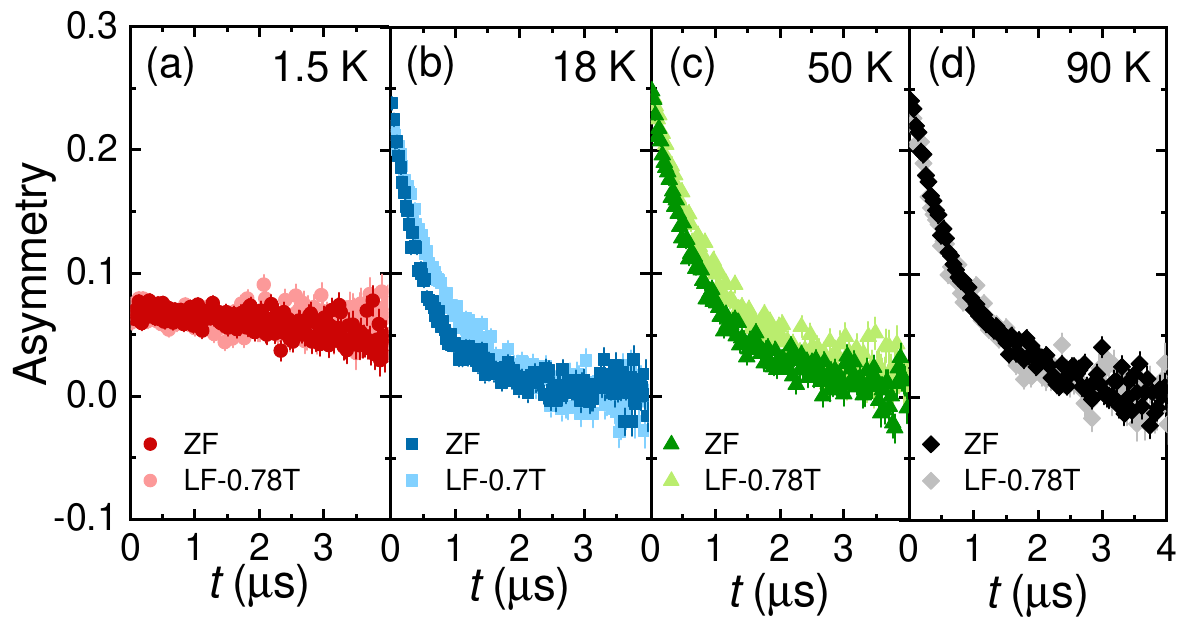}
	%\end{center}
	\centering
	\vspace{-2ex}%
	\caption{\label{fig:LF-muSR}LF-{\textmu}SR time-domain spectra collected at 1.5\,K (a), 18\,K (b), 50\,K (c), and 90\,K (d) in an applied magnetic field of 0 and 0.7 or 0.78\,T for EuAl$_4$. 
		EuGa$_4$ exhibit similar LF-{\textmu}SR results~\cite{Zhu2023}. The applied magnetic field is parallel to the muon-spin direction.  
		Data in panels (a)-(c) were taken from Ref.~\cite{Zhu2023}, while data in panel (d) are original.}
\end{figure}
%=== end figure ==========================

Figure~\ref{fig:ZF-muSR}(d)-(e) and (g)-(h) summarize the transverse and longitudinal {\textmu}SR relaxation rates, $\lambda_\mathrm{T}(T)$ and $\lambda_\mathrm{L}(T)$. 
The transverse relaxation rate $\lambda_\mathrm{T}$ is a measure of the width of the static magnetic field distribution at the muon-stopping site and is also affected by dynamical effects, as e.g., spin fluctuations. 
The longitudinal relaxation rate $\lambda_\mathrm{L}$ is determined solely by spin fluctuations. EuAl$_4$ and EuGa$_4$ exhibit completely different $\lambda_\mathrm{T}(T)$ in the AFM state. 
In EuAl$_4$ [figure~\ref{fig:ZF-muSR}(d)], $\lambda_\mathrm{T}$ is zero in the PM state, to become increasingly prominent as the temperature decreases below $T_\mathrm{N}$. 
Such a large $\lambda_\mathrm{T}$ at $T \ll$ $T_\mathrm{N}$ is unusual for an antiferromagnet and implies an increasingly inhomogeneous distribution of local fields in the AFM state of EuAl$_4$.
The enhanced local-field distribution is most likely related to the complex spatial arrangement of the Eu magnetic moments in EuAl$_4$, where the magnetic propagation vector is incommensurate with the crystal lattice~\cite{Kaneko2021}.
By contrast, in EuGa$_4$ [figure~\ref{fig:ZF-muSR}(g)], $\lambda_\mathrm{T}(T)$ follows the typical behavior of materials with a long-range (anti)ferromagnetic order~\cite{Lee1999,Yaouanc2011,Blundell2021}, i.e., diverging at $T_\mathrm{N}$ and continuously decreasing at $T < T_\mathrm{N}$.
Such $\lambda_\mathrm{T}(T)$ suggests a very homogeneous distribution of local fields, consistent with the commensurate magnetic propagation vector in EuGa$_4$~\cite{Kawasaki2016}. 
EuAl$_4$ and EuGa$_4$ exhibit a similar temperature-dependent $\lambda_\mathrm{L}(T)$ [figure~\ref{fig:ZF-muSR}(e) and (h)], which diverges near $T_\mathrm{N}$, followed by a significant drop at $T < T_\mathrm{N}$, 
indicating that spin fluctuations are the strongest close to the onset of the AFM order. In EuAl$_4$, at $T < T_\mathrm{N}$, $\lambda_\mathrm{L}$ is hundreds of times larger than in EuGa$_4$, thus suggesting 
much stronger spin fluctuations in the AFM state of EuAl$_4$ than in EuGa$_4$. Conversely, in the PM state, both EuAl$_4$ and EuGa$_4$ exhibit similar $\lambda_\mathrm{L}$ values, i.e., comparable spin fluctuations at  $T > T_\mathrm{N}$.

The vigorous spin fluctuations in both crystals are further reflected
in the LF-{\textmu}SR data (figure~\ref{fig:LF-muSR}).
Over a wide temperature range, the {\textmu}SR spectrum in a
$\sim 0.8$-T longitudinal field is mostly identical to that collected
at zero field. This invariance suggests that, in EuAl$_4$
and EuGa$_4$, the muon spins cannot be decoupled by the field
and, hence, that spin fluctuations survive even in a field close
to 1\,T. Most interestingly, the muon spins cannot be decoupled
not only in the AFM, but also in the PM state. Hence, in these materials,
spin fluctuations persist well above the AFM transition. 
The strong spin fluctuations revealed by both ZF-{\textmu}SR
and LF-{\textmu}SR might be crucial for understanding the origin of
the topological Hall effect and of the magnetic skyrmions in
EuAl$_4$ and EuGa$_4$.

We briefly mention here that the {\textmu}SR technique can also
be used to investigate the magnetic skyrmions. Since most of the skyrmion
phases appear in a field range not easily accessible by standard
{\textmu}SR instruments, to date, only a handful of results have been
reported, where LF-{\textmu}SR is used to study skyrmion-hosting
compounds. These include GaV$_4$(S,Se)$_8$~\cite{Franke2018}, Cu$_2$OSeO$_3$~\cite{Hicken2021}, and the Co-Zn-Mn alloy~\cite{Ukleev2021,Hicken2021},
whose skyrmion phases are stabilized by a relatively small field ($< 0.1$\,T). 
At the same time, in many newly discovered skyrmion-hosting
systems, such as GdRu$_2$Si$_2$, Gd$_3$Ru$_4$Al$_{12}$, or the EuAl$_4$
and EuGa$_4$ studied here,~\cite{Hirschberger2019,Khanh2020,Shang2021,Zhang2022}, the critical field required to stabilize the skyrmion phase is larger than 1\,T. 
By using transverse-field {\textmu}SR in GdRu$_2$Si$_2$, with
fields applied either along the $c$- or the $a$-axis,
Huddart \emph{et al.}\ found that the magnetic phases can be
distinguished via their different muon response, thus
providing additional evidence for the skyrmion and meron-lattice
phases~\cite{Huddart2024}. 
The muon-spin relaxation rates extracted from LF-{\textmu}SR measurements in the skyrmion phases of 
Cu$_2$OSeO$_3$ and GaV$_4$(S,Se)$_8$ are $\sim 0.2$--0.8\,{\textmu}s$^{-1}$, 
similar to those of Eu(Al,Ga)$_4$ (figure~\ref{fig:ZF-muSR}). All these skyrmion compounds exhibit similar temperature-dependent 
muon-spin relaxation rates $\lambda_\mathrm{L}(T)$, with an enhanced and broadened 
peak in $\lambda_\mathrm{L}(T)$ at temperatures just below the critical temperature.  
Since also in lon\-gi\-tu\-di\-nal magnetic fields the muon-spin
relaxation rate increases upon entering the skyrmion phase, this provides
another method for identifying the magnetic skyrmions. 
In EuAl$_4$ and EuGa$_4$, which host no skyrmion phase in zero
field, the relaxation rates diverge at $T_\mathrm{N}$, followed by a
significant drop at $T < T_\mathrm{N}$. Such drop is due to the
slowing down of spin fluctuations, a typical feature of the magnetically
ordered materials. 
A similar behavior is observed in Co$_{10}$Zn$_{10}$~\cite{Hicken2021}, 
a parent compound of the Co-Mn-Zn alloys, which lacks any skyrmion phases. 
According to magnetic-, transport- and neutron measurements (see sections~\ref{sec:trsnsport} and \ref{sec:neutron}), the skyrmion phase exists in a field range $\sim 1$--2.5\,T in EuAl$_4$ and $\sim 4$--7\,T in EuGa$_4$.
Further temperature-dependent {\textmu}SR measurements at high magnetic
fields would be an interesting pursue. %to perform. 

\begin{figure}[tp!]
	\centering %\hspace*{-3.5mm}
	\includegraphics[width=0.70\linewidth]{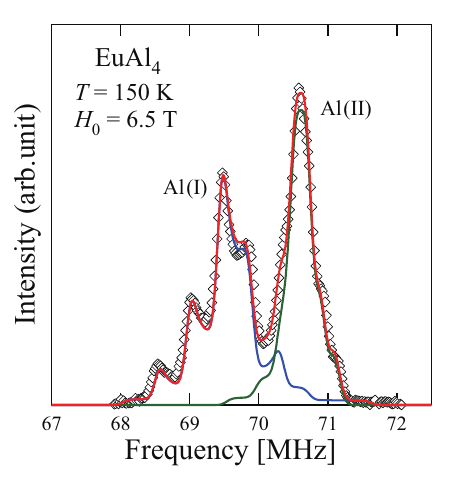}
	\vspace{-3ex}%
	\caption{\label{fig:Al_NMR_spectra}$^{27}$\!Al NMR spectra for the Al(I) and Al(II) sites in EuAl$_4$ at 150\,K. The solid lines
		are theoretical fits accounting for the quadrupole interactions
		of the Al nuclei. Figure was reproduced from Ref.~\cite{Niki2015b}.}
\end{figure}
%
%

%\section{Muon-spin rotation and relaxation}
\section{Nuclear magnetic resonance\label{sec:NMR}}

In view of the complementarity of NMR to {\textmu}SR regarding the time
scale, probe site, and/or spin, NMR has suc\-cess\-ful\-ly been used to investigate the intermetallic Eu(Al,Ga)$_4$ compounds and their equivalents.
Below we review briefly the main outcome of such studies, with particular regard to the magnetic and
electronic properties.

The nucleus of choice in these studies was the spin-$\nicefrac{5}{2}$
$^{27}$\!Al. Although more demanding, also $^{153}$Eu ($I = 5/2$) and
$^{69,71}$Ga ($I = 3/2$) have been used. Since all these nuclei have
$I > 1/2$, their nonzero quadrupole interactions provide
complementary information to {\textmu}SR. Further, since the
Al (Ga) atoms occupy two crystallographically inequivalent sites in
the BaAl$_4$-type
structure (i.e., 4$d$ and 4$e$ in Table~\ref{tab:coordinates}), in principle,
they can give rise to two distinct sets of NMR lineshapes.

\begin{figure*}[tp!]
	\centering
    \includegraphics[width=0.99\textwidth]{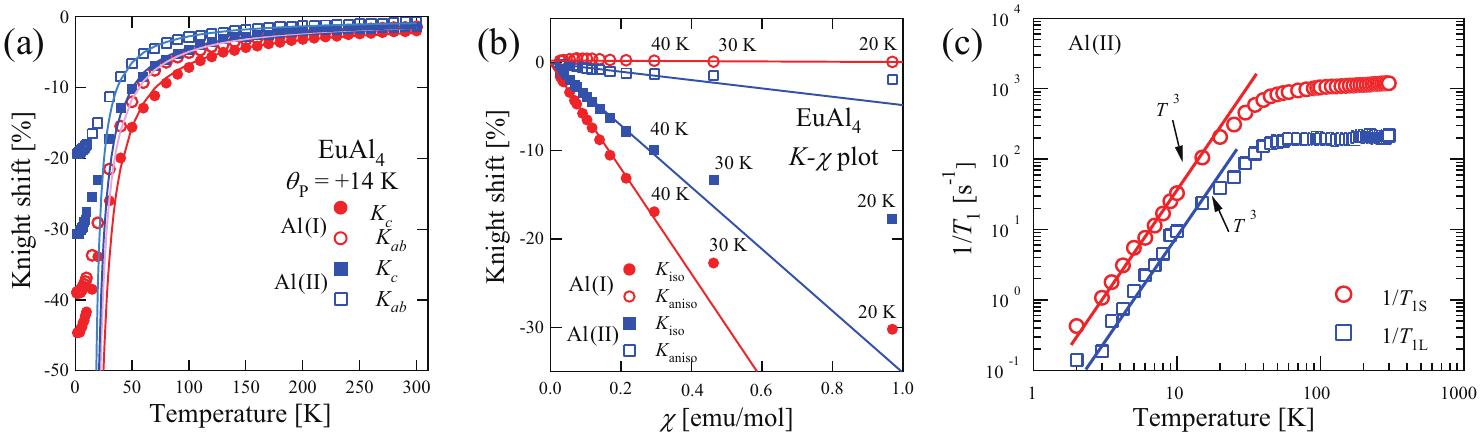}
   \vspace{-1ex}%
	\caption{\label{fig:Al_NMR_shift}(a) Temperature dependence of the
	$^{27}$\!Al Knight shifts $K_\mathrm{{iso}}$ and $K_\mathrm{{aniso}}$
	at the Al(I) and Al(II) sites in EuAl$_4$. (b) $K$-$\chi$ plots of $^{27}$\!Al Knight shifts. 
		The resulting hyperfine fields $H^\mathrm{hf}_\mathrm{{iso}}$ and
		$H^\mathrm{hf}_\mathrm{{aniso}}$ are $-3.23$ and $-0.16$\,kOe/{\textmu}$_\mathrm{B}$
		at Al(I) site, and $-1.82$ and $-0.264$\,kOe/{\textmu}$_\mathrm{B}$ at
		Al(II) site. (c) Temperature dependence of $1/T_1$ of $^{27}$\!Al NMR in EuAl$_4$ for
		the Al(II) sites [Al(I) sites show a similar behavior].
		Figures were reproduced from Ref.~\cite{Niki2015b}.}
\end{figure*}

In metallic alloys containing magnetic (Eu$^{2+}$) ions embedded in a
metallic (Al) matrix, the electronic properties are often determined
by the physics of the RKKY oscillations~\cite{Ruderman1954,Kasuya1956,Yosida1957}.
Essentially, these are the spin-density equivalent of charge-density
(Friedel) oscillations~\cite{Villain2016}. Both phenomena 
reflect the fact that an electric (magnetic) impurity in a metal is
asymptotically screened in a non-exponential oscillatory way, i.e.,
$\delta \rho(r) \sim A_0 \cos(2k_\mathrm{F}r)/r^3$.
The oscillation period is determined by the Fermi surface of the host
material (here, by the Fermi wave vector $k_\mathrm{F}$), whereas the
decay rate with distance $r$ depends on the dimensionality of the host
material (here, $1/r^3$ for 3D materials). The oscillations themselves
arise because only electrons within a limited range of wavelengths near
the Fermi level can be scattered, thus producing a ripple-like modulation
in the electronic density. As we show below, the RKKY effects are clearly
reflected in the Knight shift and are so important as to overshadow a 
possible CDW order. The latter could still be successfully
tracked from the quadrupole effects in the isostructural nonmagnetic
SrAl$_4$ compound.

The pioneering $^{27}$\!Al NMR study by van Diepen \emph{et al.}\ on EuAl$_4$
in powder form could detect only one broad resonance with a negative
Knight shift~\cite{Diepen1969}. Since EuAl$_4$ is marginally stable and
a proper annealing is problematic (due to the high vapor pressure of Eu),
this hinted at
lattice imperfections. Hence, the two expected resonance
lines merged into one due to broadening from magnetic dipole and nuclear
quadrupole interactions. In fact,
re\-cent studies by Niki \emph{et al.}\ found two sets of resonances, 
including their quadrupole satellites 
(figure~\ref{fig:Al_NMR_spectra})~\cite{Niki2015b}.
The importance of the early studies (confirmed by the new ones) was to
demonstrate the presence of RKKY effects via the nonequivalence of
$^{27}$\!Al NMR lines.
Here, the exchange interaction between the Eu$^{2+}$ rare-earth spins
and the Al conduction-electron spins, implies that the nonuniform
oscillations of magnetism affect also the polarization of conduction
electrons. In the absence of a polarizing rare-earth moment
(i.e., for Pauli paramagnets), the different resonance lines would
have the same Knight shift $K_0$ and overlap, as indeed observed in
the La$_3$Al$_{11}$ case.

More in detail, the reported $^{27}$\!Al NMR spectra
arise from the nuclear spin Hamiltonian:
\begin{equation}
	{\cal H} = -\gamma_n \hbar \boldsymbol{I} \cdot \boldsymbol{H_0}
	\left[1 + K(\theta)\right] 
	+ \frac{h \nu_Q}{6} \left[3I_z^2 - I^2\right].
	\label{eq:nmr_hamiltonian}
\end{equation}
Here, the first term represents the Zeeman interaction between the nuclear
magnetic moment $\boldsymbol{\mu_n} = \gamma_n \hbar \boldsymbol{I}$ 
and the external magnetic field $\boldsymbol{H_0}$, where $\gamma_n$ is
the nuclear gyromagnetic ratio and $\boldsymbol{I}$ is the nuclear spin.
The internal magnetic field introduces the Knight shift term $K(\theta)$. 
The second term indicates the nuclear quadrupole interaction between the
electric field gradient (EFG) and the nuclear quadrupole moment $Q$.
For symmetry reasons, the EFG is along the main principal axis. Hence,
$V_{zz}$ is parallel to the $c$-axis and $\nu_Q$, the nuclear quadrupole
frequency, is $\nu_Q = 3eQV_{zz} /2I(2I - 1)h$. Similarly, in the case of a
tetragonal symmetry, the Knight shift is only a function of $\theta$, here
corresponding to the angle between the external magnetic field and the
$c$-axis.
The calculated $\nu_Q$ values of $^{27}$\!Al in EuAl$_4$ are in good
agreement with the measured ones, 0.865 and 0.409\,MHz, assigned to
the Al(I) and Al(II) sites and are almost independent of $T$.

The isotropic- $K_\mathrm{iso}$ and the anisotropic part $K_\mathrm{aniso}$
of the Knight shift for both Al sites measured at 6.5\,T are summarized in
figure~\ref{fig:Al_NMR_shift}(a)~\cite{Niki2015b}. Upon de\-creas\-ing
the temperature, both shifts become strongly negative and follow
a Curie-Weiss law with $\theta_P = 14$\,K. This is consistent with the
magnetic susceptibility and reflects the FM state of
EuAl$_4$ above its metamagnetic transition at 2\,T. 
Considering that the magnetic susceptibility $\chi(T)$ in the
PM state is isotropic, the Knight shift can be expressed as
$K = H_\mathrm{hf} \chi(T)/ N \mu_\mathrm{B}$,  
where $H_\mathrm{hf}$ is the hyperfine field, $N$ is the Avogadro's number,
and $\mu_\mathrm{B}$ the Bohr magneton. The $K$-$\chi$ plots for
$K_\mathrm{iso}$ and $K_\mathrm{aniso}$ for both Al sites are shown in
figure~\ref{fig:Al_NMR_shift}(b). From the relevant slopes one obtains the
hyperfine field values reported in the figure caption.

The $^{27}$\!Al NMR spin-lattice relaxation times $T_1$ reflect both the
magnetic- and the nuclear quadrupole interactions. The latter arise
because of the spin-5/2 Al nucleus and the tetragonal symmetry of EuAl$_4$.
The recovery of nuclear magnetization for the central transition can
be analyzed by means of standard expressions~\cite{Narath1967}, slightly 
modified because $T_1$ is distributed. In this case, the two components
of the relaxation, a short one $1/T_\mathrm{1S}$ and a long one $1/T_\mathrm{1L}$, are
shown in figure~\ref{fig:Al_NMR_shift}(c).
Due to the fast fluctuation of the $f$-electron spins in the PM
state, the relaxation rate is almost constant close to 300\,K, but it
gradually decreases with decreasing temperature because of the slowdown
of %the
fluctuations. Below 20\,K, $1/T_1$ is almost proportional to $T^3$,
a behavior attributed to the excitation of the $f$-electron spins in the
FM state (where the NMR measurements, here at 6.5\,T, were performed).
Note that, no change in the $^{27}$\!Al NMR relaxation due to the CDW
around 140\,K could be detected, most likely because of its masking by
the fast relaxation due to the large magnetic moments of $f$-electron spins.

%trim=0px 22px 0px 0px
%
\begin{figure}[tp!]
	\centering
	\includegraphics[width=1.0\linewidth]{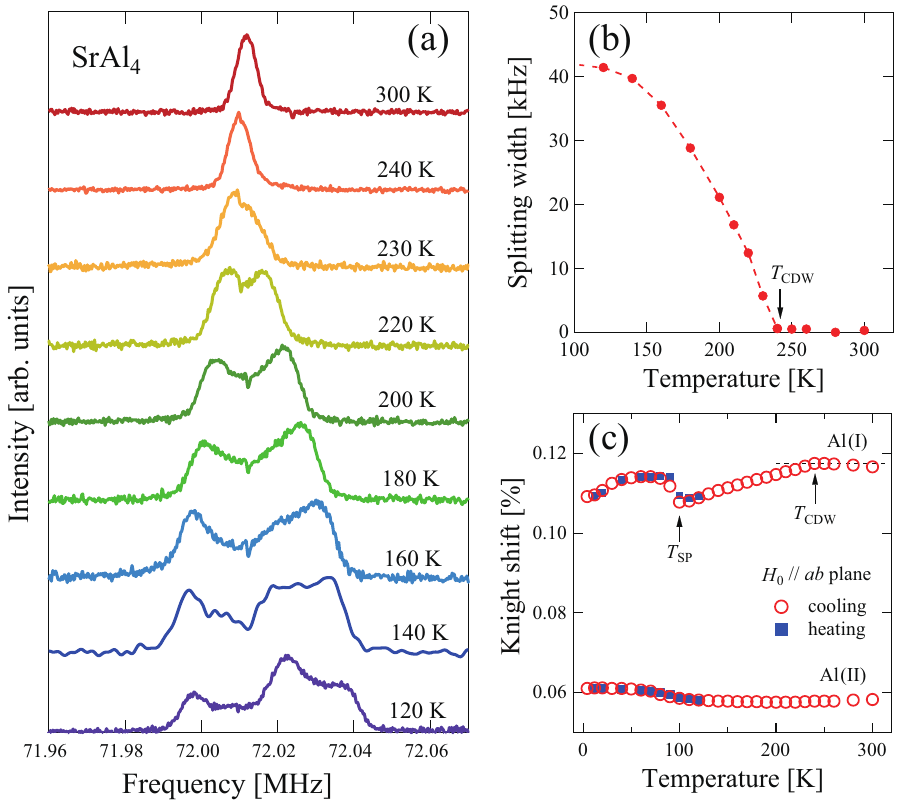}
	% [trim=left bottom right top, clip]
	\vspace{-4ex}%
	\caption{\label{fig:Al_NMR_SrAl4}(a) Temperature dependence of the
		first low-frequency satellite line of the $^{27}$\!Al NMR spectrum
		for the Al(II) site in a single crystal of SrAl$_4$. (b) Splitting between
		peaks vs temperature for the data shown in panel (a). The disappearance
		of splitting corresponds to $T_\mathrm{CDW}$. (c) Knight shift vs
		temperature for the Al(I) and Al(II) sites. In the first case, the
		core polarization of $d$-electrons dominates the shifts, here
		reflected in the clear changes at $T_\mathrm{CDW}$ and $T_\mathrm{SP}$
		(see text for details). Figures were reproduced from Ref.~\cite{Niki2015a}.}
\end{figure}

To investigate the physical properties of the CDW and the structural phase 
(SP) transitions in EuAl$_4$, $^{27}$\!Al NMR measurements were performed
in the isostructural nonmagnetic SrAl$_4$ compound~\cite{Niki2015a}. By
applying the magnetic field perpendicular to the $c$-axis one obtains a 10-peak NMR dataset, comprising
five lines for each of the two Al sites. Because of the CDW and SP
transitions (at 243 and 100\,K, respectively), the well resolved sharp
spectra at 300\,K become complex at 4\,K. The evolution of the $^{27}$\!Al
NMR lines with temperature can be followed by monitoring the first satellite
of the Al(II) site. 
Below $T_\mathrm{CDW}$, the single resonance line splits into a double-horned
shape [figure~\ref{fig:Al_NMR_SrAl4}(a)], reflecting the modulation of the electrical quadrupole interaction
by the incommensurate CDW charge distribution. The temperature dependence
of the spectral splitting, here corresponding to the CDW amplitude, is summarized 
in figure~\ref{fig:Al_NMR_SrAl4}(b).
The splitting develops rapidly below $T_\mathrm{CDW}$, to become
constant around 140\,K.
Further spectral changes were observed below 120\,K, preliminary to the
$T_\mathrm{SP}$ transition at 100\,K, with the simple incommensurate
CDW above $T_\mathrm{SP}$ changing into a different type of charge modulation
below it.

\begin{figure}[tp!]
	\centering %\hspace*{-3.5mm}
	\includegraphics[width=0.75\linewidth]{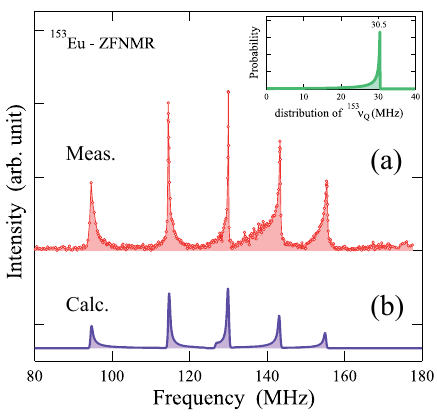}
	\vspace{-2ex}%
	\caption{\label{fig:Eu_NMR_spectra}(a) $^{153}$Eu zero-field NMR
		spectrum at $T = 4.2$\,K. (b) Cal\-cu\-lat\-ed spectrum for 
		$H_\mathrm{int}^\mathrm{Eu} \perp V_{zz}^\mathrm{Eu}$ assuming a
		log-normal distribution of $^{153}\nu_Q$ frequencies, as
		shown in the inset. The five peaks reflect the $I = 5/2$ $^{153}$Eu
		nucleus. Figure was reproduced from Ref.~\cite{Yogi2013}.}
\end{figure}

The temperature dependence of the $^{27}$\!Al NMR Knight shifts for both
Al sites in SrAl$_4$ is summarized in figure~\ref{fig:Al_NMR_SrAl4}(c).
In the Al(I) case, while 
above $T_\mathrm{CDW}$ the Knight shift is almost constant, it decreases
monotonously below $T_\mathrm{CDW}$, to show a step-like anomaly at
$T_\mathrm{SP}$. The Knight shift for the Al(II) site remains almost
constant from room temperature to $T_\mathrm{SP}$, to increase smoothly
below it. The dependency of shift with temperature for the site Al(I) can
be attributed to the core polarization of the $d$-electrons, while its
mostly constant value for the site Al(II) is compatible with a
contribution from the $s$-electrons. Such attribution is reflected also
in the $^{27}$\!Al NMR spin-lattice relaxation times $T_1$ (not shown),
where the relaxation vs temperature for each site follows closely
that of the shifts.

Since the relaxation time of the rare-earth nuclei (especially in the
fast fluctuating PM phase) is too short to detect
an NMR signal, the only possibility to observe these nuclei is in the
magnetically ordered state. For instance, the zero-field $^{153}$Eu
NMR signal could be observed in the AFM state of EuGa$_4$~\cite{Yogi2013}. 
In this case, a large internal field is
transferred to the nuclei from the adjacent magnetic moments,
producing a Zeeman splitting of the nuclear spin states, thus
giving rise to an NMR signal even in a zero magnetic field.
The hyperfine interaction Hamiltonian is again given by equation (\ref{eq:nmr_hamiltonian}), where $H_{0}$ is replaced by 
$H_\mathrm{int}$. If $H_\mathrm{int} \parallel V_{zz}$ and the 
Zeeman interaction is larger than the nuclear quadrupole interaction, the five $^{153}$Eu peaks are equidistant. On the other hand, 
if $H_\mathrm{int} \perp V_{zz}$, the peak separations become inequivalent, consistent with the observed $^{153}$Eu NMR spectrum
(figure~\ref{fig:Eu_NMR_spectra}), indicating ordered 
magnetic moments lying in the $ab$-plane [see figure~\ref{fig:n_EuGa4}(c)].
This result is in good agreement with those of magnetic susceptibility
and neutron diffraction experiments (see details in
sections~\ref{sec:trsnsport} and \ref{sec:neutron}).
%
%%%%%%%%%%%%%%%%%%%%%%%%%%%%%%%
\begin{figure*}[!htp]
	%\begin{center}
	\includegraphics[width=0.85\linewidth]{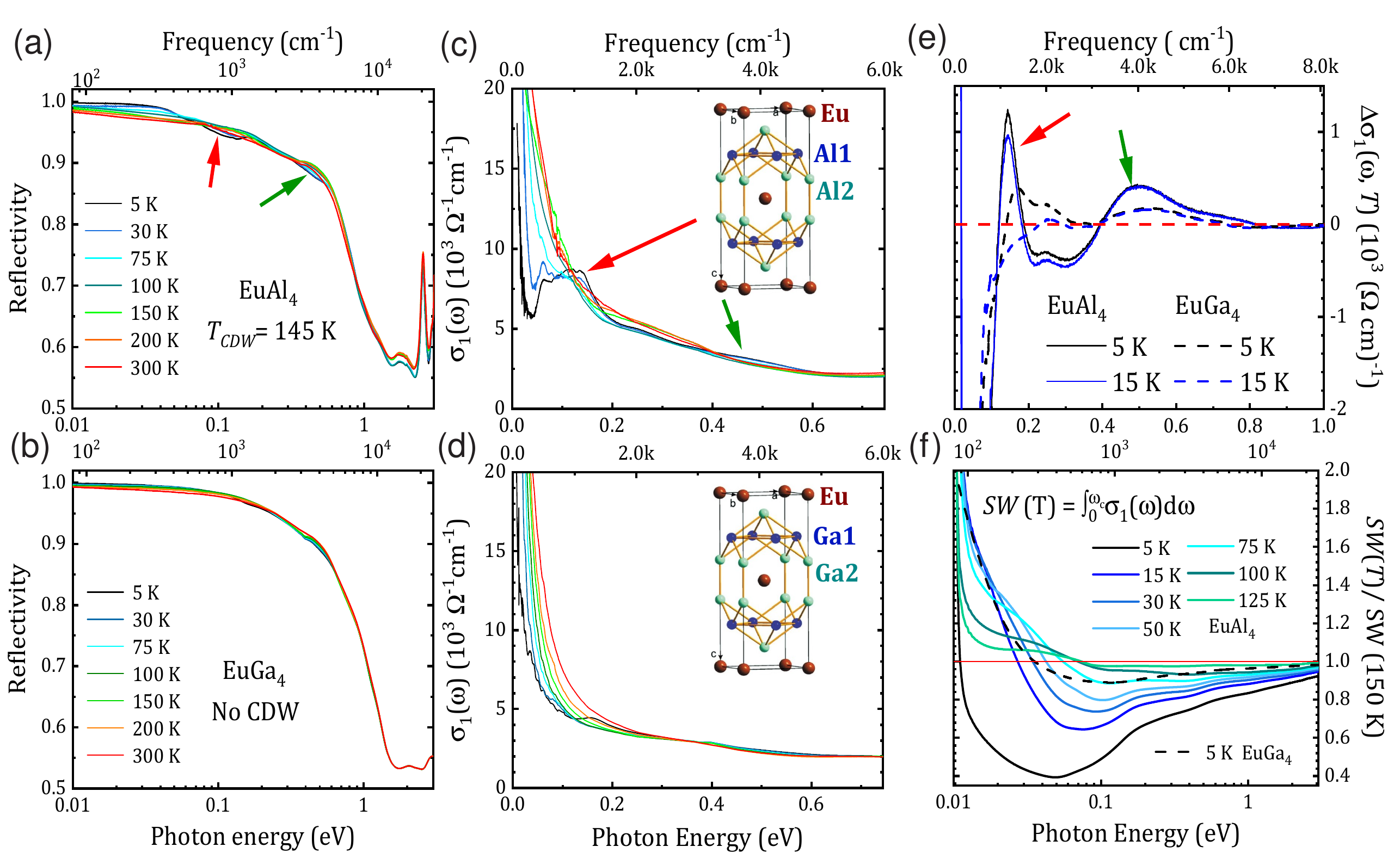}
	%\end{center}
	\centering
	%\vspace{-2ex}%
	\caption{\label{fig:optical reflectivity}Optical spectroscopy of EuAl$_4$ and EuGa$_4$. Temperature-dependent reflectivity of (a) EuAl$_4$ and (b) EuGa$_4$ with the light polarized in the $ab$-plane. Temperature evolution of the real part $\sigma_1(\omega;T)$ of the optical-conductivity spectra of EuAl$_4$ (c) and EuGa$_4$ (d) below 6\,000~cm$^{-1}$ (0.74~eV). 
		Insets display the lattice structures of EuAl$_4$ and EuGa$_4$, with the two inequivalent Al/Ga atoms. (e) Difference spectra of $\sigma_1(\omega)$ below the CDW order calculated via $\Delta\sigma_1(\omega, T)= \sigma_1(\omega, T)-\sigma_1(\omega, 150~{\rm K})$ for EuAl$_4$ (solid lines) and EuGa$_4$ (dashed lines), respectively. (f) The normalized integrated spectral weight ($SW$) of EuAl$_4$ as a function of the cutoff frequency $\omega_c$ at $T < T_\mathrm{CDW}$. 
		Figures were reproduced from Ref.~\cite{RYang2024}.}
\end{figure*}
%%%%%%%%%%%%%%%%%%%%%%%%%%%%%%%%%%%

In EuGa$_4$, calculations based on the exact diagonalization of
the nuclear-spin Hamiltonian provide an internal field at the Eu nucleus 
$H_\mathrm{int}^\mathrm{Eu} = 27.08$\,T and $^{153}\nu_Q = 30.5$\,MHz~\cite{Yogi2013}.
The very large field at the Eu nucleus suggests
an $f$-electron-derived magnetic state.  
Analogous results were found also for $^{69,71}$Ga NMR, where the
three-peak spectra reflect the $I = 3/2$ of both Ga isotopes. 
In this case, the internal field is $H_\mathrm{int}^\mathrm{Ga} = 3.03$\,T
and $^{69}\nu_Q(^{71}\nu_Q) = 5.08(3.21)$\,MHz~\cite{Yogi2013}.
Finally, the temperature dependence of the internal field, which in AFM
materials is proportional to the sublattice magnetization, can be explained
by a Brillouin function with $J=S=7/2$, here reflecting the magnetic moment of the Eu$^{2+}$ $4f^7$ ions. 

Recently, zero-field $^{153}$Eu-NMR studies were performed also in the
AFM state of EuAl$_4$~\cite{Niki2020}. Unlike in EuGa$_4$, whose Eu NMR
spectra consists of five resonance lines (a main line and two satellites
on each side), the $^{153}$Eu spectra of EuAl$_4$ show 10 lines, reflecting the
presence of two kinds of Eu$^{2+}$ magnetic moments, oriented either
along the $c$-axis or perpendicular to it. Since the signal intensities
of the $^{153}$Eu ZF-spectra produced by the two orientations are
comparable, similar numbers of magnetic moments belong to
each orientation.
The internal magnetic field at the $^{153}$Eu nucleus
is $H_\mathrm{int}^\mathrm{Eu} = 28.41$\,T and $^{153}\nu_Q = 27.0$\,MHz 
for magnetic moments parallel to the $c$-axis, and
$H_\mathrm{int}^\mathrm{Eu} = 28.11$\,T and $^{153}\nu_Q = 27.6$\,MHz 
for those perpendicular to the $c$-axis~\cite{Niki2020}. 
Thus, also the amplitudes and NQR frequencies of the
magnetic moments lying along the two directions are almost the same.
		
Finally, we recall that electron spin resonance (ESR) measurements in
EuAl$_4$ were performed already in the 1970's~\cite{Wernick1967,Taylor1975}.
These studies provide a $g$ value of 2.000(5) between 77 and 300\,K and a room temperature linewidth of 55(2)\,mT. Below 40\,K (corresponding to about 2\,$T_\mathrm{N}$), 
the ESR line moves down in frequency and broadens up to 150\,mT.
Both effects are compatible with materials that order antiferromagnetically,
for which the resonant field is expected to shift by a factor of
($2H_\mathrm{E} H_\mathrm{A})^{1/2}$, where $H_\mathrm{E}$ is the exchange
field and $H_\mathrm{A}$ the anisotropy field, if $H_\mathrm{E} \gg H_\mathrm{A}$.
This shift is so large that, below the N\'eel temperature, the resonance
drops outside the microwave frequency range and no signal is observed.
The line shape remains asymmetric over the entire temperature range, a
behavior attributed to the fact that EuAl$_4$ undergoes a metamagnetic transition just above the resonant field.
\section{Optical spectroscopy\label{sec:opt}}

As discussed in sections~\ref{sec:trsnsport} and \ref{sec:neutron},
the magnetic order can be easily tuned in EuAl$_4$. Here, a weak
magnetic field ($<$2\,T) can rotate the Eu$^{2+}$ spins, 
leading to a series of non-coplanar spin textures (including skyrmions)
and the corresponding topological Hall effect~\cite{Shang2021,takagi2022}.
These phenomena are unusual, since most of the non-coplanar spin textures occur
in materials with a noncentrosymmetric crystal structure (due to the presence of DMIs~\cite{He2022}), while EuAl$_4$ has a centrosymmetric tetragonal structure. By contrast, the magnetic order in EuGa$_4$ is quite robust.
It undergoes only one AFM transition at low temperature, and magnetic fields up to 7\,T cannot rotate the spin at 2\,K~\cite{Zhang2022}.
The two compounds exhibit the same crystal structure and share similar band structures. While EuAl$_4$ undergoes a CDW transition around 145\,K (see section~\ref{sec:structure}), EuGa$_4$ behaves as a simple metal in the PM state. The magnetism of the rare-earth metallic compounds is mostly dominated
by the local moments of $4f$ electrons. Nevertheless, the charge excitations, including itinerant carriers and local excitations, play a key role in mediating the magnetic interactions. Thus, understanding the
effects of CDW order on the band structures provides important clues 
on the mechanisms behind the complex magnetism of EuAl$_4$.

Optical spectroscopy represents one of the most pow\-er\-ful tools for 
investigating the charge excitations in solids~\cite{Dressel2002},
able to provide information on electronic gaps and band reconstructions. Recently, Yang \emph{et al.}\ con\-duct\-ed a
comparative study of EuAl$_4$ vs.\ EuGa$_4$, to elucidate the effects of CDW order on the electronic band
structures by means of optical spectroscopy and first-principles calculations~\cite{RYang2024}. The reflectivities of EuAl$_4$ and EuGa$_4$ were measured from 10\,meV to 3\,eV, at temperatures below and
above $T_\mathrm{CDW}$ [figures~\ref{fig:optical reflectivity}(a)-(b)]. Based on the Kramers-Kroning relation, the optical conductivity $\sigma_1(\omega)$
of both compounds was calculated [figures~\ref{fig:optical reflectivity}(c)-(d)]. Optical conductivity is proportional to the joint density of states~\cite{Dressel2002}. Peaks centered at zero frequency represent the intraband response (Drude peak), while the Lorentz peaks, centered at finite frequencies, come from the interband excitations. Below $T_\mathrm{CDW}$, the intraband response of EuAl$_4$ is greatly suppressed. The two interband absorptions, emerging
close to 0.1 and 0.4\,eV, indicate the opening of a CDW gap at the Fermi surface and an enhanced local excitation at 
0.4\,eV [shown by red- and green arrows in figures~\ref{fig:optical reflectivity}(a)-(e)].
In EuAl$_4$, the residual intraband response below $T_\mathrm{CDW}$ confirms the partial gap opening at the Fermi surface. 
A spectral weight analysis of the optical conductivity
indicates that, below the CDW transition, part of the itinerant
carriers become localized [figure~\ref{fig:optical reflectivity}(f)]. By contrast, the above features are absent in the optical conductivity of EuGa$_4$.
%
%%%%%%%%%%%%%%%%%%%%%%%%%%%%%%%%
\begin{figure*}[!tp]
	%\begin{center}
	\includegraphics[width=0.85\linewidth]{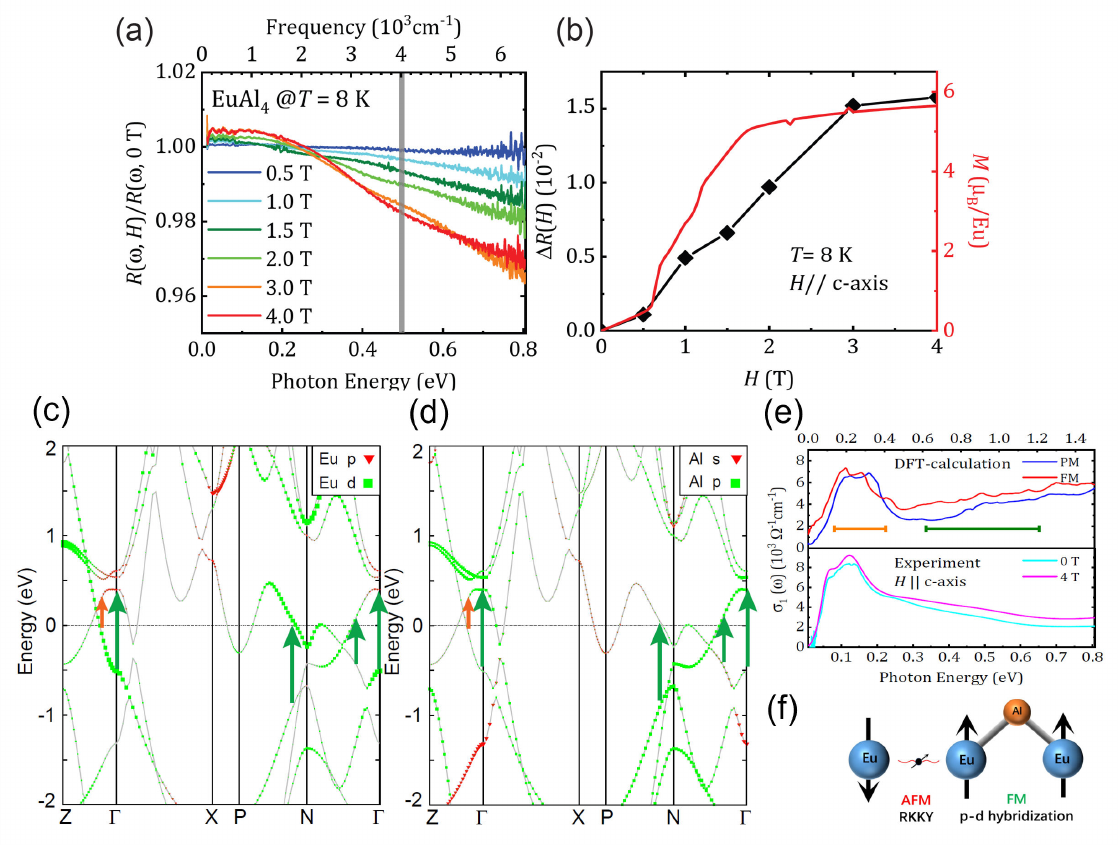}
	%\end{center}
	\centering
	\caption{\label{fig:MO}\tcr{Magneto-optical response and first-principle calculations of EuAl$_4$.
	(a) Normalized magneto-optical reflectivity spectra $R(\omega,H)$
	measured at 8\,K in a magnetic field up to 4\,T, applied along the $c$-axis. 
	(b) Comparison between the field-dependent reflectivity change
	$\Delta R(H) = [1 - R(H)/R(0\,T)]$ at 0.4\,eV [indicated by the gray
	bar in panel (a)] and the magnetization at $T = 8$\,K. 
	%First-principle calculations for EuAl$_4$. 
	Band structure of EuAl$_4$ in the PM state (without considering the CDW order), with the contributions of the Eu-5$d$ and Al-3$p$ orbitals shown in panels (c) and (d), respectively. The upper- and lower panels in (e) show the calculated and the measured interband optical conductivities. Blue and red lines in the upper panel correspond to the absorption in the PM- and FM state (where the Eu$^{2+}$ moments align along the $c$-axis). The turquoise and pink lines in the lower panel are conductivities measured at 8\,K in 0 and 8\,T. Orange and green segments represent the interband transitions, denoted by same-colour arrows in panels (c) and (d). (f) Schematic plot of the magnetic interactions in EuAl$_4$.
	Figures reproduced from Ref.~\cite{RYang2024}.}}
\end{figure*}
%%%%%%%%%%%%%%%%%%%%%%%%%%%%%%%%%%%%%%%

When entering the AFM state, the Eu$^{2+}$ moments in EuAl$_4$ can be
tuned by an external magnetic field. For instance, at 8\,K,
the absorption around 0.4\,eV is steadily enhanced and starts to saturate at $H > 3$\,T, where the Eu$^{2+}$ moments are polarized
(figure~\ref{fig:MO}).
To understand the origin of absorption at 0.4\,eV,
Yang \emph{et al.}\ calculated the band structure and the optical conductivity of
EuAl$_4$ (figure~\ref{fig:MO}). The resulting optical conductivities
in the PM- and in the polarized FM state (Eu$^{2+}$ moments aligned along
the $c$-axis) are in good agreement with the experimental data [figure~\ref{fig:MO}(e)]. A slight difference in the energy scale was  
attributed to the electron correlation, which was not included in the first-principle calculations. Considering the energy size and the possible excitations near the Fermi level, the absorption at 0.1\,eV can be ascribed to the excitations between the linear bands near the Fermi level [orange arrows in figures~\ref{fig:MO}(c)-(d) and orange segment in figure~\ref{fig:MO}(e)], while the absorption at 0.4\,eV arises mainly from the excitations between the Eu-$5d$- and Al-$4p$ bands [green arrows in figures~\ref{fig:MO}(c)-(d) and green segment figure~\ref{fig:MO}(e)]. The increased absorption at 0.4\,eV in a magnetic field could be ascribed to
the Zeeman splitting from the aligned local moments.

%%%%%%%%%%%%%%%%%%%%%%%%%%%%%%%%
%\begin{figure*}[!tp]
%	\includegraphics[width=0.6\linewidth]{Fig36_optical_band.pdf}
%	\centering
%	\caption{First-principle calculations for EuAl$_4$. Band structure of EuAl$_4$ in the PM state (without considering the CDW order), with the contributions of the Eu-5$d$ and Al-3$p$ orbitals shown in panels (a) and (b), respectively. The upper- and lower panels in (c) show the calculated and the measured interband optical conductivities. Blue and red lines in the upper panel correspond to the absorption in the PM- and FM state (where the Eu$^{2+}$ moments align along the $c$-axis). The turquoise and pink lines in the lower panel are conductivities measured at 8\,K in 0 and 8\,T. Orange and green segments represent the interband transitions denoted by the same colored arrows in panels (a) and (b). (d) Schematic plot of the magnetic interactions in EuAl$_4$. Figures were reproduced from Ref.~\cite{RYang2024}.}
%	\label{fig:band}
%\end{figure*}
%%%%%%%%%%%%%%%%%%%%%%%%%%%%%%%

By combining optical spectroscopy with first-principle calculations, Yang \emph{et al.}\ found that the CDW order in EuAl$_4$ leads to a partial deformation of the Fermi surface, while simultaneously amplifying local excitation around 0.4\,eV.
In rare-earth metallic systems, the magnetic interactions are mainly mediated by itinerant carriers through the RKKY interactions. 
In the CDW ordered state, since part of the itinerant carriers are localized, the AFM interactions are frustrated.
Moreover, below $T_\mathrm{CDW}$, the excitations between
the Eu-$5d$- and Al-$3p$ orbitals are enhanced.  
A previous study of Eu monochalcogenides pointed out that the hybridization between Eu-$5d$ and $p$ orbitals is essential to the magnetic interactions, which can be either
AFM or FM~\cite{Altman2018,Wan2011}. 
In EuAl$_4$, recent {\textmu}SR and NMR studies reveal robust magnetic fluctuations even in the magnetically ordered state~\cite{Zhu2023, Niki2015}, while they are absent in EuGa$_4$.
In SrAl$_4$, recent optical investigations also reveal an enhanced absorption around 0.4\,eV along with an amplified FM response below the CDW transition~\cite{Nakamura2016}. Since EuAl$_4$ and SrAl$_4$ share similar band structures,
in both cases the amplified $p$-$d$ hybridization can promote FM interactions [figure~\ref{fig:MO}(f)].
Therefore, Yang \emph{et al.}\ proposed that the CDW order in EuAl$_4$ could suppress the AFM interactions, but promote the FM ones. Since such interactions compete with each other, when they become of comparable intensity, it may result in a magnetic in\-sta\-bil\-i\-ty and lead to a variety of magnetic orders. At the same time, in EuGa$_4$,
there is no CDW order at ambient pressure. Hence, its AFM order
is relatively robust. This is also demonstrated by doping studies in Eu(Al$_{1-x}$Ga$_x$)$_4$. When the CDW order is suppressed upon Ga-doping, less magnetic transitions are found~\cite{Stavinoha2018}.

Moreover, since the CDW gap is partially opened on the Fermi surface, the AFM interactions mediated by the itinerant electrons are anisotropic. The incommensurate lattice distortion below $T_\mathrm{CDW}$ breaks the inversion symmetry~\cite{Ramakrishnan2022,Ni2023a}, which may split the degenerate Dirac bands into Weyl bands~\cite{Ma2020,Wang2023}.
In such an anisotropic environment, itinerant electrons from the Weyl bands can mediate the anisotropic magnetic interactions that facilitate the formation of the chiral spin textures~\cite{takagi2022,Kaneko2021,Gaudet2021}. Further experimental investigations are needed to validate this hypothesis.

%%%%%%%%%%%%
\section{Angle-resolved photoemission spectroscopy and quantum oscillations\label{sec:arpes}}
%%%%%%%%%%%%

%%%%%%%%%%%%%%%%%%%%%%%
\begin{figure*}[!tp]
	%\begin{center}
	\includegraphics[width=0.8\linewidth]{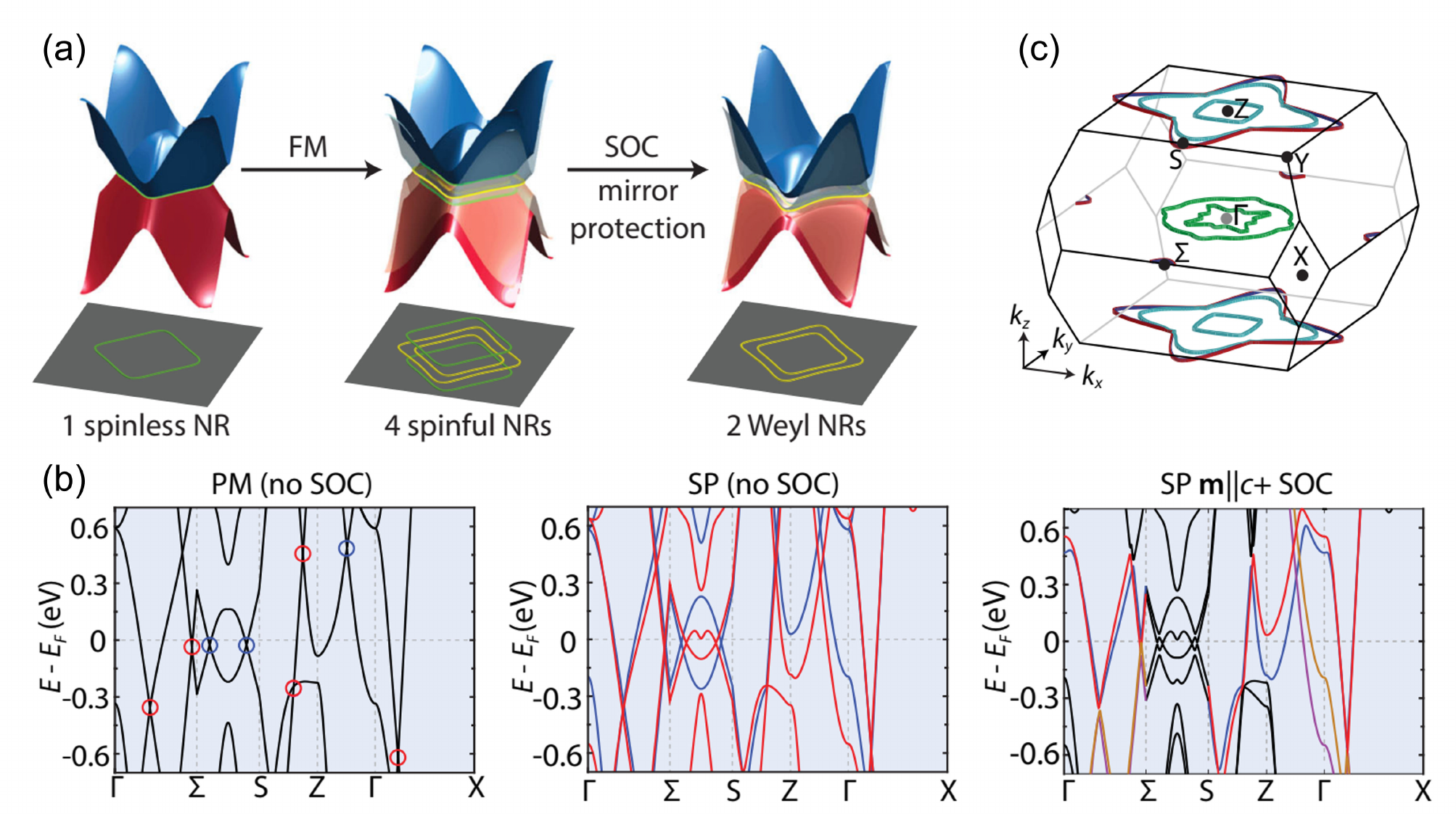}
	%\end{center}
	\centering
	%\vspace{-2ex}%
\caption{\label{fig:DFT}Calculated electronic band structures for EuGa$_4$. (a) The mechanism of formation of Weyl node rings in a square lattice. (b) Band structures of EuGa$_4$ in the PM state without SOC, SP phase without SOC, and SP phase with magnetic moment along $c$ axis and with SOC, respectively. (c) 3D view of the Weyl NRs. Three pairs of NRs are shown in green, cyan, and red/blue, respectively. Figures were reproduced from Ref.~\cite{lei2023}.}
\end{figure*}
%%%%%%%%%%%%%%%%%%%%%%%%

As mentioned in section~\ref{sec:trsnsport}, transport studies reveal
topological features in both the real- and momentum space of the Eu(Al,Ga)$_4$ system. In this section, we briefly review the theoretical and experimental
work in the search for topological features in the electronic band structure (i.e., momentum space) that allowed the identification of 
Weyl nodal rings (NR) in the spin-polarized (i.e., field\--in\-duced FM)
state of EuGa$_4$, which also provided a tentative explanation for the large MR in this regime~\cite{lei2023}.

Density-functional theory (DFT) suggests that the electronic band structures
of square-lattice compounds---to which EuAl$_4$ and EuGa$_4$ belong---feature  diamond\--shaped NRs in the mirror-invariant plane [figure~\ref{fig:DFT}(a)]. When ferromagnetism is introduced to account for the spin-polarized (SP)
state, the lifting of spin degeneracy promotes four spinful NRs to emerge from the single spinless NR at zero field, each NR now being twofold degenerate. After considering the spin-orbit coupling (SOC), only a subset of these spinful NRs protected by mirror symmetry survive, while the others are gapped out. In the PM state of EuGa$_4$, the electronic bands show multiple crossings [figure~\ref{fig:DFT}(b)]. Among them, most of those on the mirror invariant planes, at $k_z$ = 0 and $k_z = 2\pi/c$, exhibit NR geometry in the three-dimensional (3D) $k$-space. Upon entering the SP state, two sets of spin-split bands form. After further introducing the SOC and setting the magnetization along the $c$-axis (i.e., $m \parallel c$), in this case perpendicular to the mirror plane, only a subset of the crossings are retained, forming the Weyl NRs. These calculated Weyl NRs can be categorized into three pairs, colored green, cyan, and red/blue in the Brillouin zone [figure~\ref{fig:DFT}(c)]. The red/blue pair of NRs are expected to sit very close to the Fermi energy $E_\mathrm{F}$, with a small energy variation, although they span the whole Brillouin zone at the $k_z$ = $2\pi/c$ planes.

%%%%%%%%%%%%%%%%%%%%%%%%%%%%%%%
\begin{figure*}[!tp]
	%\begin{center}
	\includegraphics[width=0.88\linewidth]{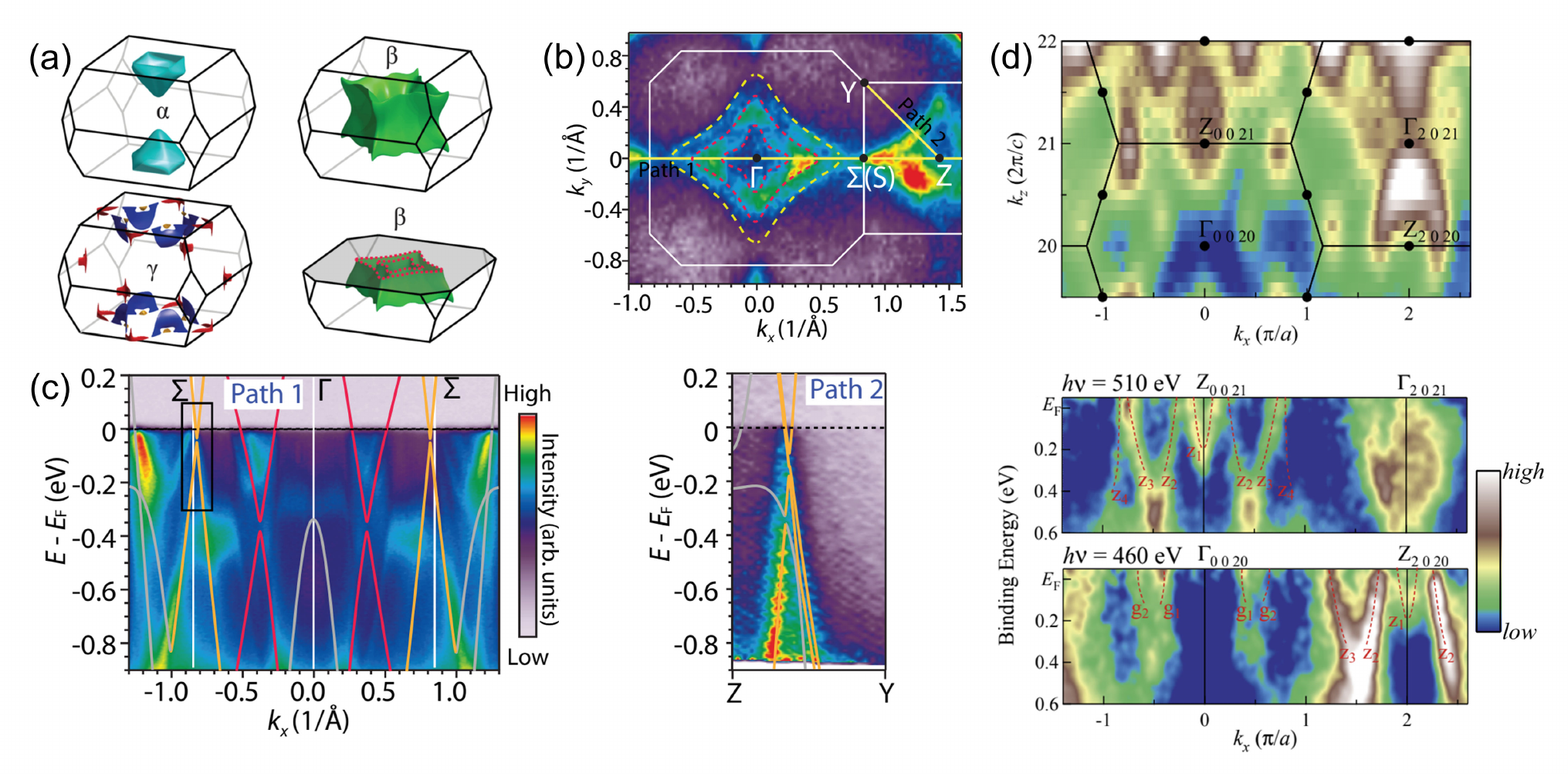}
	%\end{center}
	\centering
	%\vspace{-2ex}%
	\caption{\label{fig:ARPES}\tcr{Electronic structure of EuGa$_4$ in the PM phase measured by ARPES. (a) Three groups of FS pockets: $\alpha$, $\beta$, and $\gamma$, based on DFT calculations. (b) Fermi surfaces with $h\nu$ = 118 eV at $T$ = 25\,K. Two high-symmetry $k$-paths (yellow lines) are indicated for band dispersion analysis. 
	Solid lines represent the calculated band structures. Red and orange lines indicate the bands that form the NR1 and NR2, respectively. 
	(c) Band dispersion along path 1 and path 2 with $h\nu$ =\,120 eV. 
	(d) Fermi surface mappings and band structures of EuAl$_4$ in the vicinity of $E_\mathrm{F}$. %The dashed lines are a guide for the eyes.
	%The considerable dependence along $k_z$ suggest the three-dimensionality of the Fermi surfaces.
	Figures were reproduced from Ref.~\cite{lei2023,Kobata2016}.}}
\end{figure*}
%%%%%%%%%%%%%%%%%%%%%%%%%%%%%

Lei \emph{et al.}\ performed angle-resolved photoemission spectroscopy measurements to identify the spinless NR states in the PM state of EuGa$_4$,
which indeed confirmed the predicted spinless NRs [figure~\ref{fig:ARPES}]~\cite{lei2023}. Three groups of Fermi-surface (FS) pockets derived from the NR bands, labeled $\alpha$, $\beta$, and $\gamma$, are summarized in figure~\ref{fig:ARPES}(a). The FS cross-section measured at 25\,K (PM state),
with a photon energy roughly corresponding to the $k_z = 0$ plane,
is shown in figure~\ref{fig:ARPES}(b).
The inner- and outer portions of the $\beta$ pocket manifest themselves as two concentric diamond rings centered around $\Gamma$ (red dashed curves), while the outward warping geometry of the $\beta$ pocket along $k_z$ renders the intensity finite even outside the outer diamond (up to the yellow boundary) [figure~\ref{fig:ARPES}(b)]. The NR2, with the overlaid orange lines showing its DFT calculated structure, has its node very close to $E_\mathrm{F}$ [figure~\ref{fig:ARPES}(c)]. The suppressed spectral weight near $E_\mathrm{F}$ points to the existence of a small SOC-induced gap. The fact that the node of NR2 sits close to $E_F$ along both cuts (path 1 and path 2) indicates the low dispersion of NR2 along the ring. These observations are consistent with the theoretical
predictions for the PM state of EuGa$_4$.

\begin{figure*}[!tp]
	%\begin{center}
	\includegraphics[width=0.95\linewidth]{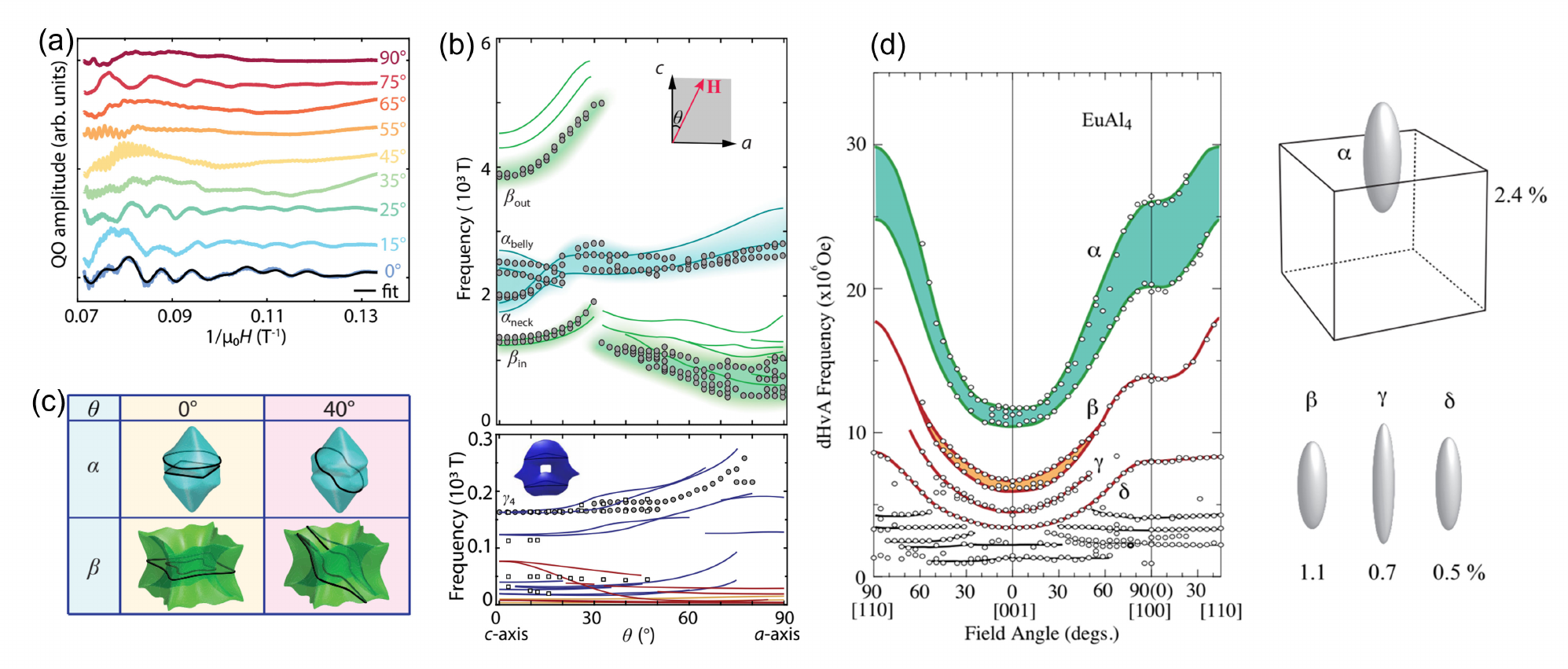}
	%\end{center}
	\centering
	\vspace{-2ex}%
	\caption{\label{fig:QO}\tcr{Fermi-surface geometry of EuGa$_4$ in the SP phase built from QO measurements. (a) A series of QO curves with $\theta$ ranging from 0$\degree$ to 90$\degree$. The solid line is a Lifshitz–Kosevich fit. (b) Angle-dependent QO frequencies (circles) above and below 300\,T. The cyan and green lines in the upper panel are theoretical predictions for the $\alpha$ and $\beta$ pockets, respectively. The red, blue, and orange lines in the lower panel represent the theoretical prediction for the $\gamma$ pockets. 
			The inset in the lower panel illustrates the extremal cyclotron orbits associated with the measured $\gamma_4$ frequency at $\theta$ = 0$\degree$.
			(c) Illustration of the extremal orbits (black lines) for the $\alpha$ and $\beta$ pockets with $\theta$ = 0$\degree$ and 40$\degree$. (d) Angular dependence of the QO frequency in EuAl$_4$ and the corresponding Fermi surfaces.
			Figures were reproduced from Ref.~\cite{lei2023,Nakamura2014}.}}
\end{figure*}

Quantum-oscillation measurements provide more quantitative information about the band splitting and the energy of the Weyl NR states. 
In the SP phase, there are also three groups of FS pockets, whose shapes are similar to those in the PM state, albeit they now appear in pairs due to the band splitting. The Shubnikov-de Haas oscillations, with the magnetic field rotating from the $c$- to the $a$-axis, are shown in figure~\ref{fig:QO}(a), while the field-angle dependence of the corresponding frequencies is
summarized in figure~\ref{fig:QO}(b).
At high frequencies, for $H \parallel c$ (i.e., $\theta$= 0$\degree$), four pairs of QO frequencies, labeled as $\beta_\mathrm{in}$, $\beta_\mathrm{out}$, $\alpha_\mathrm{neck}$, and $\alpha_\mathrm{belly}$) can be identified. The two close-lying frequencies are attributed to the spin-split bands. The field-angle dependence of the frequencies is consistent with the expected evolution of the extremal orbits on the $\alpha$ and $\beta$ FS pockets [figure~\ref{fig:QO}(c)]. Specifically, upon increasing the field angle $\theta$, the $\alpha$ frequencies merge at $\theta \sim 20 \degree$, while the $\beta$ frequencies increase first, and then show a sudden drop at $\theta \sim 30 \degree$. The $\alpha$ frequencies at small $\theta$ can be attributed to slight corrugations of the $\alpha$ pockets along the vertical axis, while the $\beta_\mathrm{in}$ and $\beta_\mathrm{out}$ frequencies are associated with the inner and outer cross-sectional areas of the torus-shaped $\beta$ pockets. The sudden drop of the $\beta$ frequencies with increasing $\theta$ beyond a critical value corresponds to a change of the extremal cross-section from the in-and-out to the sidewise pair [figure~\ref{fig:QO}(c)]. The energy of the bands that give rise to the $\beta_\mathrm{in}$ oscillations shows an excellent match with the calculated value, while the calculated energies for the $\beta_\mathrm{out}$, $\alpha_\mathrm{neck}$, and $\alpha_\mathrm{belly}$ oscillations are slightly off, allowing quantitative corrections to the calculated band structures.

The $\gamma$ pockets are composed of a series of side-by-side electron and hole pockets along the red/blue Weyl NRs. The smallness of these pockets, ascribed to the closeness of the energy of the nodes to $E_\mathrm{F}$, guarantees the low frequencies of QOs [figure~\ref{fig:QO}(a)]. Of the four obtained frequency components, $\gamma_1$ to $\gamma_4$, $\gamma_4$ is identified as the signature of the blue pocket [see inset in figure~\ref{fig:QO}(b)]. The overall $\theta$ dependence of $\gamma_4$---remaining nearly constant with increasing $\theta$ and gradually bifurcating into two branches, and eventually merging into one observable frequency---is consistent with the DFT calculations.
From the temperature-dependent QO amplitude, the effective masses of the $\gamma_4$, $\gamma_3$, and $\gamma_2$ components are estimated to be 0.7$m_e$, where $m_e$ is the free-electron mass, much higher than those of typical nonmagnetic topological semimetals, implying a sizable electronic correlation in EuGa$_4$. Furthermore, the corresponding quantum mobilities are estimated to be
$\sim 10^3$\,cm$^2$/Vs, the highest in all known magnetic topological semimetals.

%%%%%%%%%%%%%%%%%%%%%%%%%%%%%%%
%%%%%%%%%%%%%%%%%%%%%%%%%%%%%%%%%%

We also briefly mention that, prior to the work dis\-cus\-sed above, the electronic structures of EuAl$_4$ and EuGa$_4$ were
investigated by Nakamura \emph{et al.}\  and Kobata \emph{et al.}\ using QO and ARPES~\cite{Nakamura2014,Kobata2016,Nakamura2013}, with the aim to address the nesting conditions for the CDW order. \tcr{The main results regarding EuAl$_4$
are shown in figure~\ref{fig:ARPES}(d) and \ref{fig:QO}(d). The spectra show a considerable dependence along the $k_z$ direction, suggesting that the Fermi surfaces of EuAl$_4$ have three-dimensional shapes~\cite{Kobata2016}. Some dispersive features exist in the vicinity of $E_\mathrm{F}$, but their details are not well resolved in the present ARPES spectra. Moreover, the Fermi
surface of EuAl$_4$ exhibits ellipsoidal features, which are completely
different from those of EuGa$_4$ [see figure~\ref{fig:QO}(d)]~\cite{Nakamura2014}.}
In EuGa$_4$, the Fermi surfaces determined by QO consist of a small ellipsoidal hole FS and a compensated cube-like electron FS, with a vacant space in the center. These Fermi surfaces were argued to favor a CDW order, although the CDW order appears in EuGa$_4$ only under pressure. By contrast, the CDW order is present at ambient pressure in EuAl$_4$, the cube-like FS is absent at low temperatures due to FS reconstruction. It is worth mentioning that the effective masses obtained from QO measurements in these works are also in the order of tenths of the free-electron mass. The electronic structure of EuAl$_4$ in the vicinity of $E_\mathrm{F}$ was determined by ARPES to be very similar to that of its nonmagnetic counterpart SrAl$_4$. The 3D nature of the electronic structure of EuAl$_4$ was argued to disprove a simple FS nesting scenario as the origin of the CDW order. 
		
\section{Summary and outlook} 

In this review we discussed recent experimental investigations on BaAl$_4$-type Eu(Al,Ga)$_4$ topological antiferromagnets. Eu(Al,Ga)$_4$ compounds
represent one of the rare classes, which exhibit exotic physical
properties originating from both real- and momentum-space topological aspects. The electron- and charge orders coexist and compete, resulting in exotic and rich physical properties, e.g., superconductivity, topological Hall effect, giant MR, magnetic skyrmions, Weyl nodal rings, etc. Despite the numerous 
efforts aimed at understanding their origin, there are still some interesting open questions.  

SrAl$_4$ undergoes a CDW and a structural phase transition at $T_\mathrm{CDW}$ $\sim$ 243 and $T_\mathrm{S}$ $\sim$ 90\,K, respectively. Both transitions are suppressed upon substituting Al with Si, and SC emerges below 3\,K in SrAl$_{4-x}$Si$_{x}$ ($x \ge 1.5$). Although such superconductors have been investigated by magnetic-susceptibility and heat-capacity measurements, their properties at a microscopic level, in particular, the superconducting order parameter, require further investigation. 
Besides the CDW transition at $T_\mathrm{CDW} \sim 142$\,K, EuAl$_4$ exhibits multiple AFM orders below $T_\mathrm{N}$ $\sim$ 15.6\,K, and undergoes a series of metamagnetic transitions in the AFM state.
By applying pressure, the CDW order is linearly suppressed in both SrAl$_4$ and EuAl$_4$, but the SC is absent near the critical pressure. Moreover, the substitution of Al with Ga also suppresses the CDW order in Eu(Al$_{1-x}$Ga$_{x}$)$_4$, but the AFM order remains robust against chemical pressure. These observations suggest that the charge- and AFM order compete with SC, and that
carrier doping
might be one of the keys to introduce SC in the BaAl$_4$-type materials.
Electronic band-structure calculations of the electron-phonon coupling and the density of states could be useful to address this issue. In view of this, Eu$_{1-x}$Ba$_x$Al$_4$ and Eu$_{1-x}$Ba$_x$Ga$_4$ represent ideal systems to investigate the competing electronic orders and the quantum critical phenomena, and to search for possible unconventional SC. 
Furthermore, from the topological point of view, the non-trivial band topology 
of BaAl$_4$ and Eu(Al,Ga)$_4$ suggests that the BaAl$_4$-type materials also represent one of the ideal candidates to search for 
topological SC. 

In EuAl$_4$, the observed topological Hall effect is attributed to the formation of a skyrmion lattice in an applied magnetic field along the $c$-axis ($H \parallel c$), as confirmed by SANS measurements. Similar to EuAl$_4$, also Eu(Al$_{1-x}$Ga$_x$)$_4$ ($0.1 \le x \le 0.5$) and EuGa$_4$ single crystals exhibit topological Hall resistivity $\rho_\mathrm{xy}^\mathrm{THE}$ in a field range whose extension depends on Ga content. Though the same mechanism is expected to account for the THE in these crystals, the topological spin textures and their evolution with Ga content are not yet known.
According to the Hall resistivity measured in an applied magnetic field perpendicular to the $c$-axis ($H \parallel ab$) (section~\ref{sec:trsnsport}), the signature of topological contribution is clearly visible in $\rho_\mathrm{xy}(H)$. However, due to significant multiband effects, it is rather difficult to isolate the $\rho_\mathrm{xy}^\mathrm{THE}$ term. As an alternative approach, it could be interesting to check if the topological spin textures also form in an
applied magnetic field perpendicular to the $c$-axis. Currently, neither neutron- nor resonant x-ray scattering studies with $H \parallel ab$ are available. 
The magnitude of $\rho_\mathrm{xy}^\mathrm{THE}$ shows a clear dependence on the Ga content, and is almost twice larger in EuAl$_4$ than in EuGa$_4$. The $\rho_\mathrm{xy}^\mathrm{THE}$ is believed to be determined mostly by the spin polarization, the Hall coefficient, and skyrmion density. Hall-resistivity measurements reveal comparable Hall coefficients in both EuAl$_4$ and EuGa$_4$. Considering that both compounds share similar electronic band structures, the spin polarization should be comparable as well. Therefore, the difference in $\rho_\mathrm{xy}^\mathrm{THE}$ should most likely come from a different skyrmion density. To confirm this, further investigations in determining the skyrmion lattice in Eu(Al$_{1-x}$Ga$_x$)$_4$ are crucial. For instance, one can use real-space imaging techniques, including the Lorentz transmission electron microscopy, or the magnetic force microscopy to estimate the skyrmion density. 

According to recent REXS and CD XRMS studies (section~\ref{sec:neutron}), magnetic helix structures form in both phase-III ($T_\mathrm{N2} < T < T_\mathrm{N3}$) and phase-IV ($T < T_\mathrm{N4}$) in a zero-field condition.
However, the chirality of the spin helix is reversed in these phases. 
In magnetic materials, the observed THE is mostly related to a Berry-phase ac\-cu\-mu\-la\-tion in real space due to chiral spin textures. When passing through chiral spin textures, charge carriers pick up an additional Berry phase and experience a local emergent magnetic field $B_\mathrm{eff}$ that is proportional to the scalar spin chirality. As a consequence, the sign of  $\rho_\mathrm{xy}^\mathrm{THE}$ is determined by $B_\mathrm{eff}$ and, thus, by the spin chirality. In principle, the reversal of spin chirality in EuAl$_4$ would lead to a sign change in $\rho_\mathrm{xy}^\mathrm{THE}(H)$. Yet, according to Hall-resistivity measurements, the $\rho_\mathrm{xy}^\mathrm{THE}$ of EuAl$_4$ is always positive (section~\ref{sec:trsnsport}) in the studied temperature- and field range, for both $H \parallel ab$ and $H \parallel c$. Such an observation might suggest that the external magnetic field can define the spin chirality of the magnetic helix structure in EuAl$_4$. It could be interesting to extend the REXS measurements to higher magnetic fields and check the spin chirality. Considering an EuGa$_4$ single crystal, if $\rho_\mathrm{xy}^\mathrm{THE}$ can be isolated from the measured $\rho_\mathrm{xy}(H)$ for $H \parallel ab$, its sign should be opposite to that of $\rho_\mathrm{xy}^\mathrm{THE}$ for $H \parallel c$. Therefore, the identification of spin textures in EuGa$_4$ in an applied magnetic field (currently still missing) is also highly desirable.

The synchrotron x-ray diffraction and optical spectroscopy reveal a
sizable coupling among the lattice, magnetic order, and charge order in Eu(Al,Ga)$_4$.
It undergoes a tetragonal-to-orthorhombic distortion at $T_\mathrm{N3}$ $\sim$ 12.6\,K, where the magnetic structure of EuAl$_4$ changes significantly, from a SDW to a helix structure. At $T < T_\mathrm{N3}$, the unit cell is compressed and elongated along the $a$- and $b$-axis, respectively. As a consequence, the magnetic modulation vectors are parallel to the elongated $b$-axis, implying 
sizable magnetostriction and magnetovolume effects in EuAl$_4$, 
confirmed by field-dependent ther\-mal-ex\-pan\-sion measurements. The skyrmion phases, too, are found to pos\-sess
an orthorhombic structural distortion within the $ab$ plane. On the other hand, the incommensurate wave vector of CDW order $\boldsymbol{q}_\mathrm{CDW}$ = (0,0,$\delta$), appears below $T_\mathrm{CDW} = 145$\,K. The $\delta$ value decreases monotonically with decreasing temperature, from $T_\mathrm{CDW}$ to $T_\mathrm{N}$. Since it exhibits distinct anomalies at $T_\mathrm{N}$s, this implies an interplay between the magnetic- and charge order in EuAl$_4$. The REXS data also reveal that, in EuAl$_2$Ga$_2$, the intensity of the CDW order is significantly reduced below $T_\mathrm{N1}$. Thus, the formation of the AFM order suppresses the CDW order, demonstrating a strong interplay and competition between these electronic orders in Eu(Al,Ga)$_4$.  
Optical spectroscopy found that the CDW gap is partially opened on the Fermi surface, the AFM interactions mediated by the itinerant electrons should be anisotropic. In such an anisotropic environment, it is possible for itinerant electrons from the Weyl bands to mediate the anisotropic magnetic interactions that facilitate the formation of the chiral spin textures. All these observations imply couplings between the lattice-, charge-, and magnetic orders in the Eu(Al,Ga)$_4$ family. However,
the evolution of topological spin textures with structural distortion or charge order is not yet well understood,
requiring further investigations.

To conclude, recent experimental and theoretical studies of the Eu(Al,Ga)$_4$ family, establish it as a suitable platform to explore the interplay between the lattice-, charge-, and spin degrees of freedom,
and the associated emergent phenomena. Such interplay seems to underlie
the structural and electronic instabilities occurring in this family of compounds
and yields a rich variety of topological spin textures, as well as the potential to control the skyrmion phases and topological transport properties by uniaxial stress or pressure, epitaxial strain, electric and magnetic fields. These numerous appealing possibilities are open to future investigation.

\section{Acknowledgements}	
We thank
Junzhang Ma and Sailong Ju for fruitful discussions.
This work was supported by the National Natural Science foundation of China (NSFC) (Grant Nos. 12374105, 12274125, 12174103, 11874150, and 12350710785), the Natural Science Foundation of Shang\-hai (Grant Nos.\ 21ZR1420500 and 21JC1402300), the Natural Science Foundation of Chongqing (Grant No.2022NSCQ-MSX1468), and the Fundamental Research Funds for the Central Universities. This work was also financially supported by the Schweizerische Nationalfonds zur F\"{o}rderung der Wis\-sen\-schaft\-lichen For\-schung (SNF) (Grant Nos.\ 200021\_188706 and 206021\_139082).
\section{References}
	
\bibliography{EuA4_rev}
	
\end{document}